\documentclass[11pt]{article}
\usepackage{graphicx,ifthen,color,algorithm,palatino}
\usepackage{amsmath,amsthm,amssymb,amsfonts,verbatim}
\usepackage{mathrsfs,accents,setspace,soul}
\def\myand{\&\ }
\def\simind{\stackrel{{\tiny \mbox{ind.}}}{\sim}}
\def\sigeps{\sigma_{\varepsilon}}
\def\bzero{\boldsymbol{0}}
\def\bone{\boldsymbol{1}}
\def\trans{^T}
\def\bomega{\boldsymbol{\omega}}
\def\bOmega{\boldsymbol{\Omega}}
\def\bSigma{\boldsymbol{\Sigma}}
\def\bu{\boldsymbol{u}}
\def\bv{\boldsymbol{v}}

\def\bx{\boldsymbol{x}}
\def\by{\boldsymbol{y}}
\def\bX{\boldsymbol{X}}

\def\bZ{\boldsymbol{Z}}
\def\bA{\boldsymbol{A}}
\def\bB{\boldsymbol{B}}
\def\bC{\boldsymbol{C}}
\def\bD{\boldsymbol{D}}

\def\bG{\boldsymbol{G}}

\def\bI{\boldsymbol{I}}

\def\bM{\boldsymbol{M}}
\def\bO{\boldsymbol{O}}

\def\bQ{\boldsymbol{Q}}
\def\bR{\boldsymbol{R}}
\def\bbeta{\boldsymbol{\beta}}
\def\blockdiagdum{\mathop{\mbox{\rm blockdiag}}}
\def\blockdiag#1{\blockdiagdum_{#1}}
\def\stackdum{\mathop{\mbox{\rm stack}}}
\def\stack#1{\stackdum_{#1}}
\def\Cov{\mbox{\rm Cov}}
\def\bbetahat{{\widehat\bbeta}}
\def\buhat{{\widehat \bu}}

\def\bb{\boldsymbol{b}}
\def\bc{\boldsymbol{c}}
\def\bd{\boldsymbol{d}}

\def\bo{\boldsymbol{o}}
\def\tr{\mbox{tr}}
\def\biota{\boldsymbol{\iota}}
\def\smhalf{{\textstyle{\frac{1}{2}}}}
\def\infint{\int_{-\infty}^{\infty}}
\def\quarter{{\textstyle{1\over4}}}
\def\bib{\vskip12pt\par\noindent\hangindent=1 true cm\hangafter=1}

\def\bmu{\boldsymbol{\mu}}
\def\thickarrow{\longleftarrow}
\def\Ssc{{\mathcal S}}
\def\sigsqeps{\sigeps^2}
\def\bLambda{\boldsymbol{\Lambda}}
\def\diag{\mbox{diag}}

\def\smalldot{\mbox{\fontsize{0.1mm}{0.5em}\selectfont{$\bullet$}}}
\def\bBdot{\overset{\ \smalldot}{\bB}}

\def\bBdotdot{\overset{\smalldot\smalldot}{\bB}}

\def\bDdot{\overset{\ \smalldot}{\bD}}

\def\AtLev{\bA}
     \def\AUoo{\bA^{11}}
 \def\AUotCo{\bA^{12,1}} \def\AUotCoT{\bA^{12,1\,T}}
 \def\AUotCt{\bA^{12,2}} \def\AUotCtT{\bA^{12,2\,T}}
	\def\AUotCi{\bA^{12,i}} 
	\def\AUotCm{\bA^{12,m}} \def\AUotCmT{\bA^{12,m\,T}}
\def\AUotCij{\bA^{12,ij}}
\def\AUotCicommaj{\bA^{12,i,j}}
	\def\AUttCo{\bA^{22,1}}
	\def\AUttCt{\bA^{22,2}}
	\def\AUttCi{\bA^{22,i}}
	\def\AUttCm{\bA^{22,m}}
\def\AUttCij{\bA^{22,ij}}
\def\xveco{\bx_1}

\def\xvectCi{\bx_{2,i}}
\def\xvectCij{\bx_{2,ij}}

\def\bveco{\bb_1}
\def\bvect{\bb_2}
\def\bveci{\bb_i}
\def\bvecm{\bb_m}
\def\bvecij{\bb_{ij}}
\def\Bmato{\bB_1}
\def\Bmatt{\bB_2}
\def\Bmati{\bB_i}
\def\Bmatm{\bB_m}
\def\Bmatij{\bB_{ij}}
\def\Bmatdoto{\bBdot_1}
\def\Bmatdott{\bBdot_2}
\def\Bmatdoti{\bBdot_i}
\def\Bmatdotm{\bBdot_m}
\def\Bmatdotij{\bBdot_{ij}}
\def\Bmatdotdotij{\bBdotdot_{ij}}
\def\cveczi{\bc_{0i}}
\def\cvecoi{\bc_{1i}}
\def\cvecti{\bc_{2i}}
\def\Cmatzi{\bC_{0i}}
\def\Cmatoi{\bC_{1i}}
\def\Cmatti{\bC_{2i}}
\def\sumim{\sum_{i=1}^m}

\def\bigX{{\LARGE\mbox{$\times$}}}
\def\nadj{{\tilde n}}
\def\oadj{{\tilde o}}
\newboolean{ShowFigures}
\newboolean{UnBlinded}
\newboolean{DoubleSpaced}
\setboolean{ShowFigures}{true}
\setboolean{UnBlinded}{true}
\setboolean{DoubleSpaced}{false}
\newtheorem{result}{\textbf{Result}}
\makeatletter
\newcommand{\vast}{\bBigg@{4}}
\newcommand{\Vast}{\bBigg@{5}}
\makeatother
\def\iCOMMAj{\,i,\,j}
\def\sSigmaOne{s_{\mbox{\rm\tiny{$\bSigma,1$}}}}
\def\sSigmaj{s_{\mbox{\rm\tiny{$\bSigma,j$}}}}
\def\sSigmaTwo{s_{\mbox{\rm\tiny{$\bSigma,2$}}}}
\def\sSigmagOne{s_{\mbox{\rm\tiny{$\bSigmag,1$}}}}
\def\sSigmahOne{s_{\mbox{\rm\tiny{$\bSigmah,1$}}}}
\def\sSigmagTwo{s_{\mbox{\rm\tiny{$\bSigmag,2$}}}}
\def\sSigmahTwo{s_{\mbox{\rm\tiny{$\bSigmah,2$}}}}

\newcommand{\Ainv}[1]{\bA^{#1}}
\newcommand{\ATinv}[1]{\bA^{#1\,T}}
\newcommand*{\dt}[1]{\accentset{\mbox{\smalldot}}{#1}}
\newcommand*{\ddt}[1]{\accentset{\mbox{\smalldot\smalldot}}{#1}}
\newcommand{\B}[1]{\bB_{#1}}

\newcommand{\dB}[1]{\dt{\pmb{B}}_{#1}}
\newcommand{\ddB}[1]{\ddt{\pmb{B}}_{#1}}

\def\OmegaAtwoTwo{\bOmega_4}

\def\Gfull{G_{\mbox{\rm\tiny full}}}
\def\Gdiag{G_{\mbox{\rm\tiny diag}}}
\def\ASigma{\bA_{\mbox{\rm\tiny$\bSigma$}}}
\def\ASigmag{\bA_{\mbox{\rm\tiny$\bSigmag$}}}
\def\ASigmah{\bA_{\mbox{\rm\tiny$\bSigmah$}}}
\def\nuSigma{\nu_{\mbox{\rm\tiny{$\bSigma$}}}}
\def\nuSigmag{\nu_{\mbox{\rm\tiny{$\bSigmag$}}}}
\def\nuSigmah{\nu_{\mbox{\rm\tiny{$\bSigmah$}}}}
\def\nuEps{\nu_{\varepsilon}}
\def\nuGbl{\nu_{\mbox{\rm\tiny gbl}}}
\def\nuGrp{\nu_{\mbox{\rm\tiny grp}}}
\def\nuGrpg{\nu_{\mbox{\rm\tiny grp, $g$}}}
\def\nuGrph{\nu_{\mbox{\rm\tiny grp, $h$}}}
\def\sEps{s_{\varepsilon}}
\def\sGbl{s_{\mbox{\rm\tiny gbl}}}
\def\sGrp{s_{\mbox{\rm\tiny grp}}}
\def\sGrpg{s_{\mbox{\rm\tiny grp, $g$}}}
\def\sGrph{s_{\mbox{\rm\tiny grp, $h$}}}
\def\qDens{\mathfrak{q}}
\def\pDens{\mathfrak{p}}
\def\pDensUnder{\underline{\pDens}}
\def\MqASigma{\bM_{\qDens(\bA_{\bSigma}^{-1})}}
\def\uLiniz{u_{\mbox{\rm\tiny lin,$i0$}}}
\def\uLinio{u_{\mbox{\rm\tiny lin,$i1$}}}
\def\uLinijz{u_{\mbox{\rm\tiny lin,$ij0$}}}
\def\uLinijo{u_{\mbox{\rm\tiny lin,$ij1$}}}
\def\uGblk{u_{\mbox{\rm\tiny gbl,$k$}}}
\def\uGblone{u_{\mbox{\rm\tiny gbl,$1$}}}
\def\uGblKGbl{u_{\mbox{\rm\tiny gbl,$K_{\mbox{\rm\tiny gbl}}$}}}
\def\uGrpik{u_{\mbox{\rm\tiny grp,$ik$}}}
\def\uGrpijk{u_{\mbox{\rm\tiny grp,$ijk$}}}
\def\buGbl{\bu_{\mbox{\rm\tiny gbl}}}
\def\buGbli{\bu_{\mbox{\rm\tiny gbl,$i$}}}

\def\buGrpi{\bu_{\mbox{\rm\tiny grp,$i$}}}
\def\uGrpione{u_{\mbox{\rm\tiny grp,$i1$}}}
\def\uGrpiKGrp{u_{\mbox{\rm\tiny grp,$iK_{\mbox{\rm\tiny grp}}$}}}

\def\buLini{\bu_{\mbox{\rm\tiny lin,$i$}}}

\def\buHatLini{\buhat_{\mbox{\rm\tiny lin,$i$}}}

\def\buHatGbl{\buhat_{\mbox{\rm\tiny gbl}}}

\def\buHatGrpi{\buhat_{\mbox{\rm\tiny grp,$i$}}}

\def\bZgbl{\bZ_{\mbox{\rm\tiny gbl}}}
\def\bZgblo{\bZ_{\mbox{\rm\tiny gbl,$1$}}}
\def\bZgbli{\bZ_{\mbox{\rm\tiny gbl,$i$}}}
\def\bZgblm{\bZ_{\mbox{\rm\tiny gbl,$m$}}}

\def\bCgbli{\bC_{\mbox{\rm\tiny gbl,$i$}}}
\def\bCgblij{\bC_{\mbox{\rm\tiny gbl,$ij$}}}

\def\bCgrpi{\bC_{\mbox{\rm\tiny grp,$i$}}}

\def\bZgrpi{\bZ_{\mbox{\rm\tiny grp,$i$}}}

\def\Kgbl{K_{\mbox{\rm\tiny gbl}}}
\def\Kgrp{K_{\mbox{\rm\tiny grp}}}

\def\zgblo{z_{\mbox{\rm\tiny gbl,$1$}}}
\def\zgblk{z_{\mbox{\rm\tiny gbl,$k$}}}
\def\zgblKgbl{z_{\mbox{\rm\tiny gbl,$\Kgbl$}}}

\def\zgrpo{z_{\mbox{\rm\tiny grp,1}}}
\def\zgrpk{z_{\mbox{\rm\tiny grp,k}}}
\def\zgrpKgrp{z_{\mbox{\rm\tiny grp,$\Kgrp$}}}
\def\sigmaGbl{\sigma_{\mbox{\rm\tiny gbl}}}
\def\sigmaGrp{\sigma_{\mbox{\rm\tiny grp}}}
\def\sigmaGrpg{\sigma_{\mbox{\rm\tiny grp,$g$}}}
\def\sigmaGrph{\sigma_{\mbox{\rm\tiny grp,$h$}}}
\def\aGbl{a_{\mbox{\rm\tiny gbl}}}
\def\aGrp{a_{\mbox{\rm\tiny grp}}}
\def\aGrpg{a_{\mbox{\rm\tiny grp, $g$}}}
\def\aGrph{a_{\mbox{\rm\tiny grp, $h$}}}

\def\SolveTwoLevelSparseLeastSquares{\textsc{\footnotesize SolveTwoLevelSparseLeastSquares}}

\def\SolveThreeLevelSparseLeastSquares{\textsc{\footnotesize SolveThreeLevelSparseLeastSquares}}

\def\bSigmag{\bSigma_{\mbox{\rm\tiny $g$}}}
\def\bSigmah{\bSigma_{\mbox{\rm\tiny $h$}}}
\def\buLone{\bu^{\mbox{\rm\tiny $g$}}}
\def\buLtwo{\bu^{\mbox{\rm\tiny $h$}}}
\def\bZLone{\bZ^{\mbox{\rm\tiny $g$}}}
\def\bZLtwo{\bZ^{\mbox{\rm\tiny $h$}}}
\def\bCLone{\bC^{\mbox{\rm\tiny $g$}}}
\def\bCLtwo{\bC^{\mbox{\rm\tiny $h$}}}
\def\buHatLone{\buhat^{\mbox{\rm\tiny $g$}}}
\def\buHatLtwo{\buhat^{\mbox{\rm\tiny $h$}}}
\def\veryTinyBLUP{\mbox{\fontsize{1.5mm}{1em}\selectfont {\bf BLUP}}}
\def\veryTinyMFVB{\mbox{\fontsize{1.5mm}{1em}\selectfont {\bf MFVB}}}
\def\DBLUP{\bD_{\veryTinyBLUP}}
\def\DMFVB{\bD_{\veryTinyMFVB}}
\def\RBLUP{\bR_{\veryTinyBLUP}}
\def\RMFVB{\bR_{\veryTinyMFVB}}
\def\oMFVB{\bo_{\veryTinyMFVB}}
\def\aeps{a_{\varepsilon}}

\def\bASigma{\bA_{\mbox{\rm\tiny$\bSigma$}}}

\def\muq#1{\mu_{\qDens(#1)}}
\def\bmuq#1{\bmu_{\qDens(#1)}}
\def\MqSigma{\bM_{\qDens(\bSigma^{-1})}}
\def\MqSigmag{\bM_{\qDens\left((\bSigmag)^{-1}\right)}}
\def\MqSigmah{\bM_{\qDens\left((\bSigmah)^{-1}\right)}}

\def\buLoneLini{\buLone_{\mbox{\rm\tiny lin,$i$}}}

\def\buLoneGrpi{\buLone_{\mbox{\rm\tiny grp,$i$}}}

\def\buHatLoneLini{\buHatLone_{\mbox{\rm\tiny lin,$i$}}}

\def\buHatLoneGrpi{\buHatLone_{\mbox{\rm\tiny grp,$i$}}}

\def\buLtwoLinio{\buLtwo_{\mbox{\rm\tiny lin,$i$1}}}
\def\buLtwoLinini{\buLtwo_{\mbox{\rm\tiny lin,$in_i$}}}
\def\buLtwoLinij{\buLtwo_{\mbox{\rm\tiny lin,$ij$}}}
\def\buLtwoGrpij{\buLtwo_{\mbox{\rm\tiny grp,$ij$}}}

\def\buHatLtwoLinij{\buHatLtwo_{\mbox{\rm\tiny lin,$ij$}}}
\def\buHatLtwoGrpij{\buHatLtwo_{\mbox{\rm\tiny grp,$ij$}}}

\def\buLtwoGrpio{\buLtwo_{\mbox{\rm\tiny grp,$i1$}}}
\def\buLtwoGrpini{\buLtwo_{\mbox{\rm\tiny grp,$in_i$}}}
\def\bZgblij{\bZ_{\mbox{\rm\tiny gbl,$ij$}}}
\def\bZLoneGrpij{\bZLone_{\mbox{\rm\tiny grp,$ij$}}}
\def\bZLtwoGrpij{\bZLtwo_{\mbox{\rm\tiny grp,$ij$}}}
\def\bCLoneGrpij{\bCLone_{\mbox{\rm\tiny grp,$ij$}}}
\def\bCLtwoGrpij{\bCLtwo_{\mbox{\rm\tiny grp,$ij$}}}
\def\Kgrpg{K_{\mbox{\rm\tiny grp}}^{\mbox{\rm\tiny $g$}}}
\def\Kgrph{K_{\mbox{\rm\tiny grp}}^{\mbox{\rm\tiny $h$}}}
\def\zLoneGrpo{z_{\mbox{\rm\tiny grp,$1$}}^{\mbox{\rm\tiny $g$}}}
\def\zLoneGrpK{z_{\mbox{\rm\tiny grp,$\Kgrpg$}}^{\mbox{\rm\tiny $g$}}}
\def\zLtwoGrpo{z_{\mbox{\rm\tiny grp,$1$}}^{\mbox{\rm\tiny $h$}}}
\def\zLtwoGrpK{z_{\mbox{\rm\tiny grp,$\Kgrph$}}^{\mbox{\rm\tiny $h$}}}
\def\ndot{\sum_{i=1}^m n_i}
\def\ndotmh{\left(\ndot\right)^{-1/2}}

\begin{document}
\ifthenelse{\boolean{DoubleSpaced}}{\setstretch{1.5}}{}

\thispagestyle{empty}

\centerline{\Large\bf Streamlined Variational Inference for}
\vskip2mm
\centerline{\Large\bf Higher Level Group-Specific Curve Models}
\vskip7mm
\centerline{\normalsize\sc By M. Menictas$\null^1$, T.H. Nolan$\null^1$,
D.G. Simpson$\null^2$ and M.P. Wand$\null^1$}
\vskip5mm
\centerline{\textit{University of Technology Sydney$\null^1$ and University of Illinois$\null^2$}}
\vskip6mm
\centerline{11th March, 2019}

\vskip6mm
\centerline{\large\bf Abstract}
\vskip2mm

A two-level group-specific curve model is such that the mean
response of each member of a group is a separate smooth function
of a predictor of interest. The three-level extension is such
that one grouping variable is nested within another one,
and higher level extensions are analogous. Streamlined variational
inference for higher level group-specific curve models is a
challenging problem. We confront it by systematically working
through two-level and then three-level cases and making use of
the higher level sparse matrix infrastructure laid down in
Nolan \myand Wand (2018). A motivation is analysis of data from
ultrasound technology for which three-level group-specific curve
models are appropriate. Whilst extension to the number of levels exceeding
three is not covered explicitly, the pattern established by our
systematic approach sheds light on what is required for even 
higher level group-specific curve models.


\section{Introduction}\label{sec:intro}

We provide explicit algorithms for fitting and approximate Bayesian
inference for multilevel models involving, potentially, thousands of noisy
curves. The algorithms include covariance parameter estimation and
allow for pointwise credible intervals around the fitted curves.
Contrast function fitting and inference is also supported by our
approach. Both two-level and three-level situations are covered,
and a template for even higher level situations is laid down.

Models and methodology for statistical analyses of grouped data for which
the basic unit is a noisy curve continues to be an important area of research.
A driving force is rapid technological change which is resulting in the
generation of curve-type data at fine resolution levels. Examples of
such technology include accelerometers (e.g.\ Goldsmith {\it et al.}, 2015)
personal digital assistants (e.g.\ Trail {\it et al.}, 2014)
and quantitative ultrasound (e.g.\ Wirtzfeld {\it et al.}, 2015).
In some applications curve-type data have higher levels of grouping,
with groups at one level nested inside other groups. Our focus here is
streamlined variational inference for such circumstances.

\begin{figure}[!ht]
\centering
{\includegraphics[width=\textwidth]{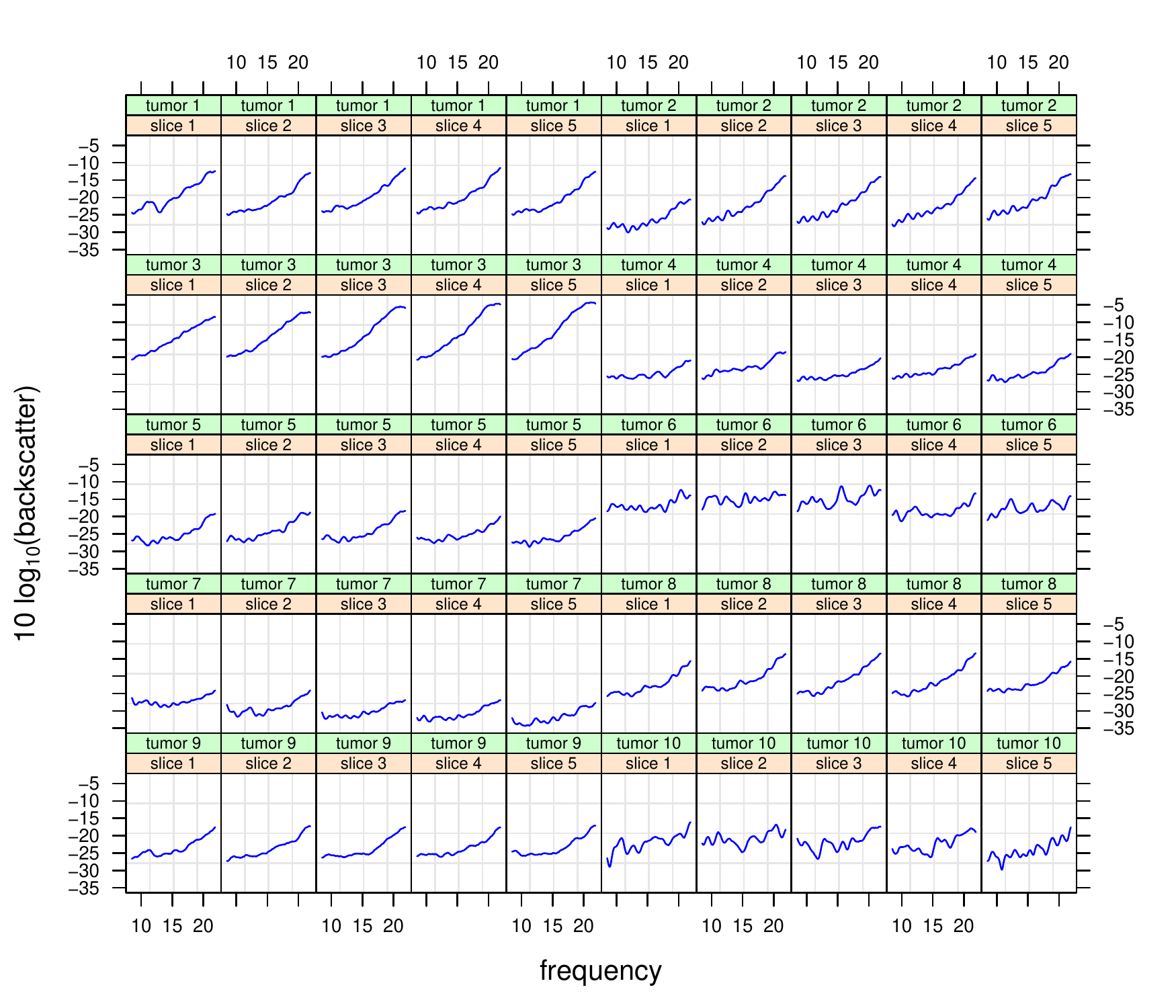}}
\caption{\it Illustrative three-level curve-type data. The response variable is
$10\log_{10}(\mbox{backscatter})$ according to ultrasound technology. Level 1 corresponds
to different ultrasound frequencies and matches the horizontal axes in each panel.
Level 2 corresponds to different slices of a tumor due to differing probe locations. Level 3
corresponds to different tumors with one tumor for each of $10$ laboratory mice.}
\label{fig:MNSWintro}
\end{figure}

Some motivating data is shown in Figure \ref{fig:MNSWintro} from
an experiment involving quantitative ultrasound technology. Each curve
corresponds to a logarithmically transformed backscatter coefficient
over a fine grid of frequency values for tumors in laboratory mice,
with exactly one tumor per mouse. The backscatter/frequency curves
are grouped according to one of 5 slices of the same tumor,
corresponding to probe locations. The slices are grouped according
to being from one of 10 tumors. We refer to such data as three-level
data with frequency measurements at level 1, slices being the 
level 2 groups and tumors constituting the level 3 groups. 
The gist of this article is efficient and flexible variational 
fitting and inference for such data, that scales well to 
much larger multilevel data sets. Indeed, our algorithms 
are linear in the number of groups at both level 2 and level 3. 
Simulation study results given later in this article 
show that curve-type data with thousands of 
groups can be analyzed quickly using our new methodology.
Depending on sample sizes and implementation language,
fitting times range from a few seconds to a few minutes.
In contrast, na\"{\i}ve implementations become infeasible
when the number of groups are in the several hundreds
due to storage and computational demands.

We work with a variant of group-specific curve models that at least
go back to Donnelly, Laird \myand Ware (1995). Other contributions
of this type include Brumback \myand Rice (1998),
Verbyla \textit{et al.} (1999), Wang (1998) and
Zhang \textit{et al.} (1998). The specific formulation that we
use is that given by Durban \textit{et al.} (2005) which involves
an embedding within the class of linear mixed models (e.g. Robinson, 1991)
with low-rank smoothing splines used for flexible function modelling
and fitting.

Even though approximate Bayesian variational inference is our overarching
goal, we also provide an important parallelism involving classical frequentist
inference. Contemporary mixed model software such as
\texttt{nlme()} (Pinheiro \textit{et al.}, 2018) and \texttt{lme4()}
(Bates \textit{et al.}, 2015) in the \textsf{R} language 
provide streamlined algorithms for obtaining the best linear 
unbiased predictions of fixed and random effects in multilevel mixed models 
with details given in, for example, Pinheiro \myand Bates (2000).
However, the sub-blocks of the covariance 
matrices required for construction of pointwise confidence interval 
bands around the estimated curves are \emph{not} provided by such software. 
In the variational Bayesian analog, these sub-blocks are required for 
covariance parameter fitting and inference which, in turn, are needed 
for curve estimation. A significant contribution of this article is 
streamlined computation for both
the best linear unbiased predictors and its corresponding covariance
computation. Similar mathematical results lead to the mean field variational
Bayesian inference equivalent. We present explicit ready-to-code
algorithms for both two-level and three-level group-specific
curve models. Extensions to higher level models could be derived using the
blueprint that we establish here. Nevertheless, the algebraic overhead is
increasingly burdensome with each increment in the number of levels.
It is prudent to treat each multilevel case separately and here we
already require several pages to cover two-level and three-level group-specific
curve models. To our knowledge, this is the first article to provide
streamlined algorithms for fitting three-level group-specific curve models.

Another important aspect of our group-specific curve fitting algorithms is the 
fact that they make use of the \SolveTwoLevelSparseLeastSquares\ and 
\SolveThreeLevelSparseLeastSquares\ algorithms developed for ordinary
linear mixed models in Nolan \textit{et al.} (2018). This realization
means that the algorithms listed in Sections \ref{sec:twoLevMods}
and \ref{sec:threeLevMods} are more concise and code-efficient:
there is no need to repeat the implementation of these two 
fundamental algorithms for stable QR-based solving of higher level
sparse linear systems. Sections \ref{sec:Solve2Lev}--\ref{sec:Solve3Lev} 
of the web-supplement provide details on the 
\SolveThreeLevelSparseLeastSquares\ and
\SolveThreeLevelSparseLeastSquares\ algorithms.

Section \ref{sec:twoLevMods} deals with the two-level case and the
three-level case is covered in Section \ref{sec:threeLevMods}.
In Section \ref{sec:accAndSpeed} we provide some assessments
concerning the accuracy and speed of the new variational
inference algorithms. 

\section{Two-Level Models}\label{sec:twoLevMods}

The simplest version of group-specific curve models involves
the pairs $(x_{ij},y_{ij})$ where $x_{ij}$ is the
$j$th value of the predictor variable within the $i$th group
and $y_{ij}$  is the corresponding value of the response variable.
We let $m$ denote the number of groups and $n_i$ denote the number
of predictor/response pairs within the $i$th group.
The Gaussian response two-level group specific curve model is
\begin{equation}
y_{ij}=f(x_{ij})+g_i(x_{ij})+\varepsilon_i,\quad\varepsilon_{ij}\simind N(0,\sigeps^2),\quad
1\le i\le m,\ 1\le j\le n_i,
\label{eq:twoLevelfg}
\end{equation}
where the smooth function $f$ is the global regression mean function
and the smooth functions $g_i$, $1\le i\le m$, allow for flexible
group-specific deviations from $f$. As in Durban \textit{et al.} (2005), we use
mixed model-based penalized basis functions to model $f$ and the $g_i$.
Specifically,
{\setlength\arraycolsep{1pt}
\begin{eqnarray*}
f(x)&=&\beta_0+\beta_1\,x+\sum_{k=1}^{\Kgbl}\,\uGblk\,\zgblk(x),\quad
\uGblk\simind N(0,\sigmaGbl^2),\ \mbox{and}\\[1ex]
g_i(x)&=&\uLiniz+\uLinio\,x+\sum_{k=1}^{\Kgrp}\,\uGrpik\,\zgrpk(x),
\ \ \left[\begin{array}{c}
\uLiniz\\[1ex]
\uLinio
\end{array}
\right]\simind N(\bzero,\bSigma),\ \
\uGrpik\simind N(0,\sigmaGrp^2),
\end{eqnarray*}
}
where $\{\zgblk(\cdot):1\le k\le \Kgbl\}$ and
$\{\zgrpk(\cdot):1\le k\le \Kgrp\}$ are suitable
sets of basis functions. Splines and wavelet families are
the most common choices for the $\zgblk(\cdot)$ and $\zgrpk(\cdot)$.
In our illustrations and simulation 
studies we use the canonical cubic O'Sullivan spline
basis as described in Section 4 of Wand \myand Ormerod (2008),
which corresponds to a low-rank version of classical smoothing splines
(e.g. Wahba, 1990). The variance parameters $\sigmaGbl^2$ and
$\sigmaGrp^2$ control the effective degrees of freedom used for
the global mean and group-specific deviation functions respectively.
Lastly, $\bSigma$ is a $2\times 2$ unstructured covariance matrix
for the coefficients of the group-specific linear deviations.

We also use the notation:
$$\bx_i\equiv\left[\begin{array}{c}
x_{i1}\\
\vdots\\
x_{in_i}
\end{array}
\right]\quad\mbox{and}\quad
\by_i\equiv\left[\begin{array}{c}
y_{i1}\\
\vdots\\
y_{in_i}
\end{array}
\right]
$$
for the vectors of predictors and responses corresponding to the $i$th group.
Notation such as $\zgblo(\bx_i)$ denotes the $n_i\times 1$ vector containing
$\zgblo(x_{ij})$ values, $1\le j\le n_i$.

\subsection{Best Linear Unbiased Prediction}

Model (\ref{eq:twoLevelfg}) is expressible as a Gaussian response linear
mixed model as follows:
\begin{equation}
\by|\bu\sim N(\bX\bbeta+\bZ\,\bu,\sigeps^2\,\bI),\quad \bu\sim N(\bzero,\bG),
\label{eq:twoLevFreq}
\end{equation}
where
$$\bX\equiv
\left[
\begin{array}{c}
\bX_1\\
\vdots \\
\bX_m \\
\end{array}
\right]
\quad\mbox{with}\quad\bX_i\equiv[\bone\ \bx_i]
\quad\mbox{and}\quad
\bbeta\equiv\left[\begin{array}{c}
\beta_0\\[1ex]
\beta_1
\end{array}
\right]
$$
are the fixed effects design matrix and coefficients,
corresponding to the linear component of $f$.
The random effects design matrix $\bZ$ and corresponding random effects vector
$\bu$ are partitioned according to
\begin{equation}
\bZ=\Big[\bZgbl\ \ \blockdiag{1\le i\le m}([\bX_i\ \bZgrpi])\Big]
\quad\mbox{and}\quad
\bu=\left[\begin{array}{l}
\ \ \ \ \ \buGbl\\[1ex]
\left[
\begin{array}{c}
\buLini\\
\buGrpi
\end{array}
\right]_{1\le i\le m}
\end{array}
\right]
\label{eq:ZanduDefn}
\end{equation}
where $\buGbl= [ \uGblone \ \hdots \ \uGblKGbl ]^T$ are the coefficients
corresponding to the non-linear component of $f$,
$\buLini=[\uLiniz \ \uLinio ]^T$ are the coefficients corresponding
to the linear component of $g_i$ and $\buGrpi=[\uGrpione \ \hdots \ \uGrpiKGrp]^T$
are the coefficients corresponding to the non-linear component of $g_i$, $1\le i\le m$.
In (\ref{eq:ZanduDefn}), $\bZgbl\equiv\stack{1\le i\le m}(\bZgbli)$ and the
matrices $\bZgbli$ and $\bZgrpi$, $1\le i\le m$, contain, respectively,
spline basis functions for the global mean function $f$ and the $i$th group
deviation functions $g_i$. Specifically,
$$
\bZgbli\equiv[\begin{array}{ccc}
\zgblo(\bx_i) & \cdots & \zgblKgbl(\bx_i)
\end{array}]\quad\mbox{and}\quad
\bZgrpi=[
\begin{array}{ccccc}
\zgrpo(\bx_i) &\cdots &\zgrpKgrp(\bx_i)
\end{array}
]
$$
for $1\le i\le m$. 
The corresponding fixed and random effects vectors are
$$
\buGbl\sim N(\bzero,\sigmaGbl^2\bI_{\Kgbl})
\quad
\mbox{and}
\quad
\left[
\begin{array}{c}
\buLini\\[1ex]
\buGrpi
\end{array}
\right]
\simind N\left(\left[\begin{array}{c}\bzero\\[1ex]\bzero\end{array}\right],
\left[
\begin{array}{cc}
\bSigma & \bO \\[1ex]
\bO     & \sigmaGrp^2\bI_{\Kgrp}
\end{array}
\right]\right),\quad 1\le i\le m.
$$
Hence, the full random effects covariance matrix is
\begin{equation}
\bG=\Cov(\bu)=\left[
\begin{array}{ccc}
\sigmaGbl^2\bI_{\Kgbl}&\bO \\[1ex]
\bO                   &
\bI_m\otimes\left[
\begin{array}{cc}
\bSigma & \bO \\[1ex]
\bO     & \sigmaGrp^2\bI_{\Kgrp}
\end{array}
\right]
\end{array}
\right].
\label{eq:Gdefn}
\end{equation}
Next define the matrices
\begin{equation}
  \begin{array}{c}
    \bC\equiv[\bX\ \bZ],\quad\DBLUP\equiv\left[
    \begin{array}{cc}
    \bO & \bO               \\[1ex]
    \bO & \bG^{-1}
    \end{array}
    \right]\quad\mbox{and}\quad\RBLUP\equiv\sigeps^2\bI.
  \end{array}
  \label{eq:CDRmatBLUPdefs}
\end{equation}
The best linear unbiased predictor of $[\bbeta\ \bu]^T$ and
corresponding covariance matrix are
\begin{equation}
{\setlength\arraycolsep{1pt}
\begin{array}{rcl}
\left[\begin{array}{c}
\bbetahat\\
\buhat
\end{array}
\right]&=&(\bC^T\RBLUP^{-1}\bC+\DBLUP)^{-1}\bC^T\RBLUP^{-1}\by\\[3ex]
\mbox{and}\quad
\mbox{Cov}\left(\left[\begin{array}{c}
\bbetahat\\
\buhat-\bu
\end{array}
\right]\right)&=&(\bC^T\RBLUP^{-1}\bC+\DBLUP)^{-1}.
\end{array}
}
\label{eq:BLUPandCov}
\end{equation}
This covariance matrix grows quadratically in $m$, so
its storage becomes infeasible for large numbers of groups.
However, only the following sub-blocks are required for
adding pointwise confidence intervals to curve estimates:

\begin{equation}
{\setlength\arraycolsep{1pt}
\begin{array}{rcl}
\Cov\left(
\left[
\begin{array}{c}
\bbetahat\\
\buHatGbl-\buGbl
\end{array}
\right]\right)
&=&\mbox{top left-hand $(2+\Kgbl)\times(2+\Kgbl)$ }\\[0ex]
&&\mbox{sub-block of $(\bC^T\RBLUP^{-1}\bC+\DBLUP)^{-1}$},\\[4ex]
\Cov\left(\left[
\begin{array}{c}
\buHatLini-\buLini\\[1ex]
\buHatGrpi-\buGrpi
\end{array}
\right]
\right)&=&\mbox{subsequent $(2+\Kgrp)\times(2+\Kgrp)$ diagonal}\\[0ex]
&&\mbox{sub-blocks of $(\bC^T\RBLUP^{-1}\bC+\DBLUP)^{-1}$}\\[1ex]
&&\mbox{below $\Cov\left(
\left[
\begin{array}{c}
\bbetahat\\
\buHatGbl-\buGbl
\end{array}
\right]\right)$,\ $1\le i\le m$, and}\\[2ex]
E\left\{
\left[
\begin{array}{c}
\bbetahat\\
\buHatGbl-\buGbl
\end{array}
\right]
\left[
\begin{array}{c}
\buHatLini-\buLini\\[1ex]
\buHatGrpi-\buGrpi
\end{array}
\right]^T
\right\}&=&\mbox{subsequent $(2+\Kgbl)\times(2+\Kgrp)$ sub-blocks}\\[0ex]
&&\mbox{of $(\bC^T\RBLUP^{-1}\bC+\DBLUP)^{-1}$ to the right of}\\[1ex]
&&\mbox{$\Cov\left(
\left[
\begin{array}{c}
\bbetahat\\
\buHatGbl-\buGbl
\end{array}
\right]\right)$,\ $1\le i\le m$.}
\end{array}
}
\label{eq:CovMain}
\end{equation}

As in Nolan, Menictas \myand Wand (2019), we define the generic two-level sparse
matrix to be determination of the vector $\bx$ which minimizes the least squares criterion
\begin{equation}
\Vert\bb-\bB\bx\Vert^2
\quad\mbox{where $\Vert\bv\Vert^2\equiv\bv^T\bv$ for any column vector $\bv$,}
\label{eqn:sparseLeastSquares}
\end{equation}
with
$\bB$ having the two-level sparse form
\begin{equation}
\bB\equiv
\left[
\arraycolsep=2.2pt\def\arraystretch{1.6}
\begin{array}{c|c|c|c|c}
\setstretch{4.5}
\Bmato            &\Bmatdoto           &\bO   &\cdots&\bO\\
\hline
\Bmatt            &\bO              &\Bmatdott&\cdots&\bO\\
\hline
\vdots            &\vdots           &\vdots           &\ddots&\vdots\\
\hline
\Bmatm &\bO       &\bO              &\cdots           &\Bmatdotm
\end{array}
\right]
\quad\mbox{and $\bb$ partitioned according to}\quad
\bb\equiv\left[
\arraycolsep=2.2pt\def\arraystretch{1.6}
\begin{array}{c}
\setstretch{4.5}
\bveco  \\
\hline
\bvect \\
\hline
\vdots \\
\hline
\bvecm \\
\end{array}
\right].
\label{eq:BandbForms}
\end{equation}
In (\ref{eq:BandbForms}), for any $1\le i\le m$, the matrices
$\bB_i$, $\Bmatdoti$ and $\bb_i$ each have the same number of rows.
The numbers of columns in $\bB_i$ and $\Bmatdoti$ are arbitrary
whereas the $\bb_i$ are column vectors. In addition to solving
for $\bx$, the sub-blocks of $(\bB^T\bB)^{-1}$ corresponding
to the non-sparse regions of $\bB^T\bB$ are included in our
definition of a two-level sparse matrix least squares problem.
Algorithm 2 of Nolan \textit{et al.} (2018) provides a
stable and efficient solution to this problem and labels it
the \SolveTwoLevelSparseLeastSquares\ algorithm. 
Section \ref{sec:Solve2Lev} of the web-supplement contains details
regarding this algorithm.
In Nolan \textit{et al.} (2018) we used \SolveTwoLevelSparseLeastSquares\   
for fitting two-level linear mixed models. However,
precisely the same algorithm can be used for fitting
two-level group-specific curve models because of:

\begin{result}
Computation of $[\bbetahat^T\ \ \buhat^T]^T$ and each
of the sub-blocks of $\mbox{\rm Cov}([\bbetahat^T\ \ (\buhat-\bu)^T]^T)$
listed in (\ref{eq:CovMain}) are expressible as solutions to
the two-level sparse matrix least squares problem:
$$\left\Vert\bb-\bB\left[
\begin{array}{c}
\bbeta\\
\bu
\end{array}
\right]
\right\Vert^2
$$
where the non-zero sub-blocks $\bB$ and $\bb$, according to the notation
in (\ref{eq:BandbForms}), are for $1\le i\le m$:
$$
\bveci\equiv
\left[
\begin{array}{c}
\sigeps^{-1}\by_i\\[1ex]
\bzero \\[1ex]
\bzero \\[1ex]
\bzero \\[1ex]
\end{array}
\right],
\quad
\Bmati\equiv
\left[
\begin{array}{cc}
\sigeps^{-1}\bX_i & \sigeps^{-1}\bZgbli\\[1ex]
\bO & m^{-1/2}\sigmaGbl^{-1}\bI_{\Kgbl}\\[1ex]
\bO & \bO   \\[1ex]
\bO & \bO   \\[1ex]
\end{array}
\right]
\quad\mbox{and}\quad
\Bmatdoti\equiv
\left[
\begin{array}{cc}
\sigeps^{-1}\bX_i  & \sigeps^{-1}\bZgrpi        \\[1ex]
\bO                & \bO            \\[1ex]
\bSigma^{-1/2}     & \bO            \\[1ex]
\bO                & \sigmaGrp^{-1}\bI_{\Kgrp}
\end{array}
\right]
$$
with each of these matrices having $\nadj_i=n_i+\Kgbl+2+\Kgrp$
rows and with $\Bmati$ having $p=2+\Kgbl$ columns and $\Bmatdoti$
having $q=2+\Kgrp$ columns. The solutions are
$$\left[
\begin{array}{c}
\bbetahat\\
\buHatGbl
\end{array}
\right]=\xveco,\quad
\Cov\left(\left[
\begin{array}{c}
\bbetahat\\
\buHatGbl-\buGbl
\end{array}
\right]\right)=\AUoo
$$
and
$$\left[
\begin{array}{c}
\buHatLini\\[1ex]
\buHatGrpi
\end{array}
\right]=\xvectCi,\quad
E\left\{
\left[
\begin{array}{c}
\bbetahat\\
\buHatGbl-\buGbl
\end{array}
\right]
\left[
\begin{array}{c}
\buHatLini-\buLini\\[1ex]
\buHatGrpi-\buGrpi
\end{array}
\right]^T
\right\}=\AUotCi,
$$
$$
\Cov\left(\left[
\begin{array}{c}
\buHatLini-\buLini\\[1ex]
\buHatGrpi-\buGrpi
\end{array}
\right]
\right)=\AUttCi,\ \ 1\le i\le m.
$$
\label{res:twoLevelBLUP}
\end{result}

A derivation of Result \ref{res:twoLevelBLUP} is given in Section \ref{sec:drvResultOne}
of the web-supplement. Algorithm \ref{alg:twoLevBLUP} encapsulates streamlined
best linear unbiased prediction computation together
with coefficient covariance matrix sub-blocks of interest.

\begin{algorithm}[!th]
  \begin{center}
    \begin{minipage}[t]{154mm}
      \begin{small}
        \begin{itemize}
          \setlength\itemsep{4pt}
          \item[] Inputs: $\by_i(n_i\times1),\ \bX_i(n_i\times 2),\ \bZgbli(n_i\times \Kgbl),\ \bZgrpi(n_i\times \Kgrp),\
                  1\le i\le m$; \ $\sigeps^2,\sigmaGbl^2,\sigmaGrp^2>0$,\\[1ex]
                  \null\ \ \ \ \ \ \ \ \ \ \ \ \ \ \ $\bSigma(q\times q), \mbox{symmetric and positive definite.}$
          \item[] For $i=1,\ldots,m$:
          \begin{itemize}
            \item[] $\begin{array}{l}
                    \bveci\thickarrow
                    \left[
                    \begin{array}{c}
                    \sigeps^{-1}\by_i\\[1ex]
                    \bzero \\[1ex]
                    \bzero \\[1ex]
                    \bzero \\[1ex]
                    \end{array}
                    \right],
                    \
                    \Bmati\thickarrow
                    \left[
                    \begin{array}{cc}
                    \sigeps^{-1}\bX_i & \sigeps^{-1}\bZgbli\\[1ex]
                    \bO & m^{-1/2}\sigmaGbl^{-1}\bI_{\Kgbl} \\[1ex]
                    \bO & \bO   \\[1ex]
                    \bO & \bO   \\[1ex]
                    \end{array}
                    \right],
                  \end{array}$
           \item[] $\begin{array}{l}
                    \Bmatdoti\thickarrow
                    \left[
                    \begin{array}{cc}
                    \sigeps^{-1}\bX_i  & \sigeps^{-1}\bZgrpi        \\[1ex]
                    \bO                & \bO            \\[1ex]
                    \bSigma^{-1/2}     & \bO            \\[1ex]
                    \bO                & \sigmaGrp^{-1}\bI_{\Kgrp}
                    \end{array}
                    \right]
                  \end{array}$
          \end{itemize}
          \item[] $\Ssc_1\thickarrow\SolveTwoLevelSparseLeastSquares
                    \Big(\big\{(\bveci,\Bmati,\Bmatdoti):1\le i\le m\big\}\Big)$
          \item[] $\left[
                  \begin{array}{c}
                  \bbetahat\\
                  \buHatGbl
                  \end{array}
                  \right]\thickarrow\mbox{$\xveco$ component of $\Ssc_1$}$\ \ \ ;\ \ \
                  $\Cov\left(\left[
                  \begin{array}{c}
                  \bbetahat\\
                  \buHatGbl-\buGbl
                  \end{array}
                  \right]\right)\thickarrow\mbox{$\AUoo$ component of $\Ssc_1$}$
          \item[] For $i=1,\ldots,m$:
          \begin{itemize}
            \setlength\itemsep{4pt}
            \item[] $\left[
                      \begin{array}{c}
                      \buHatLini\\[1ex]
                      \buHatGrpi
                      \end{array}
                      \right]\thickarrow\mbox{$\xvectCi$ component of $\Ssc_1$}$
            \item[] $\Cov\left(\left[
                    \begin{array}{c}
                    \buHatLini-\buLini\\[1ex]
                    \buHatGrpi-\buGrpi
                    \end{array}
                    \right]
                    \right)\thickarrow\mbox{$\AUttCi$ component of $\Ssc_1$}$
            \item[] $E\left\{ \left[
                      \begin{array}{c}
                      \bbetahat\\
                      \buHatGbl-\buGbl
                      \end{array}
                      \right]
                      \left[
                      \begin{array}{c}
                      \buHatLini-\buLini\\[1ex]
                      \buHatGrpi-\buGrpi
                      \end{array}
                      \right]^T
                      \right\}\thickarrow\mbox{$\AUotCi$ component of $\Ssc_1$}$
          \end{itemize}
          \item[] Output:
                $$\begin{array}{c}\Bigg(\left[
                \begin{array}{c}
                \bbetahat\\
                \buHatGbl
                \end{array}
                \right],\ \Cov\left(\left[
                \begin{array}{c}
                \bbetahat\\
                \buHatGbl-\buGbl
                \end{array}
                \right]\right),
                \Bigg\{\Bigg(
                \left[
                \begin{array}{c}
                \buHatLini\\[1ex]
                \buHatGrpi
                \end{array}
                \right],\,
                \Cov\left(\left[
                \begin{array}{c}
                \buHatLini-\buLini\\[1ex]
                \buHatGrpi-\buGrpi
                \end{array}
                \right]\right),\,\\[2ex]
                E\left\{
                \left[
                \begin{array}{c}
                \bbetahat\\
                \buHatGbl-\buGbl
                \end{array}
                \right]
                \left[
                \begin{array}{c}
                \buHatLini-\buLini\\[1ex]
                \buHatGrpi-\buGrpi
                \end{array}
                \right]^T
                \right\}
                \Bigg):\ 1\le i\le m\Bigg\}\Bigg)
                \end{array}$$
        \end{itemize}
      \end{small}
    \end{minipage}
  \end{center}
  \caption{\it Streamlined algorithm for obtaining best linear unbiased predictions
               and corresponding covariance matrix components for the two-level
               group specific curves model.}
  \label{alg:twoLevBLUP}
\end{algorithm}

\subsection{Mean Field Variational Bayes}

We now consider the following Bayesian extension of (\ref{eq:twoLevFreq})
and (\ref{eq:Gdefn}):
\begin{equation}
\begin{array}{c}
\by|\bbeta, \bu, \sigsqeps \sim N(\bX\bbeta+\bZ\,\bu,\sigeps^2\,\bI),\quad \bu|\sigmaGbl^2,\sigmaGrp^2,\bSigma\sim N(\bzero,\bG),
\quad\mbox{$\bG$ as defined in (\ref{eq:Gdefn}),}
\\[1ex]
\bbeta\sim N(\bmu_{\bbeta},\bSigma_{\bbeta}),\quad\sigeps^2|\aeps\sim\mbox{Inverse-$\chi^2$}(\nuEps,1/\aeps),
\quad\aeps\sim\mbox{Inverse-$\chi^2$}(1,1/(\nuEps\sEps^2)),\\[2ex]
\quad\sigmaGbl^2|\aGbl\sim\mbox{Inverse-$\chi^2$}(\nuGbl,1/\aGbl),
\quad\aGbl\sim\mbox{Inverse-$\chi^2$}(1,1/(\nuGbl\sGbl^2)),\\[2ex]
\quad\sigmaGrp^2|\aGrp\sim\mbox{Inverse-$\chi^2$}(\nuGrp,1/\aGrp),
\quad\aGrp\sim\mbox{Inverse-$\chi^2$}(1,1/(\nuGrp\sGrp^2)),\\[2ex]
\bSigma|\ASigma\sim\mbox{Inverse-G-Wishart}\big(\Gfull,\nuSigma+2,\ASigma^{-1}\big),\\[2ex]
\ASigma\sim\mbox{Inverse-G-Wishart}(\Gdiag,1,\bLambda_{\ASigma}),\quad
\bLambda_{\ASigma}\equiv\{\nuSigma\diag(\sSigmaOne^2,\sSigmaTwo^2)\}^{-1}.
\end{array}
\label{eq:twoLevBayes}
\end{equation}
Here the $2\times1$ vector $\bmu_{\bbeta}$ and $2\times2$ symmetric positive definite
matrix $\bSigma_{\bbeta}$ are hyperparameters corresponding to the prior distribution
on $\bbeta$ and
$$\nuEps,\sEps,\nuGbl,\sGbl,\nuGrp,\sGrp,\nuSigma,\sSigmaOne,\sSigmaTwo>0$$
are hyperparameters for the variance and covariance matrix parameters.
Details on the Inverse G-Wishart distribution, and the Inverse-$\chi^2$
special case, are given in Section \ref{sec:IGWandICS} of the web-supplement.
The auxiliary variable $\aeps$ is defined so that $\sigeps$ has a
Half-$t$ distribution with degrees of freedom parameter $\nuEps$
and scale parameter $\sEps$, with larger values of $\sEps$ corresponding
to greater noninformativity. Analogous comments apply to the other standard
deviation parameters. Setting $\nuSigma=2$ leads to the correlation parameter
in $\bSigma$ having a Uniform distribution on $(-1,1)$ (Huang \myand Wand, 2013).

Throughout this article we use $\pDens$ generically to denote a density function
corresponding to random quantities in Bayesian models such as  (\ref{eq:twoLevBayes}).
For example, $\pDens(\bbeta)$ denotes the prior density function of $\bbeta$ and
$\pDens(\bu|\sigmaGbl^2,\sigmaGrp^2,\bSigma)$ denotes the density function of $\bu$
conditional on $(\sigmaGbl^2,\sigmaGrp^2,\bSigma)$.
Now consider the following mean field restriction on the joint posterior
density function of all parameters in (\ref{eq:twoLevBayes}):
\begin{equation}
\pDens(\bbeta,\bu,\aeps,\aGbl,\aGrp,\ASigma,\sigeps^2,\sigmaGbl^2,\sigmaGrp^2,\bSigma|\by)
\approx \qDens(\bbeta,\bu,\aeps,\aGbl,\aGrp,\ASigma)\,\qDens(\sigeps^2,\sigmaGbl^2,\sigmaGrp^2,\bSigma).
\label{eq:producRestrict}
\end{equation}
Here, generically, each $\qDens$ denotes an approximate posterior density function of the
random vector indicated by its argument according to the mean field restriction
(\ref{eq:producRestrict}). Then application of the minimum Kullback-Leibler divergence
equations (e.g. equation (10.9) of Bishop, 2006) leads to the optimal $\qDens$-density
functions for the parameters of interest being as follows:
\begin{equation*}
  \begin{array}{c}
    \begin{array}{ll}
      &\qDens^*(\bbeta,\bu)\ \mbox{has a $N\big(\bmu_{\qDens(\bbeta,\bu)},\bSigma_{\qDens(\bbeta,\bu)}\big)$ distribution,}
      \\[1.5ex]
      &\qDens^*(\sigeps^2)\ \mbox{has an $\mbox{Inverse-$\chi^2$}
      \big(\xi_{\qDens(\sigeps^2)},\lambda_{\qDens(\sigeps^2)}\big)$ distribution,}
      \\[1.5ex]
      &\qDens^*(\sigmaGbl^2)\ \mbox{has an $\mbox{Inverse-$\chi^2$}
      \big(\xi_{\qDens(\sigmaGbl^2)},\lambda_{\qDens(\sigmaGbl^2)}\big)$ distribution,}
      \\[1.5ex]
      &\qDens^*(\sigmaGrp^2)\ \mbox{has an $\mbox{Inverse-$\chi^2$}
      \big(\xi_{\qDens(\sigmaGrp^2)},\lambda_{\qDens(\sigmaGrp^2)}\big)$ distribution}
      \\[1.5ex]
      \mbox{and} & \qDens^*(\bSigma)\ \mbox{has an
      $\mbox{Inverse-G-Wishart}(\Gfull,\xi_{\qDens(\bSigma)},\bLambda_{\qDens(\bSigma)})$ distribution.}\\
    \end{array}
  \end{array}
\end{equation*}
The optimal $\qDens$-density parameters are determined via an
iterative coordinate ascent algorithm, with details given
in Section \ref{sec:drvAlgTwo} of this article's web-supplement.
The stopping criterion is based on the variational lower bound on the
marginal likelihood (e.g. Bishop, 2006; Section 10.2.2) and denoted
$\underline{\pDens}(\by;\qDens)$. Its logarithmic form and derivation are given
in Section \ref{sec:lowerBound} of the web-supplement.

Note that updates for
$\bmu_{\qDens(\bbeta,\bu)}$ and $\bSigma_{\qDens(\bbeta,\bu)}$ may be written
\begin{equation}
\bmu_{\qDens(\bbeta,\bu)}\leftarrow(\bC^T\RMFVB^{-1}\bC+\DMFVB)^{-1}(\bC^T\RMFVB^{-1}\by + \oMFVB)
\quad \mbox{and}\quad
\bSigma_{\qDens(\bbeta,\bu)}\leftarrow(\bC^T\RMFVB^{-1}\bC+\DMFVB)^{-1}
\label{eq:muSigmaMFVBupd}
\end{equation}
where
\begin{equation}
\begin{array}{l}
\RMFVB\equiv\mu_{\qDens(1/\sigeps^2)}^{-1}\bI,
\quad
\DMFVB\equiv
\left[
\begin{array}{ccc}
\bSigma_{\bbeta}^{-1} & \bO &\bO  \\[1ex]
\bO             &\mu_{\qDens(1/\sigmaGbl^2)}\bI    &  \bO       \\[1ex]
\bO             & \bO  & \displaystyle{\blockdiag{1\le i\le m}}
                     \left[
                   \begin{array}{cc}
                   \bM_{\qDens(\bSigma^{-1})} & \bO                   \\
                   \bO       &\mu_{\qDens(1/\sigmaGrp^2)}\bI    \\
                   \end{array}
                   \right]
\end{array}
\right]\\[3ex]
\quad\mbox{and}\quad
\oMFVB\equiv\left[
\begin{array}{c}
\bSigma_{\bbeta}^{-1}\bmu_{\bbeta}\\[1ex]
\bzero
\end{array}
\right].
\end{array}
\label{eq:MFVBmatDefns}
\end{equation}
For increasingly large numbers of groups the matrix $\bSigma_{\qDens(\bbeta,\bu)}$
approaches a size that is untenable for random access memory storage
on standard 2020s workplace computers. However, only the following relatively
small sub-blocks of $\bSigma_{\qDens(\bbeta,\bu)}$ are required for
variational inference concerning the variance and covariance matrix
parameters:
\begin{equation}
{\setlength\arraycolsep{1pt}
\begin{array}{rcl}
&&\bSigma_{\qDens(\bbeta,\buGbl)}=\mbox{top left-hand $(2+\Kgbl)\times(2+\Kgbl)$ sub-block
of $(\bC^T\RMFVB^{-1}\bC+\DMFVB)^{-1}$},\\[2ex]
&&\bSigma_{\qDens(\buLini,\buGrpi)}=\mbox{subsequent $(2+\Kgrp)\times(2+\Kgrp)$ diagonal sub-blocks of}
\\
&&\qquad\qquad\qquad\quad\mbox{$(\bC^T\RMFVB^{-1}\bC+\DMFVB)^{-1}$ below
$\bSigma_{\qDens(\bbeta,\buGbl)}$,\ $1\le i\le m$, and}\\[2ex]
&&
E_\qDens \left\{\left(\left[\begin{array}{c}\bbeta\\ \buGbl\end{array}\right]-\bmu_{\qDens(\bbeta,\buGbl)}\right)
             \left(\left[\begin{array}{c}\buLini\\ \buGrpi\end{array}\right]
                -\bmuq{\buLini,\buGrpi)}\right)^T\right\}
=
\mbox{subsequent}\\[1ex]
&&\qquad\qquad\qquad\qquad\mbox{$(2+\Kgbl)\times(2+\Kgrp)$ sub-blocks of $(\bC^T\RMFVB^{-1}\bC+\DMFVB)^{-1}$}\\[1ex]
&&\qquad\qquad\qquad\qquad\mbox{to the right of $\bSigma_{\qDens(\bbeta,\buGbl)}$,\ $1\le i\le m$.}
\end{array}
}
\label{eq:CovMFVB}
\end{equation}

\noindent
For a streamlined mean field variational Bayes algorithm, we appeal to:

\begin{result}
The mean field variational Bayes updates of $\bmu_{\qDens(\bbeta,\bu)}$ and each
of the sub-blocks of $\bSigma_{\qDens(\bbeta,\bu)}$ in (\ref{eq:CovMFVB}) are expressible as a
two-level sparse matrix least squares problem of the form:
$$\left\Vert\bb-\bB\bmu_{\qDens(\bbeta,\bu)}
\right\Vert^2
$$
where the non-zero sub-blocks $\bB$ and $\bb$, according to the notation
in (\ref{eq:BandbForms}), are, for $1\le i\le m$,
$$\bveci\equiv\left[\begin{array}{c}
\mu_{\qDens(1/\sigeps^2)}^{1/2}\by_i\\[2ex]
m^{-1/2}\bSigma_{\bbeta}^{-1/2}\bmu_{\bbeta}\\[2ex]
\bzero\\[1ex]
\bzero\\[1ex]
\bzero
\end{array}
\right],
\quad\Bmati\equiv\left[\begin{array}{cc}
\mu_{\qDens(1/\sigeps^2)}^{1/2}\bX_i & \mu_{\qDens(1/\sigeps^2)}^{1/2}\bZgbli\\[2ex]
m^{-1/2}\bSigma_{\bbeta}^{-1/2}& \bO  \\[2ex]
\bO                            &  m^{-1/2}\mu_{\qDens(1/\sigmaGbl^2)}^{1/2}\bI_{\Kgbl}       \\[2ex]
\bO                            &  \bO        \\[2ex]
\bO                            &  \bO        \\[2ex]
\end{array}
\right]
$$
and
$$
\Bmatdoti\equiv
\left[\begin{array}{cc}
\mu_{\qDens(1/\sigeps^2)}^{1/2}\bX_i & \mu_{\qDens(1/\sigeps^2)}^{1/2}\bZgrpi  \\[2ex]
\bO                             &   \bO                              \\[2ex]
\bO                             &   \bO                              \\[2ex]
\bM_{\qDens(\bSigma^{-1})}^{1/2}     &   \bO                              \\[2ex]
\bO                             &  \mu_{\qDens(1/\sigmaGrp^2)}^{1/2}\bI_{\Kgrp}
\end{array}
\right]
$$
with each of these matrices having $\nadj_i=n_i+2+\Kgbl+2+\Kgrp$
rows and with $\Bmati$ having $p=2+\Kgbl$ columns and $\Bmatdoti$
having $q=2+\Kgrp$ columns. The solutions are
$$
\bmu_{\qDens(\bbeta,\buGbl)}
=\xveco,\quad
\bSigma_{\qDens(\bbeta,\buGbl)}
=\AUoo,
$$
$$
\bmu_{\qDens(\buLini,\buGrpi)}
=\xvectCi,\quad
\bSigma_{\qDens(\buLini,\buGrpi)}
=\AUttCi,\
$$
and
$$
E_q\left\{
\left[
\begin{array}{c}
\bbeta-\bmu_{\qDens(\bbeta)}\\
\buGbl-\bmu_{\qDens(\buGbl)}
\end{array}
\right]
\left[
\begin{array}{c}
\buLini-\bmu_{\qDens(\buLini)}\\[1ex]
\buGrpi-\bmu_{\qDens(\buGrpi)}
\end{array}
\right]^T
\right\}=\AUotCi, 1\le i\le m.
$$
\label{res:twoLevelMFVB}
\end{result}

%
\begin{algorithm}[!th]
  \begin{center}
    \begin{minipage}[t]{159mm}
      \begin{small}
        \begin{itemize}
          \setlength\itemsep{0pt}
          \item[] Data Inputs: $\by_i(n_i\times1),\ \bX_i(n_i\times 2),\ \bZgbli(n_i\times \Kgbl),\ \bZgrpi(n_i\times \Kgrp),\
                  1\le i\le m$;
          \item[] Hyperparameter Inputs: $\bmu_{\bbeta}(2\times1)$,
                  $\bSigma_{\bbeta}(2\times 2)\ \mbox{symmetric and positive definite}$,
          \item[] \qquad $s_{\varepsilon},\nu_{\varepsilon},s_{\mbox{\rm\tiny gbl}},
                        \nu_{\mbox{\rm\tiny gbl}},\sSigmaOne,\sSigmaTwo,\nu_{\bSigma},
                        s_{\mbox{\rm\tiny grp}},\nu_{\mbox{\rm\tiny grp}}>0$.
          \item[] For $i=1,\ldots,m$:
          \begin{itemize}
            \item[]$\bCgbli\thickarrow[\bX_i\ \bZgbli]$\ \ \ ;\ \ \ $\bCgrpi\thickarrow[\bX_i\ \bZgrpi]$
          \end{itemize}
          \item[] Initialize: $\muq{1/\sigsqeps}$, $\muq{1/\sigma_{\mbox{\rm\tiny gbl}}^{2}}$,
                  $\muq{1/\sigma_{\mbox{\rm\tiny grp}}^{2}}$, $\muq{1/\aeps}$,
                  $\muq{1/a_{\mbox{\rm\tiny gbl}}}$, $\muq{1/a_{\mbox{\rm\tiny grp}}} > 0$,
          \item[] \qquad \qquad $\MqSigma (2 \times 2), \MqASigma (2 \times 2)$ both symmetric and
                  positive definite.
          \item[] $\xi_{\qDens(\sigeps^2)}\thickarrow \nu_{\varepsilon} + \sumim n_i$\ \ \ ;\ \ \
                  $\xi_{\qDens(\sigmaGbl^2)}\thickarrow\nu_{\mbox{\rm\tiny gbl}}+\Kgbl$\ \ \ ;\ \ \
                  $\xi_{\qDens(\bSigma)}\thickarrow\nu_{\bSigma}+2+m$
          \item[] $\xi_{\qDens(\sigmaGrp^2)}\thickarrow \nu_{\mbox{\rm\tiny grp}} + m\Kgrp$ \ \ \ ; \ \ \
                  $\xi_{\qDens(a_{\varepsilon})}\thickarrow \nuEps + 1$\ \ \ ; \ \ \
                  $\xi_{\qDens(a_{\mbox{\rm\tiny gbl}})}\thickarrow \nuGbl + 1$\ \ \ ; \ \ \
                  $\xi_{\qDens(a_{\mbox{\rm\tiny grp}})}\thickarrow \nuGrp + 1$
          \item[] $\xi_{\qDens(\bA_{\bSigma})}\thickarrow \nuSigma + 2$
          \item[] Cycle:
          \begin{itemize}
            \item[] For $i = 1,\ldots, m$:
            \begin{itemize}
              \item[] $\bveci\thickarrow\left[
                        \begin{array}{c}
                          \mu_{\qDens(1/\sigeps^2)}^{1/2}\by_i\\[1.5ex]
                          m^{-1/2}\bSigma_{\bbeta}^{-1/2}\bmu_{\bbeta} \\[1ex]
                          \bzero \\[1ex]
                          \bzero \\[1ex]
                          \bzero \\[1ex]
                        \end{array}
                        \right],\
                        \Bmati\thickarrow
                        \left[
                        \begin{array}{cc}
                          \mu_{\qDens(1/\sigeps^2)}^{1/2}\bX_i & \mu_{\qDens(1/\sigeps^2)}^{1/2}\bZgbli\\[1.5ex]
                          m^{-1/2}\bSigma_{\bbeta}^{-1/2} & \bO \\[1ex]
                          \bO & m^{-1/2}\mu_{\qDens(1/\sigmaGbl^2)}^{1/2}\bI_{\Kgbl}   \\[1ex]
                          \bO & \bO   \\[1ex]
                          \bO & \bO   \\[1ex]
                        \end{array}
                        \right],$
            \item[] $\Bmatdoti\thickarrow
                        \left[
                        \begin{array}{cc}
                          \mu_{\qDens(1/\sigeps^2)}^{1/2}\bX_i & \mu_{\qDens(1/\sigeps^2)}^{1/2}\bZgrpi \\[1ex]
                          \bO & \bO \\[1ex]
                          \bO & \bO \\[1ex]
                          \bM_{\qDens(\bSigma^{-1})}^{1/2} & \bO \\[1ex]
                          \bO & \mu_{\qDens(1/\sigmaGrp^2)}^{1/2}\bI_{\Kgrp}
                        \end{array}
                        \right]$
            \end{itemize}
            \item[] $\Ssc_2\thickarrow\SolveTwoLevelSparseLeastSquares\Big(\big\{(
                      \bveci,\Bmati,\Bmatdoti):1\le i\le m\big\}\Big)$
            \item[] $\bmu_{\qDens(\bbeta,\buGbl)}\thickarrow\mbox{$\xveco$ component of $\Ssc_2$}$
                    \ \ \ ;\ \ \ $\bSigma_{\qDens(\bbeta,\buGbl)}\thickarrow\mbox{$\AUoo$ component of $\Ssc_2$}$
            \item[] $\bmu_{\qDens(\buGbl)}\thickarrow\mbox{last $\Kgbl$ rows of $\bmu_{\qDens(\bbeta,\buGbl)}$}$
            \item[] $\bSigma_{\qDens(\buGbl)}\thickarrow\mbox{bottom-right $\Kgbl\times\Kgbl$ sub-block of $\bSigma_{\qDens(\bbeta,\buGbl)}$}$
            \item[] $\lambda_{\qDens(\sigsqeps)}\thickarrow\muq{1/\aeps}$\ \ ;\ \
                    $\Lambda_{\qDens(\bSigma)}\thickarrow \MqASigma$\ \ ;\ \
                    $\lambda_{\qDens(\sigma^{2}_{\mbox{\rm\tiny grp}})}\thickarrow\mu_{\qDens(1/a_{\mbox{\rm\tiny grp}})}$
            \item[] For $i = 1,\ldots, m$:
            \begin{itemize}
              \item[] $\bmu_{\qDens(\buLini,\buGrpi)}\thickarrow\mbox{$\xvectCi$ component of $\Ssc_2$}$
              \item[] $\bSigma_{\qDens(\buLini,\buGrpi)}\thickarrow\mbox{$\AUttCi$ component of $\Ssc_2$}$
              \item[] $\bmu_{\qDens(\buLini)}\thickarrow\mbox{first $2$ rows of $\bmu_{\qDens(\buLini,\buGrpi)}$}$
              \item[] $\bSigma_{\qDens(\buLini)}\thickarrow\mbox{top left $2\times 2$ sub-block of
                      $\bSigma_{\qDens(\buLini,\buGrpi)}$}$
              \item[] $\bmu_{\qDens(\buGrpi)}\thickarrow\mbox{last $\Kgrp$ rows of $\bmu_{\qDens(\buLini,\buGrpi)}$}$
              \item[] $\bSigma_{\qDens(\buGrpi)}\thickarrow\mbox{bottom right $\Kgrp \times \Kgrp$ sub-block of
                      $\bSigma_{\qDens(\buLini,\buGrpi)}$}$
              \item[] \textsl{continued on a subsequent page}\ $\ldots$
            \end{itemize}
          \end{itemize}
        \end{itemize}
      \end{small}
    \end{minipage}
  \end{center}
  \caption{\it QR-decomposition-based streamlined algorithm for obtaining mean field variational
             Bayes approximate posterior density functions for the parameters in the
             Bayesian two-level group-specific curves model (\ref{eq:twoLevBayes}) with
             product density restriction (\ref{eq:producRestrict}).}
  \label{alg:twoLevMFVB}
\end{algorithm}

\setcounter{algorithm}{1}
\begin{algorithm}[!th]
  \begin{center}
    \begin{minipage}[t]{159mm}
      \begin{small}
        \begin{itemize}
          \item[]
            \begin{itemize}
              \item[]
                \begin{itemize}
              \item[] $E_{\qDens}\left\{\left(\left[\begin{array}{c}\bbeta\\ \buGbl\end{array}\right]
                       -\bmu_{\qDens(\bbeta,\buGbl)}\right)
                        \left(\left[\begin{array}{c}\buLini\\ \buGrpi\end{array}\right]
                        -\bmuq{\buLini,\buGrpi)}\right)^T\right\}$
              \item[] \qquad\qquad\qquad$\thickarrow\mbox{$\AUotCi$ component of $\Ssc_2$}$
                  \item[] $\lambda_{\qDens(\sigsqeps)}\thickarrow \lambda_{\qDens(\sigsqeps)}
                           +\big\Vert\by_i-\bCgbli\bmu_{\qDens(\bbeta,\buGbl)}
                           -\bCgrpi\bmu_{\qDens(\buLini,\buGrpi)}\big\Vert^2$
                  \item[] $\lambda_{\qDens(\sigsqeps)}\thickarrow \lambda_{\qDens(\sigsqeps)}
                           +\mbox{tr}(\bCgbli^T\bCgbli\bSigma_{\qDens(\bbeta,\buGbl)})
                           +\mbox{tr}(\bCgrpi^T\bCgrpi\bSigma_{\qDens(\buLini,\buGrpi)})$
                  \item[] $\lambda_{\qDens(\sigsqeps)}\thickarrow \lambda_{\qDens(\sigsqeps)}$
                  \item[] \hspace{2mm}$+2\,\mbox{tr}\left[\bCgrpi^T\bCgbli\,E_{\qDens}\left\{\left(
                          \left[\begin{array}{c}\bbeta \\ \buGbl\end{array}\right]
                          -\bmu_{\qDens(\bbeta,\buGbl)}\right)\left(\left[\begin{array}{c}\buLini\\
                          \buGrpi\end{array}\right]
                         -\bmuq{\buLini,\buGrpi)}\right)^T\right\}\right]$
                  \item[] $\bLambda_{\qDens(\bSigma)}\thickarrow\bLambda_{\qDens(\bSigma)}+
                           \bmu_{\qDens(\buLini)}\bmu_{\qDens(\buLini)}^T+ \bSigma_{\qDens(\buLini)}$
                  \item[] $\lambda_{\qDens(\sigma^{2}_{\mbox{\rm\tiny grp}})} \thickarrow
                           \lambda_{\qDens(\sigma^{2}_{\mbox{\rm\tiny grp}})} +
                           \Vert\bmu_{\qDens(\bu_{\mbox{\rm\tiny grp,i}})}\Vert^2 +
                           \tr\big(\bSigma_{\qDens(\bu_{\mbox{\rm\tiny grp,i}})}\big)$
                \end{itemize}
              \item[] $\lambda_{\qDens(\sigma^{2}_{\mbox{\rm\tiny gbl}})} \thickarrow \mu_{\qDens(1/a_{\mbox{\rm\tiny gbl}})} +
                       \Vert\bmu_{\qDens(\bu_{\mbox{\rm\tiny gbl}})}\Vert^2 +
                       \tr\big(\bSigma_{\qDens(\bu_{\mbox{\rm\tiny gbl}})}\big)$
              \item[] $\muq{1/\sigsqeps} \leftarrow \xi_{\qDens(\sigeps)}/\lambda_{\qDens(\sigsqeps)}$ \ \ \ ; \ \ \
                      $\muq{1/\sigma^{2}_{\mbox{\rm\tiny gbl}}} \leftarrow \xi_{\qDens(\sigma^{2}_{\mbox{\rm\tiny gbl}})}/
                       \lambda_{\qDens(\sigma^{2}_{\mbox{\rm\tiny gbl}})}$
              \item[] $\MqSigma \leftarrow(\xi_{\qDens(\bSigma)}-1)\bLambda^{-1}_{\qDens(\bSigma)}$ \ \ \ ; \ \ \
                      $\muq{1/\sigma^{2}_{\mbox{\rm\tiny grp}}} \leftarrow \xi_{\qDens(\sigma^{2}_{\mbox{\rm\tiny grp}})}/
                       \lambda_{\qDens(\sigma^{2}_{\mbox{\rm\tiny grp}})}$
              \item[] $\lambda_{\qDens(a_{\varepsilon})}\thickarrow\muq{1/\sigma_{\varepsilon}^{2}}
                       +1/(\nu_{\varepsilon} s_{\varepsilon}^2)$\ \ \ ;\ \ \
                       $\muq{1/a_{\varepsilon}} \thickarrow \xi_{\qDens(a_{\varepsilon})}/
                       \lambda_{\qDens(a_{\varepsilon})}$
              \item[] $\bLambda_{\qDens(\ASigma)}\thickarrow
               \diag\big\{\mbox{diagonal}\big(\bM_{\qDens(\bSigma^{-1})}\big)\big\}+\{\nuSigma\diag(\sSigmaOne^2,\sSigmaTwo^2)\}^{-1}$
              \item[] $\bM_{\qDens(\ASigma^{-1})}\thickarrow \xi_{\qDens(\ASigma)}\bLambda_{\qDens(\ASigma)}^{-1}$
              \item[] $\lambda_{\qDens(a_{\mbox{\rm\tiny gbl}})}\thickarrow\muq{1/\sigma_{\mbox{\rm\tiny gbl}}^{2}}
                      +1/(\nu_{\mbox{\rm\tiny gbl}} s_{\mbox{\rm\tiny gbl}}^2)$\ \ \ ;\ \ \
                      $\muq{1/a_{\mbox{\rm\tiny gbl}}} \thickarrow \xi_{\qDens(a_{\mbox{\rm\tiny gbl}})}/
                      \lambda_{\qDens(a_{\mbox{\rm\tiny gbl}})}$
              \item[] $\lambda_{\qDens(a_{\mbox{\rm\tiny grp}})}\thickarrow\muq{1/\sigma_{\mbox{\rm\tiny grp}}^{2}}
                       +1/(\nu_{\mbox{\rm\tiny grp}} s_{\mbox{\rm\tiny grp}}^2)$\ \ \ ;\ \ \
                       $\muq{1/a_{\mbox{\rm\tiny grp}}} \thickarrow \xi_{\qDens(a_{\mbox{\rm\tiny grp}})}/
                       \lambda_{\qDens(a_{\mbox{\rm\tiny grp}})}$
            \end{itemize}
          \item[] until the increase in $\underline{\pDens}(\by;\qDens)$ is negligible.
          \item[] Outputs: $\bmu_{\qDens(\bbeta,\buGbl)}$,\ $\bSigma_{\qDens(\bbeta,\buGbl)}$,\
                           $\Big\{\bmu_{\qDens(\buLini,\buGrpi)}, \bSigma_{\qDens(\buLini,\buGrpi)},$
          \item[] \qquad\qquad $E_{\qDens}\left\{\left(\left[\begin{array}{c}\bbeta\\ \buGbl\end{array}\right]
                   -\bmu_{\qDens(\bbeta,\buGbl)}\right)
                   \left(\left[\begin{array}{c}\buLini\\ \buGrpi\end{array}\right]
                   -\bmuq{\buLini,\buGrpi)}\right)^T\right\}:1\le i\le m\Big\},$
          \item[] \qquad \qquad $\xi_{\qDens(\sigeps)},\lambda_{\qDens(\sigsqeps)},\xi_{\qDens(\sigmaGbl^2)},
                  \lambda_{\qDens(\sigmaGbl^2)},\xi_{\qDens(\bSigma)},\bLambda^{-1}_{\qDens(\bSigma)},
                  \xi_{\qDens(\sigmaGrp^2)},\lambda_{\qDens(\sigmaGrp^2)}.$
        \end{itemize}
      \end{small}
    \end{minipage}
  \end{center}
  \caption{\textbf{continued.} \textit{This is a continuation of the description of this algorithm that commences
on a preceding page.}}
\end{algorithm}
%
%
Algorithm \ref{alg:twoLevMFVB} utilizes Result \ref{res:twoLevelMFVB} to facilitate
streamlined computation of the variational parameters.

Lastly, we note that Algorithm \ref{alg:twoLevMFVB} is loosely related to Algorithm 2
of Lee \myand Wand (2016). One difference is that we are treating the Gaussian,
rather than Bernoulli, response situation here. In addition, we are using the
recent sparse multilevel matrix results of Nolan \myand Wand (2018) which are
amenable to higher level extensions, such as the three-level group specific
curve model treated in Section \ref{sec:threeLevMods}.

\subsection{Contrast Function Extension}

In many curve-type data applications the data can be categorized as being from
two or more types. Of particular interest in such circumstances are
contrast function estimates and accompanying standard errors. The streamlined
approaches used in Algorithms \ref{alg:twoLevBLUP} and \ref{alg:twoLevMFVB} still
apply for the contrast function extension regardless of the number of categories.
The two category situation, where there is a single contrast function, is
described here. The extension to higher numbers of categories is straightforward.

Suppose that the $(x_{ij},y_{ij})$ pairs are from one of two categories,
labeled $A$ and $B$, and introduce the indicator variable data:
$$\iota^A_{ij}\equiv\left\{
\begin{array}{ll}
1 & \mbox{if $(x_{ij},y_{ij})$ is from category $A$},\\
0 & \mbox{if $(x_{ij},y_{ij})$ is from category $B$}.\\
\end{array}
\right.
$$
Then penalized spline models for the global mean and deviation functions for each category
are
$$
\left.
\begin{array}{rcl}
f^A(x)&=&\beta_0^{\mbox{\rm\tiny A}}
+\beta_1^{\mbox{\rm\tiny A}}\,x+\displaystyle{\sum_{k=1}^{\Kgbl}}\,\uGblk^A\zgblk(x)\\[1ex]
g^A_i(x)&=&\uLiniz^A+\uLinio^A\,x
+\displaystyle{\sum_{k=1}^{\Kgrp}}\,\uGrpik^A\zgrpk(x)
\end{array}
\right\}\ \mbox{for category A}
$$
and
$$
\left.
\begin{array}{rcl}
f^B(x)&=&\beta_0^{\mbox{\rm\tiny A}}+\beta_0^{\mbox{\rm\tiny BvsA}}+
(\beta_1^{\mbox{\rm\tiny A}}+\beta_1^{\mbox{\rm\tiny BvsA}})\,x
+\displaystyle{\sum_{k=1}^{\Kgbl}}\,\uGblk^B\zgblk(x)\\[1ex]
g^B_i(x)&=&\uLiniz^B+\uLinio^B\,x
+\displaystyle{\sum_{k=1}^{\Kgrp}}\,\uGrpik^B\zgrpk(x)
\end{array}
\right\}\ \mbox{for category B.}
$$
This allows us to estimate the global contrast function\index{contrast function}
\begin{equation}
c(x)\equiv f^B(x) - f^A(x)
=\beta_0^{\mbox{\rm\tiny BvsA}}+\beta_1^{\mbox{\rm\tiny BvsA}}\,x
+\sum_{k=1}^{\Kgbl}(\uGblk^B-\uGblk^A)\zgblk(x).
\label{eq:contrastCurve}
\end{equation}
The distributions on the random coefficients are
$$[\uLiniz^A\ \uLinio^A\ \uLiniz^B\ \uLinio^B]^T\simind N(\bzero,\bSigma)$$
and
$$\uGblk^A\simind N\big(0,(\sigmaGbl^A)^2\big),\ \ \uGblk^B\simind N\big(0,(\sigmaGbl^B)^2\big),
\ \ \uGrpik^A\simind N\big(0,\sigmaGrp^2\big)\ \ \mbox{and}\ \ \uGrpik^B\simind N\big(0,\sigmaGrp^2\big)
$$
independently of each other. In this two-category extension, the matrix $\bSigma$ is
an unstructured $4\times4$ covariance matrix.

Algorithms \ref{alg:twoLevBLUP} and \ref{alg:twoLevMFVB} can be used
to achieve streamlined fitting and inference for the contrast curve
extension, but with key matrices having new definitions. Firstly,
the $\bX_i$, $\bZgbli$ and $\bZgrpi$ matrices need to become:
$$\bX_i=[\begin{array}{ccccc}
\bone & \bx_i & \bone-\biota^A_i & (\bone-\biota^A_i)\odot\bx_i
\end{array}],
$$
$$\bZgbli=[{\setlength\arraycolsep{2pt}
\begin{array}{cccccc}
\biota^A_i\odot\zgblo(\bx_i) & \cdots & \biota^A_i\odot\zgblKgbl(\bx_i)&
(\bone-\biota^A_i)\odot\zgblo(\bx_i) & \cdots & (\bone-\biota^A_i)\odot\zgblKgbl(\bx_i)
\end{array}}]
$$
and
$$\bZgrpi=[
{\setlength\arraycolsep{2pt}
\begin{array}{cccccc}
\biota^A_i\odot\zgrpo(\bx_i)& \cdots& \biota^A_i\odot\zgrpKgrp(\bx_i)&
(\bone-\biota^A_i)\odot\zgrpo(\bx_i) & \cdots & (\bone-\biota^A_i)\odot\zgrpKgrp(\bx_i)
\end{array}}]
$$
where $\biota^A_i$ is the $n_i\times1$ vector of $\iota^A_{ij}$ values.
In the case of best linear unbiased prediction the updates for the
$\bB_i$ and $\bBdot_i$ matrices in Algorithm \ref{alg:twoLevBLUP}
need to be replaced by:
$$
\Bmati\thickarrow
\left[
\begin{array}{cc}
\sigeps^{-1}\bX_i & \sigeps^{-1}\bZgbli\\[1ex]
\bO & m^{-1/2}\left[\begin{array}{cc}
(\sigmaGbl^A)^{-1}\bI_{\Kgbl}& \bzero \\
\bzero & (\sigmaGbl^B)^{-1}\bI_{\Kgbl}
\end{array}
\right]\\[1ex]
\bO & \bO   \\[1ex]
\bO & \bO   \\[1ex]
\end{array}
\right]\ \mbox{and}\
\Bmatdoti\thickarrow
\left[
\begin{array}{cc}
\sigeps^{-1}\bX_i  & \sigeps^{-1}\bZgrpi        \\[1ex]
\bO                & \bO            \\[1ex]
\bSigma^{-1/2}     & \bO            \\[1ex]
\bO                & \sigmaGrp^{-1}\bI_{2\Kgrp}
\end{array}
\right]
$$
and the output coefficient vectors change to
$$\left[
\begin{array}{c}
\bbetahat\\[1ex]
\buHatGbl^A\\[1ex]
\buHatGbl^B
\end{array}
\right]
\quad\mbox{and}\quad
\left[
\begin{array}{c}
\buHatLini^A\\[1ex]
\buHatLini^B\\[1ex]
\buHatGrpi^A\\[1ex]
\buHatGrpi^B
\end{array}
\right].
$$

In the case of mean field variational Bayes the updates of the $\bB_i$ and $\bBdot_i$
matrices in Algorithm \ref{alg:twoLevMFVB} need to be replaced by:
$$
\Bmati\thickarrow
\left[
\begin{array}{cc}
\mu_{\qDens(1/\sigeps^2)}^{1/2}\bX_i & \mu_{\qDens(1/\sigeps^2)}^{1/2}\bZgbli\\[1.5ex]
m^{-1/2}\bSigma_{\bbeta}^{-1/2} & \bO \\[1ex]
\bO & m^{-1/2}\left[
\begin{array}{cc}
\mu_{\qDens(1/(\sigmaGbl^A)^2)}^{1/2}\bI_{\Kgbl} & \bzero \\
\bzero     &  \mu_{\qDens(1/(\sigmaGbl^B)^2)}^{1/2}\bI_{\Kgbl}
\end{array}
\right]\\[1ex]
\bO & \bO   \\[1ex]
\bO & \bO   \\[1ex]
\end{array}
\right],
$$
and
$$
\Bmatdoti\thickarrow
\left[
\begin{array}{cc}
\mu_{\qDens(1/\sigeps^2)}^{1/2}\bX_i  & \mu_{\qDens(1/\sigeps^2)}^{1/2}\bZgrpi        \\[1ex]
\bO                & \bO            \\[1ex]
\bO                & \bO            \\[1ex]
\bM_{\qDens(\bSigma^{-1})}^{1/2}     & \bO            \\[1ex]
\bO                & \mu_{\qDens(1/\sigmaGrp^2)}^{1/2}\bI_{2\Kgrp}
\end{array}
\right].
$$

A contrast curves adjustment to the mean field variational
Bayes updates is also required for some of 
the covariance matrix parameters. However, these 
calculations are comparatively simple and analogous
to those given in Section \ref{sec:drvAlgTwo}.

\begin{figure}
\begin{tabular}{cc}
  \includegraphics[width=72mm]{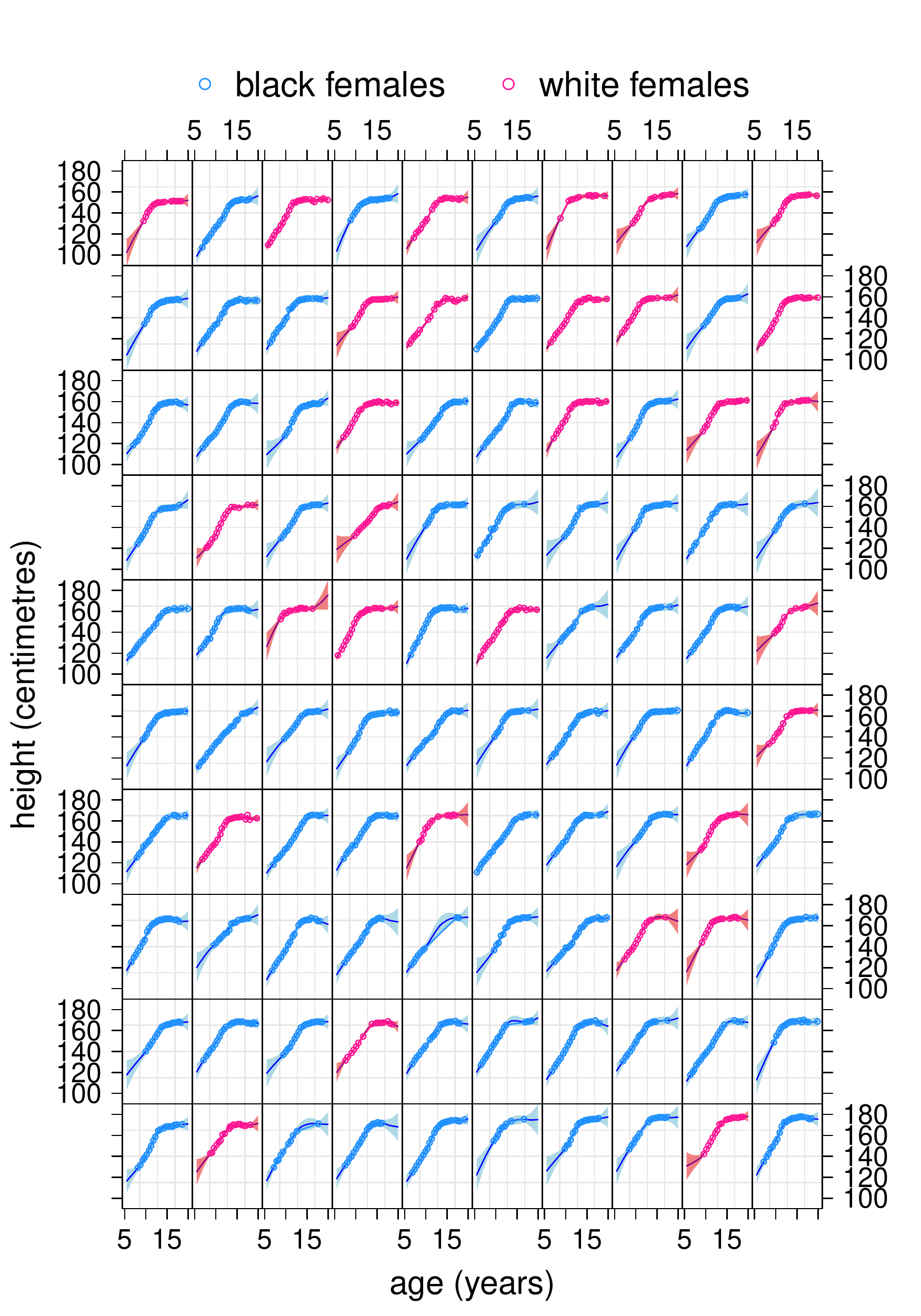} &
  \includegraphics[width=72mm]{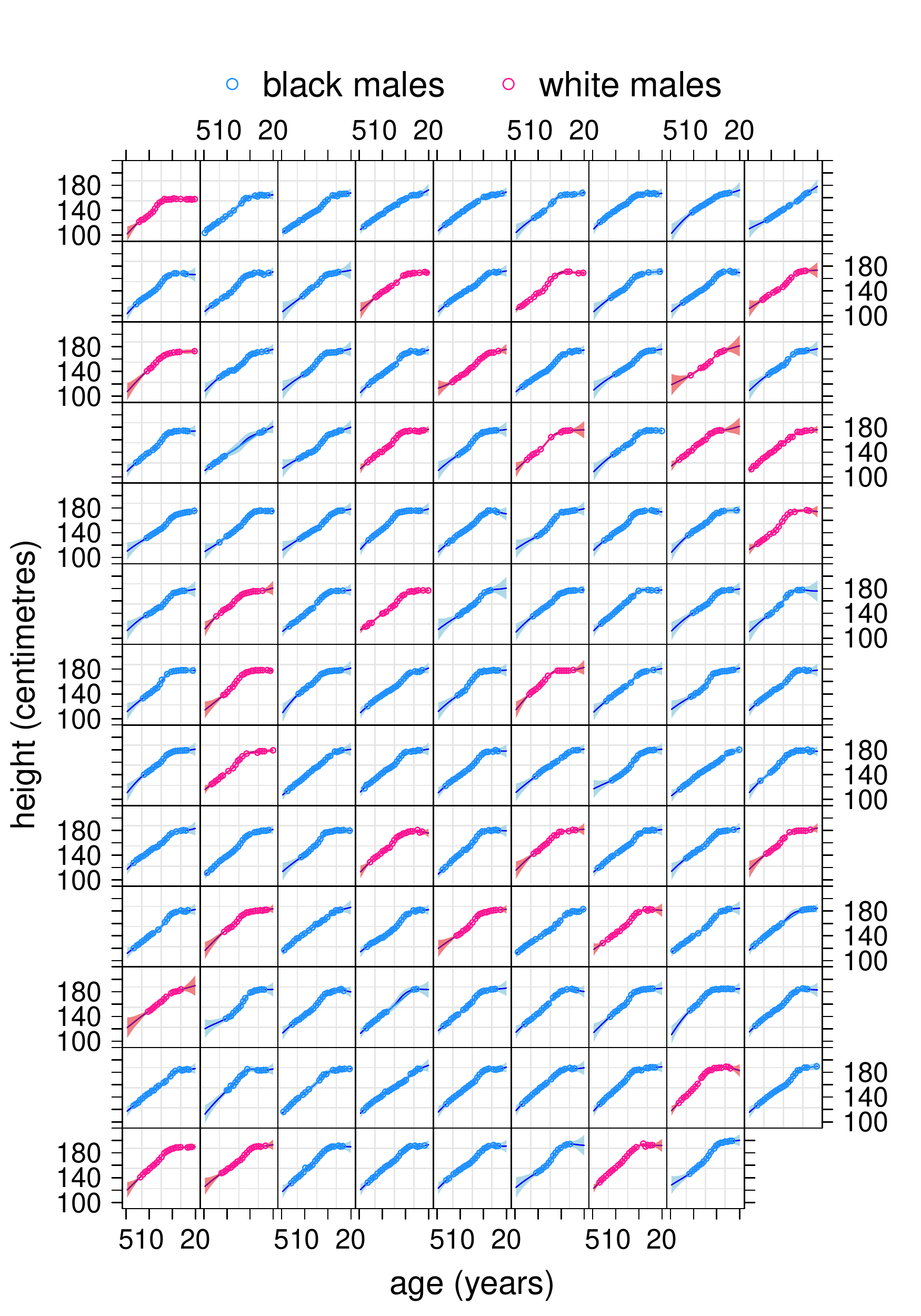} \\
  \includegraphics[width=72mm]{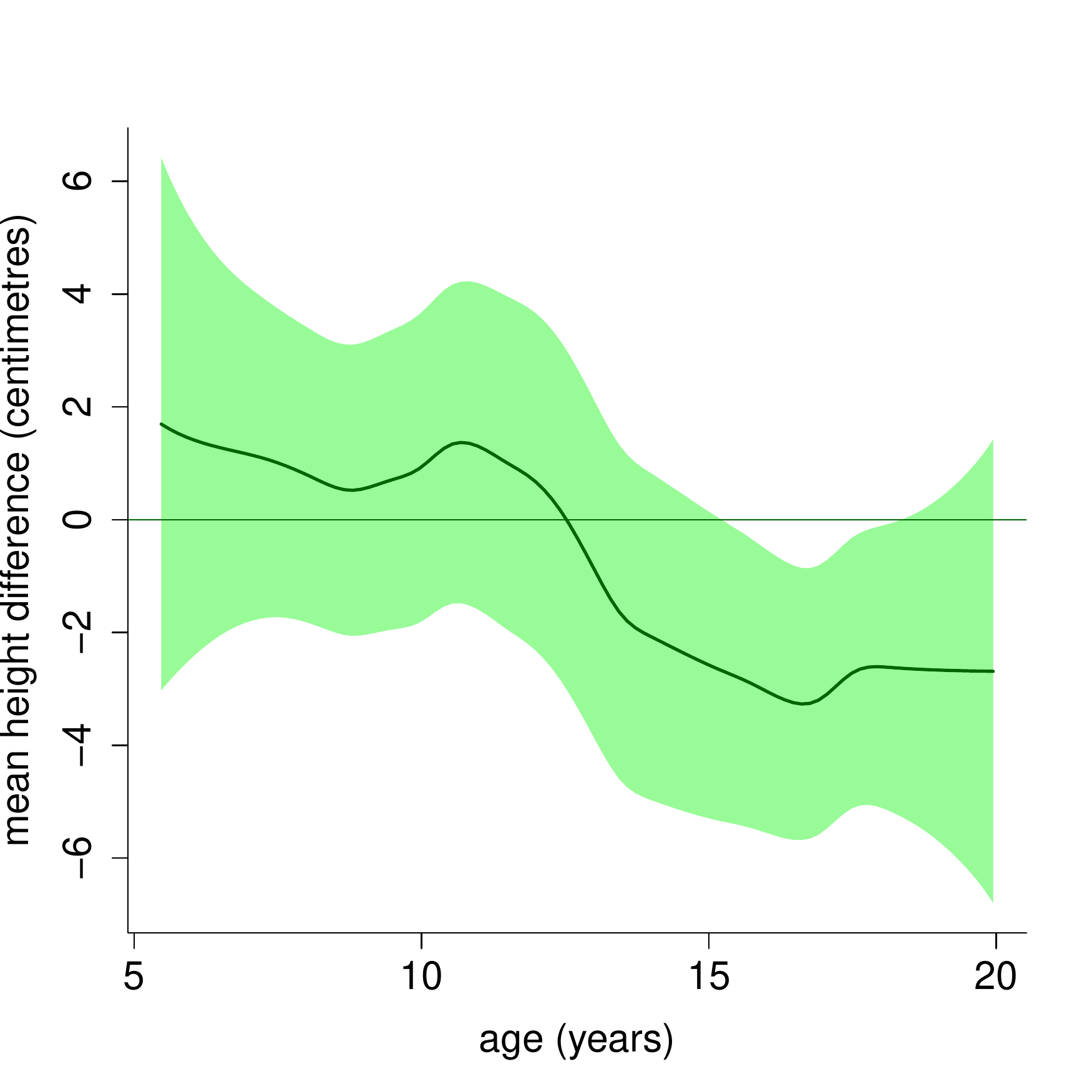} &
  \includegraphics[width=72mm]{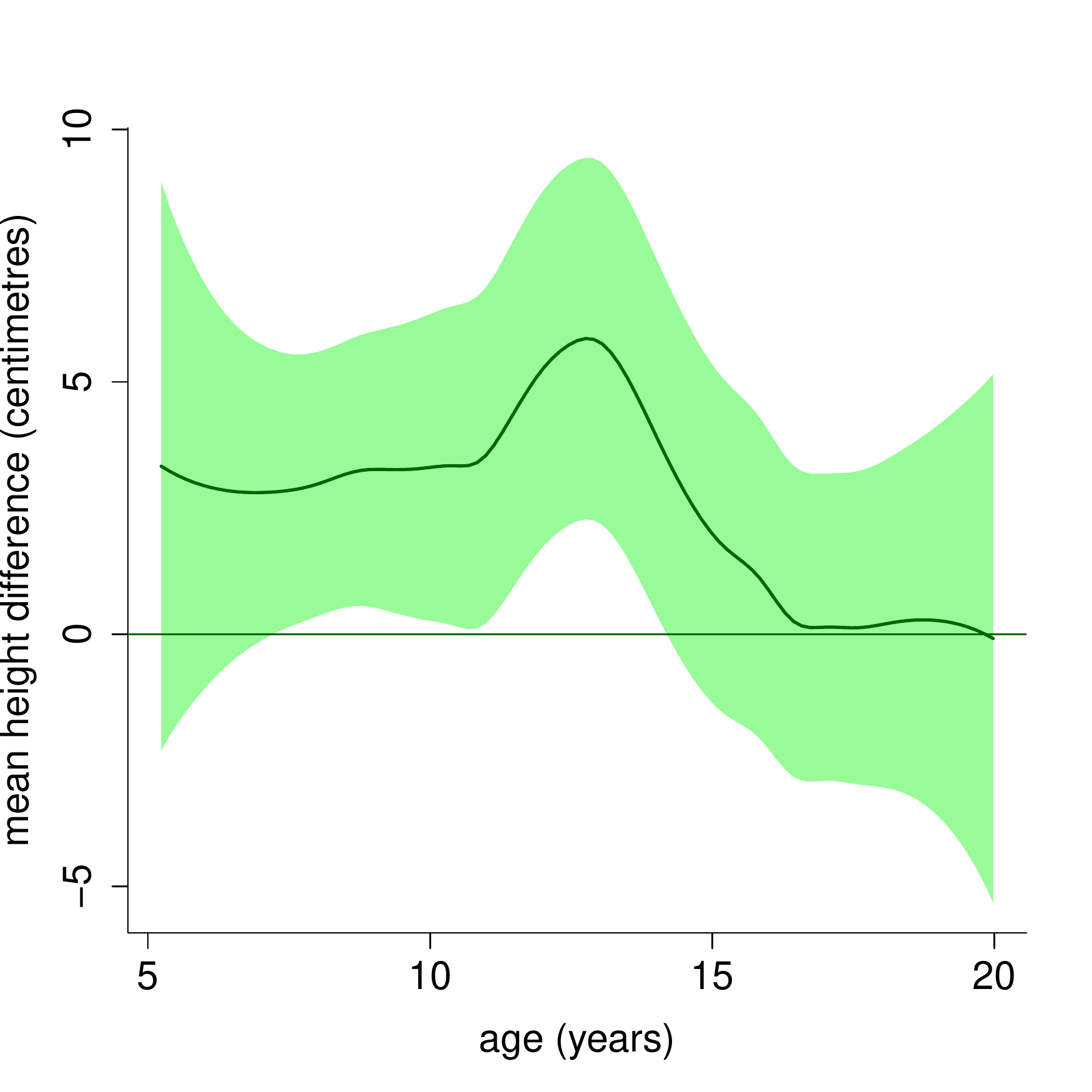} \\
\end{tabular}
\caption{
  \textit{Top panels: fitted group-specific curves for 100 female subjects (left) and
  116 male subjects (right) from the data on adolescent somatic growth 
(Pratt {\it et al.} 1989). The shading corresponds to approximate pointwise 99\% credible intervals.
  Bottom panels: similar to the top panels but for the estimated contrast curve. The shaded
  regions correspond to approximate pointwise 95\% credible intervals.}}
\label{fig:growthIndianaFits}
\end{figure}

We demonstrate the use of Algorithm \ref{alg:twoLevMFVB} in this setting for data from a
longitudinal study on adolescent somatic growth. More detail on this data can be found in
Pratt {\it et al.} (1989). The variables of interest are
\begin{equation*}
   \begin{array}{lcl}
      y_{ij} & = & \mbox{$j$th height measurement (centimetres) of subject $i$, and}
      \\
      x_{ij} & = & \mbox{age (years) of subject $i$ when $y_{ij}$ is recorded,}
   \end{array}
\end{equation*}
for $1 \le i \le m$ and $1 \le j \le n_{i}$.
The subjects are categorized into black ethnicity and white ethnicity
and comparison of mean height between the two populations is of interest.
Algorithm \ref{alg:twoLevMFVB} is seen to have good agreement with
the data in each sub-panel of the top two plots in Figure \ref{fig:growthIndianaFits}.
The bottom panels of Figure \ref{fig:growthIndianaFits} show the estimated height
gap between black and white adolescents as a function of age.
For the females, there is a significant height difference only at 16-17 years old. Between
5 and 15 years, there is no obvious height difference.
For the males, it is highest and (marginally) statistically
significant up to about 14 years of age, peaking at 13 years of
age. Between 17 and 20 years old there is no discernible height
difference between the two populations.
%

\section{Three-Level Models}\label{sec:threeLevMods}

The three-level version of group-specific curve models corresponds
to curve-type data having two nested groupings. For example, the
data in each panel of Figure \ref{fig:MNSWintro} are first grouped
according to slice, which is the level 2 group, and
the slices are grouped according to tumor which is the 
level 3 group. We denote predictor/response pairs as $(x_{ijk},y_{ijk})$
where $x_{ijk}$ is the $k$th value of the predictor variable in the $i$th level 3 group
and $(i,j)$th level 2 group and $y_{ijk}$ is the corresponding value of
the response variable. 
We let $m$ denote the number of level 3 groups,
$n_i$ denote that number of level 2 groups in the $i$th level 3 group
and $o_{ij}$ denote the the number of units within the 
$(i,j)$th level 2 group. The Figure \ref{fig:MNSWintro} data, which happen to
be balanced, are such that
{\setlength\arraycolsep{1pt}
\begin{eqnarray*}
m&=&\mbox{number of tumors}=10,\\
n_i&=&\mbox{number of slices for the $i$th tumor}=5\\
\mbox{and}\ \ o_{ij}&=&\mbox{number of predictor/response pairs for the $i$th tumor and $j$th slice}=128.
\end{eqnarray*}
}

The Gaussian response three-level group specific curve model for such data is
\begin{equation}
\begin{array}{l}
y_{ijk}=f(x_{ijk})+g_i(x_{ijk})+h_{ij}(x_{ijk})+\varepsilon_{ijk},
\quad\varepsilon_{ijk}\simind N(0,\sigeps^2),\\[2ex]
\quad
1\le i\le m,\ 1\le j\le n_i,\ 1\le k\le o_{ij},
\end{array}
\label{eq:threeLevelfg}
\end{equation}
where the smooth function $f$ is the global mean function, the $g_i$ functions, $1\le i\le m$,
allow for group-specific deviations according to membership of the $i$th level 3 group
and the $h_{ij}$, $1\le i\le m$ and $1\le j\le n_i$ allow for an additional level of
group-specific deviations according to membership of the $j$th level 2 group within the $i$th level 3
group. The mixed model-based penalized spline models for these functions are
{\setlength\arraycolsep{1pt}
\begin{eqnarray*}
f(x)&=&\beta_0+\beta_1\,x+\sum_{k=1}^{\Kgbl}\,\uGblk\,\zgblk(x),\quad
\uGblk\simind N(0,\sigmaGbl^2),\\[1ex]
g_i(x)&=&\uLiniz^g+\uLinio^g\,x+\sum_{k=1}^{\Kgrp^g}\,\uGrpik^g\,\zgrpk^g(x),
\ \left[\begin{array}{c}
\uLiniz^g\\[1ex]
\uLinio^g
\end{array}
\right]\simind N(\bzero,\bSigmag),\
\uGrpik^g\simind N\big(0,\sigmaGrpg^2\big)\\[1ex]
\mbox{and}&&\\[1ex]
h_{ij}(x)&=&\uLinijz^h+\uLinijo^h\,x+\sum_{k=1}^{\Kgrp^h}\,\uGrpijk^h\,\zgrpk^h(x),
\ \left[\begin{array}{c}
\uLinijz^h\\[1ex]
\uLinijo^h
\end{array}
\right]\simind N(\bzero,\bSigmah),\
\uGrpijk^h\simind N\big(0,\sigmaGrph^2\big),
\end{eqnarray*}
}
with all random effect distributions independent of each other.
For this three-level case we have three bases:
$$\{\zgblk(\cdot):1\le k\le \Kgbl\},\quad \{\zgrpk^g(\cdot):1\le k\le \Kgrp^g\}
\quad\mbox{and}\quad\{\zgrpk^h(\cdot):1\le k\le \Kgrp^h\}.
$$
The variance and covariance matrix parameters are analogous to the
two-level model. For example, $\bSigmag$ and $\bSigmah$ are both
unstructured $2\times2$ matrices corresponding to the linear components
of the $g_i$ and $h_{ij}$ respectively.

The following notation is useful for setting up the required
design matrices: if $\bM_1,\ldots,\bM_d$ is a set of matrices
each having the same number of columns then
$$\stack{1\le i\le d}(\bM_i)\equiv
\left[
\begin{array}{c}
\bM_1\\
\vdots\\
\bM_d
\end{array}
\right].
$$
We then define, for $1\le i\le m$ and $1\le j\le n_i$,
$$\bx_i\equiv\stack{1\le j\le n_i}\big(\bx_{ij}\big)
\quad\mbox{and}\quad
\bx_{ij}\equiv\stack{1\le k\le\, o_{ij}}\big(x_{ijk}\big).
$$

\subsection{Best Linear Unbiased Prediction}

Model (\ref{eq:threeLevelfg}) is expressible as a Gaussian response linear
mixed model as follows:
\begin{equation}
\by|\bu\sim N(\bX\bbeta+\bZ\,\bu,\sigeps^2\,\bI),\quad \bu\sim N(\bzero,\bG),
\label{eq:threeLevFreq}
\end{equation}
where the design matrices are
$$\bX=\stack{1 \le i \le m}\big(\bX_i\big)
\quad\mbox{with}\quad
\bX_i=\stack{1 \le j \le n_i}\big(\bX_{ij}\big)
\quad\mbox{and}\quad
\bX_{ij}\equiv[\bone\ \bx_{ij}]
$$
and
$$
\bZ\equiv\Big[\bZgbl\,
\blockdiag{1\le i\le m}
\Big[\stack{1\le j\le n_i}([\bX_{ij}\ \bZLoneGrpij])\,
\blockdiag{1\le j\le n_i}([\bX_{ij}\ \bZLtwoGrpij])
\Big]\Big].
$$
where
$$\bZgbl\equiv\stack{1\le i\le m}\big(\stack{1\le j\le n_i}(\bZgblij)\big)$$
and the matrices $\bZgblij$, $\bZLoneGrpij$ and $\bZLtwoGrpij$,  $1\le i\le m$, $1\le j\le n_i$,
contain, respectively, spline basis functions for the global mean function $f$, the $i$th
level one group deviation functions $g_i$ and $(i,j)$th level two group deviation
functions $h_{ij}$. Specifically,
{\setlength\arraycolsep{1pt}
\begin{eqnarray*}
\bZgblij&\equiv&[\begin{array}{ccc}
\zgblo(\bx_{ij}) & \cdots & \zgblKgbl(\bx_{ij})
\end{array}],\quad\bZLoneGrpij=
[\zLoneGrpo(\bx_{ij}) \cdots \zLoneGrpK(\bx_{ij})]\\[2ex]
\mbox{and}\ \bZLtwoGrpij&\equiv&
[\zLtwoGrpo(\bx_{ij}) \cdots \zLtwoGrpK(\bx_{ij})]
\quad\mbox{for $1\le i\le m$ and $1\le j\le n_i$.}
\end{eqnarray*}
}
The fixed and random effects vectors are
$$
\bbeta\equiv\left[\begin{array}{c}
\beta_0\\[1ex]
\beta_1
\end{array}
\right]
\quad
\mbox{and}
\quad
\bu\equiv
\left[
\begin{array}{c}
\buGbl\\[1ex]
{\displaystyle\stack{1\le i\le m}}\left(
\left[
\begin{array}{c}
\left[
\begin{array}{c}
\buLoneLini\\
\buLoneGrpi
\end{array}
\right]
\\[2ex]
\left[
\begin{array}{c}
\buLtwoLinio\\
\buLtwoGrpio
\end{array}
\right]
\\[2ex]
\vdots
\\[2ex]
\left[
\begin{array}{c}
\buLtwoLinini\\
\buLtwoGrpini
\end{array}
\right]
\end{array}
\right]
\right)
\end{array}
\right]
\quad\mbox{where}\quad
\buLoneLini\equiv
\left[\begin{array}{c}
\uLiniz^g\\[1ex]
\uLinio^g
\end{array}
\right]
$$
with $\buLoneGrpi$, $\buLtwoLinij$ and $\buLtwoGrpij$ defined similarly
and the covariance matrix of $\bu$ is
\begin{equation}
  \begin{array}{c}
    \bG=\Cov(\bu)
    =\left[
    \begin{array}{cc}
    \sigmaGbl^2\bI    &  \bO                             \\[1ex]
    \bO             &  \displaystyle{\blockdiag{1\le i\le m}}
                       \left[
                       \begin{array}{ccc}
                       \bSigmag & \bO            & \bO       \\
                       \bO       &\sigmaGrpg^2\bI & \bO       \\
                       \bO       &  \bO           & \bI_{n_i}\otimes
                       \left[
                       \begin{array}{cc}
                       \bSigmah & \bO \\
                       \bO       & \sigmaGrph^2\bI
                       \end{array}
                       \right]
                       \end{array}
                       \right]
    \end{array}
    \right].
  \end{array}
  \label{eqn:threeLevBLUPCov}
\end{equation}

\noindent
We define matrices in a similar way to what is given in (\ref{eq:CDRmatBLUPdefs}).
The best linear unbiased predictor of $[ \bbeta \ \bu ]$ and corresponding covariance matrix are
as shown in (\ref{eq:BLUPandCov}), but, with entries as described in this section.  This covariance
matrix grows quadratically in both $m$ and the $n_{i}$s, and so, storage becomes impractical
for large numbers of level 2 and level 3 groups. However, only certain sub-blocks are required for
the addition of pointwise confidence intervals to curve estimates. In particular, we only require the
non-zero sub-blocks of the general three-level sparse matrix given in Section 3 of Nolan
\myand Wand (2018) that correspond to $(\bC^T\RBLUP^{-1}\bC+\DBLUP)^{-1}$. In the case of the
three-level Gaussian response linear model, Nolan \myand Wand's
\begin{equation*}
  \begin{array}{l}
    \bA_{11} \mbox{ sub-block corresponds to a } (2 + \Kgbl) \times (2 + \Kgbl) \mbox{ matrix }
      \Cov\left(\left[
      \begin{array}{c}
      \bbetahat\\
      \buHatGbl-\buGbl
      \end{array}
      \right]\right); \\[2ex]
    \bA_{22,i} \mbox{ sub-block corresponds to a } (2 + \Kgrpg) \times (2 + \Kgrpg) \mbox{ matrix }
    \Cov\left(\left[
    \begin{array}{c}
    \buHatLoneLini-\buLoneLini\\[1ex]
    \buHatLoneGrpi-\buLoneGrpi
    \end{array}
    \right]
    \right); \\[1ex]
    \bA_{12,i} \mbox{ sub-block corresponds to a } (2 + \Kgbl) \times (2 + \Kgrpg) \mbox{ matrix } \\[1ex]
    \qquad \qquad \qquad
      E\left\{
      \left[
      \begin{array}{c}
      \bbetahat\\
      \buHatGbl-\buGbl
      \end{array}
      \right]
      \left[
      \begin{array}{c}
      \buHatLoneLini-\buLoneLini\\[1ex]
      \buHatLoneGrpi-\buLoneGrpi
      \end{array}
      \right]^T
      \right\}, \ 1\le i \le m; \\[1ex]
    \bA_{22,ij} \mbox{ sub-block corresponds to a } (2 + \Kgrph) \times (2 + \Kgrph) \mbox{ matrix }
      \Cov
      \left(
      \left[
      \begin{array}{c}
      \buHatLtwoLinij-\buLtwoLinij\\[1ex]
      \buHatLtwoGrpij-\buLtwoGrpij
      \end{array}
      \right]
      \right); \\[1ex]
    \bA_{12,ij} \mbox{ sub-block corresponds to a } (2 + \Kgbl) \times (2 + \Kgrph) \mbox{ matrix } \\[1ex]
      \qquad \qquad \qquad E\left\{
      \left[
      \begin{array}{c}
      \bbetahat\\
      \buHatGbl-\buGbl
      \end{array}
      \right]
      \left[
      \begin{array}{c}
      \buHatLtwoLinij-\buLtwoLinij\\[1ex]
      \buHatLtwoGrpij-\buLtwoGrpij
      \end{array}
      \right]^T
      \right\}; \\[1ex]
    \bA_{12,\iCOMMAj} \mbox{ sub-block corresponds to a } (2 + \Kgrpg) \times (2 + \Kgrph) \mbox{ matrix } \\[1ex]
      \qquad \qquad \qquad E\left\{
      \left[
      \begin{array}{c}
      \buHatLoneLini-\buLoneLini\\[1ex]
      \buHatLoneGrpi-\buLoneGrpi
      \end{array}
      \right]
      \left[
      \begin{array}{c}
      \buHatLtwoLinij-\buLtwoLinij\\[1ex]
      \buHatLtwoGrpij-\buLtwoGrpij
      \end{array}
      \right]^T
      \right\}, \ 1 \le i \le m, \ 1 \le j \le n_{i}.
  \end{array}
\end{equation*}
As described in Nolan, Menictas \myand Wand (2019), the $\SolveThreeLevelSparseLeastSquares$
algorithm arises in the special case where $\bx$ is the minimizer of the least squares
problem given in equation (\ref{eqn:sparseLeastSquares}), where $\bB$ has the three-level sparse
form and $\bb$ is partitioned according to that shown in equation (7) of Nolan \myand Wand (2018).
This algorithm can be used for fitting three-level group-specific curve models by making use
of Result \ref{res:threeLevelBLUP}.

\begin{result}
Computation of $[\bbetahat^T\ \ \buhat^T]^T$ and each
of the sub-blocks of $\mbox{\rm Cov}([\bbetahat^T\ \ (\buhat-\bu)^T]^T)$
listed in (\ref{eq:CovMain}) are expressible as the three-level sparse
matrix least squares form:
$$\left\Vert\bb-\bB\left[
\begin{array}{c}
\bbeta\\
\bu
\end{array}
\right]
\right\Vert^2
$$
where the non-zero sub-blocks $\bB$ and $\bb$, according to the notation
in Section 3.1 of Nolan \myand Wand (2018), are
for $1\le i\le m$ and $1\le j\le n_i$:
$$
\bvecij \equiv \left[
  \begin{array}{c}
    \sigeps^{-1}\by_{ij}\\[1ex]
    \bzero \\[1ex]
    \bzero \\[1ex]
    \bzero \\[1ex]
    \bzero \\[1ex]
    \bzero \\[1ex]
  \end{array} \right],
\qquad
\Bmatij\equiv \left[
\begin{array}{cc}
  \sigeps^{-1}\bX_{ij} & \sigeps^{-1}\bZgblij\\[1ex]
  \bO & \ndotmh \sigmaGbl^{-1}\bI_{\Kgbl} \\[1ex]
  \bO & \bO   \\[1ex]
  \bO & \bO   \\[1ex]
  \bO & \bO   \\[1ex]
  \bO & \bO   \\[1ex]
\end{array} \right],
$$
$$
\Bmatdotij\equiv \left[
   \begin{array}{cc}
     \sigeps^{-1}\bX_{ij} & \sigeps^{-1}\bZLoneGrpij \\[1ex]
     \bO & \bO \\[1ex]
     n_{i}^{-1/2}\bSigmag^{-1/2} & \bO \\[1ex]
     \bO & n_{i}^{-1/2} \sigmaGrpg^{-1} \bI_{\Kgrpg} \\[1ex]
     \bO & \bO \\[1ex]
     \bO & \bO \\[1ex]
   \end{array}\right]
\quad and\quad
\Bmatdotdotij\equiv \left[
\begin{array}{cc}
  \sigeps^{-1}\bX_{ij} & \sigeps^{-1}\bZLtwoGrpij \\[1ex]
  \bO & \bO \\[1ex]
  \bO & \bO \\[1ex]
  \bO & \bO \\[1ex]
  \bSigmah^{-1/2} & \bO \\[1ex]
  \bO & \sigmaGrph^{-1}\bI_{\Kgrph}
\end{array}\right]
$$
with each of these matrices having $\oadj_{ij}=o_{ij}+\Kgbl+2+\Kgrpg+2+\Kgrph$
rows and with $\Bmati$ having $p=2+\Kgbl$ columns, $\Bmatdoti$ having $q_1=2+\Kgrpg$
columns and $\Bmatdotdotij$ having $q_2=2+\Kgrph$ columns. The solutions are
$$\left[
\begin{array}{c}
\bbetahat\\
\buHatGbl
\end{array}
\right]=\xveco,\quad
\Cov\left(\left[
\begin{array}{c}
\bbetahat\\
\buHatGbl-\buGbl
\end{array}
\right]\right)=\AUoo,
$$
$$\left[
\begin{array}{c}
\buHatLoneLini\\[1ex]
\buHatLoneGrpi
\end{array}
\right]=\xvectCi,\quad
E\left\{
\left[
\begin{array}{c}
\bbetahat\\
\buHatGbl-\buGbl
\end{array}
\right]
\left[
\begin{array}{c}
\buHatLoneLini-\buLoneLini\\[1ex]
\buHatLoneGrpi-\buLoneGrpi
\end{array}
\right]^T
\right\}=\AUotCi,
$$
$$
\Cov\left(\left[
\begin{array}{c}
\buHatLoneLini-\buLoneLini\\[1ex]
\buHatLoneGrpi-\buLoneGrpi
\end{array}
\right]
\right)=\AUttCi,\ \ 1\le i\le m,
$$
$$\left[
\begin{array}{c}
\buHatLtwoLinij\\[1ex]
\buHatLtwoGrpij
\end{array}
\right]=\bx_{2,ij},\quad
E\left\{
\left[
\begin{array}{c}
\bbetahat\\
\buHatGbl-\buGbl
\end{array}
\right]
\left[
\begin{array}{c}
\buHatLtwoLinij-\buLtwoLinij\\[1ex]
\buHatLtwoGrpij-\buLtwoGrpij
\end{array}
\right]^T
\right\}=\bA^{12,ij},
$$
$$
E\left\{
\left[
\begin{array}{c}
\buHatLoneLini-\buLoneLini\\[1ex]
\buHatLoneGrpi-\buLoneGrpi
\end{array}
\right]
\left[
\begin{array}{c}
\buHatLtwoLinij-\buLtwoLinij\\[1ex]
\buHatLtwoGrpij-\buLtwoGrpij
\end{array}
\right]^T
\right\}=\bA^{12,\iCOMMAj}
$$
and
$$
\Cov
\left(
\left[
\begin{array}{c}
\buHatLtwoLinij-\buLtwoLinij\\[1ex]
\buHatLtwoGrpij-\buLtwoGrpij
\end{array}
\right]
\right)=\bA^{22,ij},
\quad 1\le i\le m,\ 1\le j\le n_i.
$$
\label{res:threeLevelBLUP}
\end{result}

%
\begin{algorithm}[!th]
  \begin{center}
    \begin{minipage}[t]{154mm}
      \begin{small}
        \begin{itemize}
        \setlength\itemsep{2pt}
        \item[] Inputs: $\by_{ij}(o_{ij}\times1),\ \bX_{ij}(o_{ij}\times 2),\ \bZgblij(o_{ij}\times \Kgbl),\
                \bZLoneGrpij(o_{ij}\times \Kgrpg),$
        \item[] \qquad\qquad $\bZLtwoGrpij(o_{ij}\times \Kgrph), 1\le i\le m, \ 1\le j\le n_{i}$;
                $\quad\sigeps^2,\sigmaGbl^2,\sigmaGrpg^2,\sigmaGrph^2>0$,
        \item[] \qquad\qquad $\bSigmag(2\times2), \ \bSigmah(2\times 2),
                \mbox{symmetric and positive definite.}$
        \item[] For $i=1,\ldots,m$:
        \begin{itemize}
          \item[] For $j=1,\ldots,n_{i}$:
          \begin{itemize}
            \setlength\itemsep{4pt}
            \item[] $\begin{array}{l}
                      \bvecij\thickarrow \left[
                        \begin{array}{c}
                          \sigeps^{-1}\by_{ij}\\[1ex]
                          \bzero \\[1ex]
                          \bzero \\[1ex]
                          \bzero \\[1ex]
                          \bzero \\[1ex]
                          \bzero \\[1ex]
                        \end{array} \right],
                        \
                        \Bmatij\thickarrow \left[
                        \begin{array}{cc}
                          \sigeps^{-1}\bX_{ij} & \sigeps^{-1}\bZgblij\\[1ex]
                          \bO & \ndotmh \sigmaGbl^{-1}\bI_{\Kgbl} \\[1ex]
                          \bO & \bO   \\[1ex]
                          \bO & \bO   \\[1ex]
                          \bO & \bO   \\[1ex]
                          \bO & \bO   \\[1ex]
                        \end{array} \right]
                      \end{array}$
             \item[] $\begin{array}{l}
                      \Bmatdotij\thickarrow \left[
                         \begin{array}{cc}
                           \sigeps^{-1}\bX_{ij} & \sigeps^{-1}\bZLoneGrpij \\[1ex]
                           \bO & \bO \\[1ex]
                           n_{i}^{-1/2}\bSigmag^{-1/2} & \bO \\[1ex]
                           \bO & n_{i}^{-1/2} \sigmaGrpg^{-1} \bI_{\Kgrpg} \\[1ex]
                           \bO & \bO \\[1ex]
                           \bO & \bO \\[1ex]
                         \end{array}\right],
                         \
                         \Bmatdotdotij\thickarrow \left[
                         \begin{array}{cc}
                           \sigeps^{-1}\bX_{ij} & \sigeps^{-1}\bZLtwoGrpij \\[1ex]
                           \bO & \bO \\[1ex]
                           \bO & \bO \\[1ex]
                           \bO & \bO \\[1ex]
                           \bSigmah^{-1/2} & \bO \\[1ex]
                           \bO & \sigmaGrph^{-1}\bI_{\Kgrph}
                         \end{array}\right]
                      \end{array}$
          \end{itemize}
        \end{itemize}
        \item[] $\Ssc_3\thickarrow\SolveThreeLevelSparseLeastSquares\Big(\big\{(
                  \bvecij,\Bmatij,\Bmatdotij,\Bmatdotdotij):1\le i\le m, $
        \item[] $\qquad \qquad 1 \le j\le n_{i} \big\}\Big)$
        \item[] $\left[
                  \begin{array}{c}
                    \bbetahat\\
                    \buHatGbl
                  \end{array} \right]\thickarrow\mbox{$\xveco$ component of $\Ssc_3$}$\ \ \ ;\ \ \
                  $\Cov\left(\left[
                  \begin{array}{c}
                    \bbetahat\\
                    \buHatGbl-\buGbl
                  \end{array} \right]\right)\thickarrow\mbox{$\AUoo$ component of $\Ssc_3$}$
        \item[] For $i=1,\ldots,m$:
                \begin{itemize}
                  \setlength\itemsep{4pt}
                  \item[] $\left[
                            \begin{array}{c}
                            \buHatLoneLini\\[1ex]
                            \buHatLoneGrpi
                            \end{array}
                            \right]\thickarrow\mbox{$\xvectCi$ component of $\Ssc_3$}$
                  \item[] $\Cov\left(\left[ \begin{array}{c}
                                              \buHatLoneLini-\buLoneLini\\[1ex]
                                              \buHatLoneGrpi-\buLoneGrpi
                                            \end{array} \right] \right)\thickarrow\mbox{$\AUttCi$ component of $\Ssc_3$}$
                  \item[] $E\left\{ \left[ \begin{array}{c}
                                            \bbetahat\\
                                            \buHatGbl-\buGbl
                                          \end{array} \right] \left[
                                          \begin{array}{c}
                                            \buHatLoneLini-\buLoneLini\\[1ex]
                                            \buHatLoneGrpi-\buLoneGrpi
                                          \end{array} \right]^T \right\}\thickarrow\mbox{$\AUotCi$ component of $\Ssc_3$}$
                  \item[] \textsl{continued on a subsequent page}\ $\ldots$
                \end{itemize}
        \end{itemize}
      \end{small}
    \end{minipage}
  \end{center}
  \caption{\it Streamlined algorithm for obtaining best linear unbiased predictions
  and corresponding covariance matrix components for the two-level group specific curves
  model.}
  \label{alg:threeLevBLUP}
\end{algorithm}

\setcounter{algorithm}{2}
\begin{algorithm}[!th]
  \begin{center}
    \begin{minipage}[t]{154mm}
      \begin{small}
        \begin{itemize}
          \item[]
          \begin{itemize}
            \item[]
            \begin{itemize}
                  \item[] For $j=1,\ldots,n_{i}$:
                  \item[] \begin{itemize}
                            \item[] $\left[
                                      \begin{array}{c}
                                      \buHatLtwoLinij\\[1ex]
                                      \buHatLtwoGrpij
                                      \end{array}
                                      \right]\thickarrow\mbox{$\bx_{2,ij}$ component of $\Ssc_3$}$
                            \item[] $E\left\{
                                      \left[
                                      \begin{array}{c}
                                      \bbetahat\\
                                      \buHatGbl-\buGbl
                                      \end{array}
                                      \right]
                                      \left[
                                      \begin{array}{c}
                                      \buHatLtwoLinij-\buLtwoLinij\\[1ex]
                                      \buHatLtwoGrpij-\buLtwoGrpij
                                      \end{array}
                                      \right]^T
                                      \right\}\thickarrow\mbox{$\bA^{12,ij}$ component of $\Ssc_3$}$
                          \item[] $E\left\{
                        \left[
                        \begin{array}{c}
                        \buHatLoneLini-\buLoneLini\\[1ex]
                        \buHatLoneGrpi-\buLoneGrpi
                        \end{array}
                        \right]
                        \left[
                        \begin{array}{c}
                        \buHatLtwoLinij-\buLtwoLinij\\[1ex]
                        \buHatLtwoGrpij-\buLtwoGrpij
                        \end{array}
                        \right]^T
                        \right\}\thickarrow\mbox{$\bA^{12,\iCOMMAj}$ component of $\Ssc_3$}$
                        \item[] $\Cov
                        \left(
                        \left[
                        \begin{array}{c}
                        \buHatLtwoLinij-\buLtwoLinij\\[1ex]
                        \buHatLtwoGrpij-\buLtwoGrpij
                        \end{array}
                        \right]
                        \right)\thickarrow\mbox{$\bA^{22,ij}$ component of $\Ssc_3$}$
            \end{itemize}
          \end{itemize}
          \end{itemize}
          \item[] Output: $$\begin{array}{c}\Bigg(\left[
                            \begin{array}{c}
                            \bbetahat\\
                            \buHatGbl
                            \end{array}
                            \right],\ \Cov\left(\left[
                            \begin{array}{c}
                            \bbetahat\\
                            \buHatGbl-\buGbl
                            \end{array}
                            \right]\right),
                            \Bigg\{\Bigg(
                            \left[
                            \begin{array}{c}
                            \buHatLini\\[1ex]
                            \buHatGrpi
                            \end{array}
                            \right],\,
                            \Cov\left(\left[
                            \begin{array}{c}
                            \buHatLini-\buLini\\[1ex]
                            \buHatGrpi-\buGrpi
                            \end{array}
                            \right]\right),\,\\[3ex]
                            E\left\{
                            \left[
                            \begin{array}{c}
                            \bbetahat\\
                            \buHatGbl-\buGbl
                            \end{array}
                            \right]
                            \left[
                            \begin{array}{c}
                            \buHatLini-\buLini\\[1ex]
                            \buHatGrpi-\buGrpi
                            \end{array}
                            \right]^T
                            \right\}
                            \Bigg):\ 1\le i\le m, \\[3ex]
                            \Bigg( \left[
                            \begin{array}{c}
                            \buHatLtwoLinij\\[1ex]
                            \buHatLtwoGrpij
                            \end{array}
                            \right],\,
                            E\left\{
                            \left[
                            \begin{array}{c}
                            \bbetahat\\
                            \buHatGbl-\buGbl
                            \end{array}
                            \right]
                            \left[
                            \begin{array}{c}
                            \buHatLtwoLinij-\buLtwoLinij\\[1ex]
                            \buHatLtwoGrpij-\buLtwoGrpij
                            \end{array}
                            \right]^T
                            \right\}, \\[3ex]
                            E\left\{
                            \left[
                            \begin{array}{c}
                            \buHatLoneLini-\buLoneLini\\[1ex]
                            \buHatLoneGrpi-\buLoneGrpi
                            \end{array}
                            \right]
                            \left[
                            \begin{array}{c}
                            \buHatLtwoLinij-\buLtwoLinij\\[1ex]
                            \buHatLtwoGrpij-\buLtwoGrpij
                            \end{array}
                            \right]^T
                            \right\}, \,
                            \Cov
                            \left(
                            \left[
                            \begin{array}{c}
                            \buHatLtwoLinij-\buLtwoLinij\\[1ex]
                            \buHatLtwoGrpij-\buLtwoGrpij
                            \end{array}
                            \right]
                            \right) \Bigg): \\[3ex]
                            1 \le i \le m, \, 1 \le j \le n_{i} \Big\} \Big)
                            \end{array}$$
        \end{itemize}
      \end{small}
    \end{minipage}
  \end{center}
  \caption{\textbf{continued.} \textit{This is a continuation of the description of this algorithm that commences
on a preceding page.}}
\end{algorithm}
%

A derivation of Result \ref{res:threeLevelBLUP} is 
given in Section \ref{sec:drvResultThree} of the web-supplement.
Result \ref{res:threeLevelBLUP} combined with Theorem 4 of
Nolan \myand Wand (2018) leads to Algorithm \ref{alg:threeLevBLUP}.
The \SolveThreeLevelSparseLeastSquares\ algorithm is given 
in Section \ref{sec:Solve3Lev}.
\subsection{Mean Field Variational Bayes}
A Bayesian extension of (\ref{eq:threeLevFreq}) and (\ref{eqn:threeLevBLUPCov}) is:
\begin{equation}
  \begin{array}{c}
    \by|\bbeta, \bu, \sigsqeps\sim N(\bX\bbeta+\bZ\,\bu,\sigeps^2\,\bI),\quad \bu|\sigmaGbl^2,\sigmaGrpg^2,\bSigmag,
    \sigmaGrph^2,\bSigmah\sim N(\bzero,\bG),
    \quad\mbox{$\bG$ as defined in (\ref{eqn:threeLevBLUPCov}),} \\[1ex]
    \bbeta\sim N(\bmu_{\bbeta},\bSigma_{\bbeta}),\quad\sigeps^2|\aeps\sim\mbox{Inverse-$\chi^2$}(\nuEps,1/\aeps),
    \quad\aeps\sim\mbox{Inverse-$\chi^2$}(1,1/(\nuEps\sEps^2)),\\[2ex]
    \quad\sigmaGbl^2|\aGbl\sim\mbox{Inverse-$\chi^2$}(\nuGbl,1/\aGbl),
    \quad\aGbl\sim\mbox{Inverse-$\chi^2$}(1,1/(\nuGbl\sGbl^2)),\\[2ex]
    \quad\sigmaGrpg^2|\aGrpg\sim\mbox{Inverse-$\chi^2$}(\nuGrpg,1/\aGrpg),
    \quad\aGrpg\sim\mbox{Inverse-$\chi^2$}(1,1/(\nuGrpg\sGrpg^2)),\\[2ex]
    \quad\sigmaGrph^2|\aGrph\sim\mbox{Inverse-$\chi^2$}(\nuGrph,1/\aGrph),
    \quad\aGrph\sim\mbox{Inverse-$\chi^2$}(1,1/(\nuGrph\sGrph^2)),\\[2ex]
    \bSigmag|\ASigmag\sim\mbox{Inverse-G-Wishart}\big(\Gfull,\nuSigmag+2,\ASigmag^{-1}\big),\\[2ex]
    \ASigmag\sim\mbox{Inverse-G-Wishart}(\Gdiag,1,\bLambda_{\ASigmag}),\quad
    \bLambda_{\ASigmag}\equiv\{\nuSigmag\diag(\sSigmagOne^2,\sSigmagTwo^2)\}^{-1},\\[2ex]
    \bSigmah|\ASigmah\sim\mbox{Inverse-G-Wishart}\big(\Gfull,\nuSigmah+2,\ASigmah^{-1}\big),\\[2ex]
    \ASigmah\sim\mbox{Inverse-G-Wishart}(\Gdiag,1,\bLambda_{\ASigmah}),\quad
    \bLambda_{\ASigmah}\equiv\{\nuSigmah\diag(\sSigmahOne^2,\sSigmahTwo^2)\}^{-1}.
  \end{array}
  \label{eq:threeLevBayes}
\end{equation}
The following mean field restriction is imposed on the joint posterior density function of all
parameters in (\ref{eq:threeLevBayes}):
\begin{equation}
\begin{array}{l}
  \pDens(\bbeta,\bu,\aeps,\aGbl,\aGrpg,\ASigmag,\aGrph,\ASigmah,\sigeps^2,\sigmaGbl^2,\sigmaGrpg^2,\bSigmag,
  \sigmaGrph^2,\bSigmah|\by) \\
  \qquad \qquad \approx \qDens(\bbeta,\bu,\aeps,\aGbl,\aGrpg,\ASigmag,\aGrph,\ASigmah)\,
  \qDens(\sigeps^2,\sigmaGbl^2,\sigmaGrpg^2,\bSigmag,\sigmaGrph^2,\bSigmah).
\end{array}
\label{eq:producRestrict3lev}
\end{equation}
The optimal $\qDens$-density functions for the parameters of interest are
{\setlength\arraycolsep{1pt}
\begin{eqnarray*}
&&\qDens^*(\bbeta,\bu)\ \mbox{has a $N\big(\bmu_{\qDens(\bbeta,\bu)},\bSigma_{\qDens(\bbeta,\bu)}\big)$ distribution,}\\[1ex]
&&\qDens^*(\sigeps^2)\ \mbox{has an $\mbox{Inverse-$\chi^2$}
\big(\xi_{\qDens(\sigeps^2)},\lambda_{\qDens(\sigeps^2)}\big)$ distribution,}\\[1ex]
&&\qDens^*(\sigmaGbl^2)\ \mbox{has an $\mbox{Inverse-$\chi^2$}
\big(\xi_{\qDens(\sigmaGbl^2)},\lambda_{\qDens(\sigmaGbl^2)}\big)$ distribution,}\\[1ex]
&&\qDens^*(\sigmaGrpg^2)\ \mbox{has an $\mbox{Inverse-$\chi^2$}
\big(\xi_{\qDens(\sigmaGrpg^2)},\lambda_{\qDens(\sigmaGrpg^2)}\big)$ distribution}\\[1ex]
&&\qDens^*(\sigmaGrph^2)\ \mbox{has an $\mbox{Inverse-$\chi^2$}
\big(\xi_{\qDens(\sigmaGrph^2)},\lambda_{\qDens(\sigmaGrph^2)}\big)$ distribution}\\[1ex]
&&\qDens^*(\bSigmag)\ \mbox{has an
$\mbox{Inverse-G-Wishart}(\Gfull,\xi_{\qDens(\bSigmag)},\bLambda_{\qDens(\bSigmag)})$ distribution}\\[1ex]
\mbox{and}&&\qDens^*(\bSigmah)\ \mbox{has an
$\mbox{Inverse-G-Wishart}(\Gfull,\xi_{\qDens(\bSigmah)},\bLambda_{\qDens(\bSigmah)})$ distribution.}
\end{eqnarray*}
}
The optimal $\qDens$-density parameters are determined through an iterative coordinate ascent algorithm,
details of which are given in Section \ref{sec:drvAlgFour} of the web-supplement.
As in the two-level case, the updates for $\bmu_{\qDens(\bbeta,\bu)}$ and $\bSigma_{\qDens(\bbeta,\bu)}$ may
be written in the same form as (\ref{eq:muSigmaMFVBupd}) but with a three-level version of the $\bC$ matrix
and
\begin{equation}
  \begin{array}{l}
  \DMFVB \equiv \\[1ex]
  \left[
      {\setlength\arraycolsep{1pt}
      \begin{array}{ccc}
        \bSigma_{\bbeta}^{-1}   & \bO & \bO \\[1ex]
        \bO                     & \bmu_{\qDens(1/{\sigmaGbl^{2}})}\bI    &  \bO    \\[1ex]
        \bO                     & \bO               &  \bI_{m} \otimes
                                                       \left[
                                                         {\setlength\arraycolsep{1pt}
                                                         \begin{array}{ccc}
                                                           \bM_{\qDens(\bSigmag^{-1})} & \bO            & \bO       \\
                                                           \bO       &\bmu_{\qDens(1/\sigmaGrpg^2)}\bI & \bO       \\
                                                           \bO       &  \bO           & \bI_{n_i}\otimes
                                                           \left[
                                                           \begin{array}{cc}
                                                             \bM_{\qDens(\bSigmah^{-1})} & \bO \\
                                                             \bO       & \bmu_{\qDens(1/\sigmaGrph^2)}\bI
                                                           \end{array}
                                                           \right]
                                                         \end{array}
                                                         }
                                                       \right]
      \end{array}
}
    \right].
  \end{array}
  \label{eqn:DMFVB}
\end{equation}
For large numbers of level 2 and level 3 groups, $\bSigma_{\qDens(\bbeta, \bu)}$'s size becomes
infeasible to deal with. However, only relatively small sub-blocks of $\bSigma_{\qDens(\bbeta,\bu)}$
are needed for variational inference regarding the variance and covariance parameters. These
sub-block positions correspond to the non-zero sub-block positions of a general three-level
sparse matrix defined in Section 3 of Nolan \myand Wand (2018). Here, Nolan \myand Wand's
\begin{equation}
  \begin{array}{l}
    \bA_{11} \mbox{ sub-block corresponds to a } (2 + \Kgbl) \times (2 + \Kgbl) \mbox{ matrix }
      \bSigma_{\qDens(\bbeta, \buGbl)}; \\[2ex]
    \bA_{22,i} \mbox{ sub-block corresponds to a } (2 + \Kgrpg) \times (2 + \Kgrpg) \mbox{ matrix }
    \bSigma_{\qDens(\buLoneLini, \buLoneGrpi)}; \\[1ex]
    \bA_{12,i} \mbox{ sub-block corresponds to a } (2 + \Kgbl) \times (2 + \Kgrpg) \mbox{ matrix } \\[1ex]
    \qquad \qquad \qquad
      E\left\{ \left(
      \left[
      \begin{array}{c}
      \bbeta\\
      \buGbl
      \end{array}
      \right] - \bmu_{\qDens(\bbeta,\buGbl)} \right)
      \left( \left[
      \begin{array}{c}
      \buLoneLini\\[1ex]
      \buLoneGrpi
      \end{array}
      \right] - \bmu_{\qDens(\buLoneLini,\buLoneGrpi)} \right)^T
      \right\}, \ 1\le i \le m; \\[1ex]
    \bA_{22,ij} \mbox{ sub-block corresponds to a } (2 + \Kgrph) \times (2 + \Kgrph) \mbox{ matrix }
      \bSigma_{\qDens(\buLtwoLinij,\buLtwoGrpij)}; \\[1ex]
    \bA_{12,ij} \mbox{ sub-block corresponds to a } (2 + \Kgbl) \times (2 + \Kgrph) \mbox{ matrix } \\[1ex]
      \qquad \qquad \qquad E\left\{
      \left( \left[
      \begin{array}{c}
      \bbeta\\
      \buGbl
      \end{array}
      \right] - \bmu_{\qDens(\bbeta, \buGbl)} \right)
      \left( \left[
      \begin{array}{c}
      \buLtwoLinij\\[1ex]
      \buLtwoGrpij
      \end{array}
      \right] - \bmu_{\qDens(\buLtwoLinij,\buLtwoGrpij)} \right)^T
      \right\}; \\[1ex]
    \bA_{12,\iCOMMAj} \mbox{ sub-block corresponds to a } (2 + \Kgrpg) \times (2 + \Kgrph) \mbox{ matrix } \\[1ex]
      \qquad \qquad \qquad E\left\{
      \left( \left[
      \begin{array}{c}
      \buLoneLini\\[1ex]
      \buLoneGrpi
      \end{array}
      \right] - \bmu_{\qDens(\buLoneLini, \buLoneGrpi)} \right)
      \left( \left[
      \begin{array}{c}
      \buLtwoLinij\\[1ex]
      \buLtwoGrpij
      \end{array}
      \right] - \bmu_{\qDens(\buLtwoLinij, \buLtwoGrpij)} \right)^T
      \right\}, \\[1ex]
      1 \le i \le m, \ 1 \le j \le n_{i}.
  \end{array}
  \label{eq:subBlocksThreeLevMfvb}
\end{equation}
We appeal to Result \ref{res:threeLevelMFVB} for a streamlined mean field variational Bayes algorithm.

\begin{result}
The mean field variational Bayes updates of $\bmu_{\qDens(\bbeta,\bu)}$ and each
of the sub-blocks of $\bSigma_{\qDens(\bbeta,\bu)}$ in (\ref{eq:subBlocksThreeLevMfvb})
are expressible as a three-level sparse matrix least squares problem of the form:
$$\left\Vert\bb-\bB\left[
\begin{array}{c}
\bbeta\\
\bu
\end{array}
\right]
\right\Vert^2
$$
where the non-zero sub-blocks $\bB$ and $\bb$, according to the notation
in Section 3.1 of Nolan \myand Wand (2018), are
for $1\le i\le m$ and $1\le j\le n_i$.
$$
\bb_{ij}\equiv
\left[
\begin{array}{c}
\mu_{\qDens(1/\sigeps^2)}^{1/2}\by_{ij}\\[1ex]
\ndotmh \bSigma_{\bbeta}^{-1/2} \bmu_{\bbeta}\\[1ex]
\bzero \\[1ex]
\bzero \\[1ex]
\bzero \\[1ex]
\bzero \\[1ex]
\bzero \\[1ex]
\end{array}
\right],
\ \
\bB_{ij}\equiv
\left[
\begin{array}{cc}
\mu_{\qDens(1/\sigeps^2)}^{1/2}\bX_{ij} & \mu_{\qDens(1/\sigeps^2)}^{1/2}\bZgblij\\[1ex]
\ndotmh \bSigma_{\bbeta}^{-1/2} & \bO \\[1ex]
\bO &\ndotmh \mu_{\qDens(1/\sigmaGbl^2)}^{1/2}\bI_{\Kgbl}\\[1ex]
\bO & \bO   \\[1ex]
\bO & \bO   \\[1ex]
\bO & \bO   \\[1ex]
\bO & \bO   \\[1ex]
\end{array}
\right],
$$
$$
\bBdot_{ij}\equiv
\left[
\begin{array}{cc}
\mu_{\qDens(1/\sigeps^2)}^{1/2}\bX_{ij}  & \mu_{\qDens(1/\sigeps^2)}^{1/2}\bZLoneGrpij  \\[1ex]
\bO                & \bO            \\[1ex]
\bO                & \bO            \\[1ex]
n_i^{-1/2}\bM_{\qDens(\bSigmag^{-1})}^{1/2}     & \bO            \\[1ex]
\bO                            &n_i^{-1/2}\mu_{\qDens(1/\sigmaGrpg^2)}^{1/2}\bI_{\Kgrpg}\\[1ex]
\bO                & \bO            \\[1ex]
\bO                & \bO            \\
\end{array}
\right]
\ \ and\ \
\bBdotdot_{ij}\equiv
\left[
\begin{array}{cc}
\mu_{\qDens(1/\sigeps^2)}^{1/2}\bX_{ij}  & \mu_{\qDens(1/\sigeps^2)}^{1/2}\bZLtwoGrpij  \\[1ex]
\bO                & \bO            \\[1ex]
\bO                & \bO            \\[1ex]
\bO                & \bO            \\[1ex]
\bO                & \bO            \\[1ex]
\bM_{\qDens(\bSigmah^{-1})}^{1/2}   & \bO            \\[1ex]
\bO                &\mu_{\qDens(1/\sigmaGrph^2)}^{1/2}\bI_{\Kgrph}\\
\end{array}
\right]
$$
with each of these matrices having $\oadj_{ij}=o_{ij}+2+\Kgbl+2+\Kgrpg+2+\Kgrph$
rows and with $\Bmati$ having $p=2+\Kgbl$ columns, $\Bmatdoti$ having $q_1=2+\Kgrpg$
columns and $\Bmatdotdotij$ having $q_2=2+\Kgrph$ columns. The solutions are
$$\bmu_{\qDens(\bbeta,\buGbl)}
=\xveco,\quad
\bSigma_{\qDens(\bbeta,\buGbl)}
=\AUoo,
$$
$$
\bmu_{\qDens(\buLoneLini,\buLoneGrpi)}
=\xvectCi,\quad
E_q\left\{
\left[
\begin{array}{c}
\bbeta-\bmu_{\qDens(\bbeta)}\\
\buGbl-\bmu_{\qDens(\buGbl)}
\end{array}
\right]
\left[
\begin{array}{c}
\buLoneLini-\bmu_{\qDens(\buLoneLini)}\\[1ex]
\buLoneGrpi-\bmu_{\qDens(\buLoneGrpi)}
\end{array}
\right]^T
\right\}=\AUotCi,
$$
$$
\bSigma_{\qDens(\buLoneLini,\buLoneGrpi)}
=\AUttCi,\ \ 1\le i\le m,
$$
$$\bmu_{\qDens(\buLtwoLinij,\buLtwoGrpij)}
=\bx_{2,ij},\quad
E_q\left\{
\left[
\begin{array}{c}
\bbeta-\bmu_{\qDens(\bbeta)}\\
\buGbl-\bmu_{\qDens(\buGbl)}
\end{array}
\right]
\left[
\begin{array}{c}
\buLtwoLinij-\bmu_{\qDens(\buLtwoLinij)}\\[1ex]
\buLtwoGrpij-\bmu_{\qDens(\buLtwoGrpij)}
\end{array}
\right]^T
\right\}=\bA^{12,ij},
$$
$$
E_q\left\{
\left[
\begin{array}{c}
\buLoneLini-\bmu_{\qDens(\buLoneLini)}\\[1ex]
\buLoneGrpi-\bmu_{\qDens(\buLoneGrpi)}
\end{array}
\right]
\left[
\begin{array}{c}
\buLtwoLinij-\bmu_{\qDens(\buLtwoLinij)}\\[1ex]
\buLtwoGrpij-\bmu_{\qDens(\buLtwoGrpij)}
\end{array}
\right]^T
\right\}=\bA^{12,\iCOMMAj}
$$
and
$$
\bSigma_{\qDens(\buLtwoLinij,\buLtwoGrpij)}
=\bA^{22,ij},
\quad 1\le i\le m,\ 1\le j\le n_i.
$$
\label{res:threeLevelMFVB}
\end{result}

\begin{algorithm}[!th]
  \begin{center}
    \begin{minipage}[t]{159mm}
      \begin{small}
        \begin{itemize}
          \setlength\itemsep{0pt}
          \item[] Data Inputs: $\by_{ij}(o_{ij}\times1),\ \bX_{ij}(o_{ij}\times 2),\
                  \bZgblij (o_{ij} \times \Kgbl), \bZLoneGrpij(o_{ij}\times \Kgrpg),$
          \item[] \qquad$\bZLtwoGrpij(o_{ij}\times \Kgrph)\ 1\le i\le m,\ 1\le j\le n_{i}.$
          \item[] Hyperparameter Inputs: $\bmu_{\bbeta}(2\times1)$,
                  $\bSigma_{\bbeta}(2\times 2)\ \mbox{symmetric and positive definite}$,
          \item[] \qquad $s_{\varepsilon},\nu_{\varepsilon},s_{\mbox{\rm\tiny gbl}},
                        \nu_{\mbox{\rm\tiny gbl}},\sSigmagOne,\sSigmagTwo,\nuSigmag,
                        \sGrpg, \nuGrpg, \sSigmahOne,\sSigmahTwo,\nuSigmah, \sGrph, \nuGrph>0$.
         \item[] For $i=1,\ldots,m:$
         \item[]
           \begin{itemize}
             \item[] For $j=1,\ldots,n_{i}:$
             \item[]
               \begin{itemize}
                 \item[]$\bCgblij\thickarrow[\bX_{ij}\ \bZgblij]$\ \ \ ;\ \ \ $\bCLoneGrpij\thickarrow[\bX_{ij}\ \bZLoneGrpij]$
                          \ \ \ ;\ \ \ $\bCLtwoGrpij\thickarrow[\bX_{ij}\ \bZLtwoGrpij]$
               \end{itemize}
           \end{itemize}
          \item[] Initialize: $\muq{1/\sigsqeps}$, $\muq{1/\sigma_{\mbox{\rm\tiny gbl}}^{2}}$,
                  $\muq{1/\sigma_{\mbox{\rm\tiny grp, g}}^{2}}$, $\muq{1/\sigma_{\mbox{\rm\tiny grp, h}}^{2}}$, $\muq{1/\aeps}$,
                  $\muq{1/a_{\mbox{\rm\tiny gbl}}}$,
          \item[] \qquad \qquad $\muq{1/a_{\mbox{\rm\tiny grp, g}}}$, $\muq{1/a_{\mbox{\rm\tiny grp, h}}} > 0$,
                   $\bM_{\qDens(\bSigma_{\mbox{\rm\tiny g}}^{-1})} (2 \times 2),
                  \bM_{\qDens(\bSigma_{\mbox{\rm\tiny h}}^{-1})} (2 \times 2),$
          \item[] \qquad \qquad $\bM_{\qDens(\bA^{-1}_{\mbox{\rm\tiny g}})} (2 \times 2),
                  \bM_{\qDens(\bA^{-1}_{\mbox{\rm\tiny h}})} (2 \times 2)$ symmetric and
                  positive definite.
          \item[] $\xi_{\qDens(\sigeps^2)}\thickarrow \nu_{\varepsilon} + \sumim \sum_{j=1}^{n_{i}} o_{ij}$\ \ \ ;\ \ \
                  $\xi_{\qDens(\sigmaGbl^2)}\thickarrow\nu_{\mbox{\rm\tiny gbl}}+\Kgbl$\ \ \ ;\ \ \
                  $\xi_{\qDens(\bSigma_{\mbox{\rm\tiny g}})}\thickarrow\nuSigmag+2+m$
          \item[] $\xi_{\qDens(\bSigma_{\mbox{\rm\tiny h}})}\thickarrow\nuSigmah+2+\sum_{i=1}^{m} n_{i}$ \ \ \ ; \ \ \
                  $\xi_{\qDens(\sigmaGrpg^2)}\thickarrow \nu_{\mbox{\rm\tiny grp, g}} + m\Kgrpg$
          \item[] $\xi_{\qDens(\sigmaGrph^2)}\thickarrow \nu_{\mbox{\rm\tiny grp, h}} + \Kgrph \sum_{i=1}^{m}n_{i}$\ \ \ ; \ \ \
                  $\xi_{\qDens(a_{\varepsilon})}\thickarrow \nuEps + 1$\ \ \ ; \ \ \
                  $\xi_{\qDens(a_{\mbox{\rm\tiny gbl}})}\thickarrow \nuGbl + 1$
          \item[] $\xi_{\qDens(a_{\mbox{\rm\tiny grp, g}})}\thickarrow \nuGrpg + 1$ \ \ \ ; \ \ \
                  $\xi_{\qDens(a_{\mbox{\rm\tiny grp, h}})}\thickarrow \nuGrph + 1$ \ \ \ ; \ \ \ 
                  $\xi_{\qDens(\bA_{\bSigmag})}\thickarrow \nuSigmag + 2$\ \ \ ;\ \ \ 
                  $\xi_{\qDens(\bA_{\bSigmah})}\thickarrow \nuSigmah + 2$
          \item[] Cycle:
          \begin{itemize}
            \item[] For $i = 1,\ldots, m$:
            \item[]
              \begin{itemize}
                \item[] For $j = 1, \ldots, n_{i}$:
                \item[]
                \begin{itemize}
                  \item[] $\bvecij\thickarrow\left[
                            \begin{array}{c}
                              \mu_{\qDens(1/\sigeps^2)}^{1/2}\by_{ij}\\[1.5ex]
                              n_{\boldsymbol{\cdot}}^{-1/2} \bSigma_{\bbeta}^{-1/2}\bmu_{\bbeta} \\[1ex]
                              \bzero \\[1ex]
                              \bzero \\[1ex]
                              \bzero \\[1ex]
                              \bzero \\[1ex]
                              \bzero
                            \end{array}
                            \right],\
                            \Bmatij\thickarrow
                            \left[
                            \begin{array}{cc}
                            \mu_{\qDens(1/\sigeps^2)}^{1/2}\bX_{ij} & \mu_{\qDens(1/\sigeps^2)}^{1/2}\bZgblij\\[1ex]
                            \ndotmh \bSigma_{\bbeta}^{-1/2} & \bO \\[1ex]
                            \bO &\ndotmh \mu_{\qDens(1/\sigmaGbl^2)}^{1/2}\bI_{\Kgbl}\\[1ex]
                            \bO & \bO   \\[1ex]
                            \bO & \bO   \\[1ex]
                            \bO & \bO   \\[1ex]
                            \bO & \bO   \\[1ex]
                            \end{array}
                            \right],$
                \item[] $\Bmatdotij\thickarrow
                          \left[
                          \begin{array}{cc}
                          \mu_{\qDens(1/\sigeps^2)}^{1/2}\bX_{ij}  & \mu_{\qDens(1/\sigeps^2)}^{1/2}\bZLoneGrpij  \\[1ex]
                          \bO                & \bO            \\[1ex]
                          \bO                & \bO            \\[1ex]
                          n_i^{-1/2}\bM_{\qDens(\bSigmag^{-1})}^{1/2}     & \bO            \\[1ex]
                          \bO                            &n_i^{-1/2}\mu_{\qDens(1/\sigmaGrpg^2)}^{1/2}\bI_{\Kgrpg}\\[1ex]
                          \bO                & \bO            \\[1ex]
                          \bO                & \bO            \\
                          \end{array}
                          \right]$
              \end{itemize}
            \end{itemize}
            \item[] \textsl{continued on a subsequent page}\ $\ldots$
          \end{itemize}
        \end{itemize}
      \end{small}
    \end{minipage}
  \end{center}
  \caption{\it QR-decomposition-based streamlined algorithm for obtaining mean field variational
             Bayes approximate posterior density functions for the parameters in the
             Bayesian three-level group-specific curves model (\ref{eq:threeLevBayes}) with
             product density restriction (\ref{eq:producRestrict3lev})}
  \label{alg:threeLevMFVB}
\end{algorithm}
%

\setcounter{algorithm}{3}
\begin{algorithm}[!th]
  \begin{center}
    \begin{minipage}[t]{159mm}
      \begin{small}
        \begin{itemize}
            \item[]

           \begin{itemize}
              \begin{itemize}
                  \begin{itemize}
                \item[] $\Bmatdotdotij\thickarrow
                          \left[
                          \begin{array}{cc}
                          \mu_{\qDens(1/\sigeps^2)}^{1/2}\bX_i  & \mu_{\qDens(1/\sigeps^2)}^{1/2}\bZLtwoGrpij  \\[1ex]
                          \bO                & \bO            \\[1ex]
                          \bO                & \bO            \\[1ex]
                          \bO                & \bO            \\[1ex]
                          \bO                & \bO            \\[1ex]
                          \bM_{\qDens(\bSigmah^{-1})}^{1/2}   & \bO            \\[1ex]
                          \bO                &\mu_{\qDens(1/\sigmaGrph^2)}^{1/2}\bI_{\Kgrph}\\
                          \end{array}
                          \right]$
                    \end{itemize}
              \end{itemize}
            \end{itemize}
    
            \item[]
            \begin{itemize}
            \item[] $\Ssc_4\thickarrow\SolveThreeLevelSparseLeastSquares\Big(\big\{(
                      \bvecij,\Bmatij,\Bmatdotij,\Bmatdotdotij):1\le i\le m, $
            \item[] $\qquad  1 \le j\le n_{i} \big\}\Big)$
            \item[] $\bmu_{\qDens(\bbeta,\buGbl)}\thickarrow\mbox{$\xveco$ component of $\Ssc_4$}$
                    \ \ \ ;\ \ \ $\bSigma_{\qDens(\bbeta,\buGbl)}\thickarrow\mbox{$\AUoo$ component of $\Ssc_4$}$
            \item[] $\bmu_{\qDens(\buGbl)}\thickarrow\mbox{last $\Kgbl$ rows of $\bmu_{\qDens(\bbeta,\buGbl)}$}$
            \item[] $\bSigma_{\qDens(\buGbl)}\thickarrow\mbox{bottom-right $\Kgbl\times\Kgbl$ sub-block of 
                    $\bSigma_{\qDens(\bbeta,\buGbl)}$}$
            \item[] $\lambda_{\qDens(\sigsqeps)}\thickarrow\muq{1/\aeps}$\ \ ;\ \
                    $\Lambda_{\qDens(\bSigma_{\mbox{\rm\tiny g}})}\thickarrow \bM_{\qDens(\ASigmag^{-1})}$\ \ ;\ \
                    $\Lambda_{\qDens(\bSigma_{\mbox{\rm\tiny h}})}\thickarrow \bM_{\qDens(\ASigmah^{-1})}$
            \item[] $\lambda_{\qDens(\sigma^{2}_{\mbox{\rm\tiny grp, g}})}\thickarrow\mu_{\qDens(1/a_{\mbox{\rm\tiny grp, g}})}$\ \ ;\ \
                    $\lambda_{\qDens(\sigma^{2}_{\mbox{\rm\tiny grp, g}})}\thickarrow\mu_{\qDens(1/a_{\mbox{\rm\tiny grp, g}})}$
            \item[] For $i = 1,\ldots, m$:
            \begin{itemize}
              \item[] $\bmu_{\qDens(\buLoneLini,\buLoneGrpi)}
                       \thickarrow\mbox{$\xvectCi$ component of $\Ssc_4$}$
              \item[] $\bSigma_{\qDens(\buLoneLini,\buLoneGrpi)}
                       \thickarrow\mbox{$\AUttCi$ component of $\Ssc_4$}$
              \item[] $\bmu_{\qDens(\buLoneLini)}\thickarrow\mbox{first $2$ rows of
                      $\bmu_{\qDens(\buLoneLini,\buLoneGrpi)}$}$
              \item[] $\bSigma_{\qDens(\buLoneLini)}\thickarrow\mbox{top left $2 \times 2$ sub-block of
                      $\bSigma_{\qDens(\buLoneLini,\buLoneGrpi)}$}$
              \item[] $\bmu_{\qDens(\buLoneGrpi)}\thickarrow\mbox{last $\Kgrpg$ rows of
                      $\bmu_{\qDens(\buLoneLini,\buLoneGrpi)}$}$
              \item[] $\bSigma_{\qDens(\buLoneGrpi)}\thickarrow\mbox{bottom right
                      $\Kgrpg \times \Kgrpg$ sub-block of
                      $\bSigma_{\qDens(\buLoneLini,\buLoneGrpi)}$}$
              \item[] $E_{\qDens}\left\{\left(\left[\begin{array}{c}\bbeta\\ \buGbl\end{array}\right]
                       -\bmu_{\qDens(\bbeta,\buGbl)}\right)
                        \left(\left[\begin{array}{c}\buLoneLini\\ \buLoneGrpi\end{array}\right]
                        -\bmuq{\buLoneLini,\buLoneGrpi)}\right)^T\right\}$
              \item[] \qquad\qquad\qquad$\thickarrow\mbox{$\AUotCi$ component of $\Ssc_4$}$
              \item[] For $j = 1,\ldots, n_{i}$:
              \begin{itemize}
                \item[] $\bmu_{\qDens(\buLtwoLinij,\buLtwoGrpij)}
                         \thickarrow\mbox{$\xvectCij$ component of $\Ssc_4$}$
                \item[] $\bSigma_{\qDens(\buLtwoLinij,\buLtwoGrpij)}
                         \thickarrow\mbox{$\AUttCij$ component of $\Ssc_4$}$
                \item[] $\bmu_{\qDens(\buLtwoLinij)}\thickarrow\mbox{first $2$ rows of
                        $\bmu_{\qDens(\buLtwoLinij,\buLtwoGrpij)}$}$
                \item[] $\bSigma_{\qDens(\buLtwoLinij)}\thickarrow\mbox{top left
                        $2\times 2$ sub-block of
                        $\bSigma_{\qDens(\buLtwoLinij,\buLtwoGrpij)}$}$
                \item[] $\bmu_{\qDens(\buLtwoGrpij)}\thickarrow\mbox{last $\Kgrph$ rows of
                        $\bmu_{\qDens(\buLtwoLinij,\buLtwoGrpij)}$}$
                \item[] $\bSigma_{\qDens(\buLtwoGrpij)}\thickarrow\mbox{bottom right
                        $\Kgrph \times \Kgrph$ sub-block of
                        $\bSigma_{\qDens(\buLtwoLinij,\buLtwoGrpij)}$}$
                \item[] $E_{\qDens}\left\{\left(\left[\begin{array}{c}\bbeta\\ \buGbl\end{array}\right]
                         -\bmu_{\qDens(\bbeta,\buGbl)}\right)
                          \left(\left[\begin{array}{c}
                             \buLtwoLinij\\ 
                             \buLtwoGrpij
                          \end{array}\right]
                          -\bmuq{\buLoneLini,\buLoneGrpi)}\right)^T\right\}$
                \item[] \qquad\qquad\qquad$\thickarrow\mbox{$\AUotCij$ component of $\Ssc_4$}$
                \item[] \textsl{continued on a subsequent page}\ $\ldots$
              \end{itemize}
            \end{itemize}
          \end{itemize}
        \end{itemize}
      \end{small}
    \end{minipage}
  \end{center}
  \caption{\textbf{continued.} \textit{This is a continuation of the description of this algorithm that commences
on a preceding page.}}
\end{algorithm}

\setcounter{algorithm}{3}
\begin{algorithm}[!th]
  \begin{center}
    \begin{minipage}[t]{159mm}
      \begin{small}
          \begin{itemize}
            \item[]
              \begin{itemize}
                \item[]
                \begin{itemize}
                  \item[]
                  \begin{itemize}
                \item[] $E_{\qDens}\left\{\left(\left[\begin{array}{c}\buLoneLini\\
                         \buLoneGrpi \end{array}\right]
                         -\bmuq{\buLoneLini,\buLoneGrpi)}\right)
                          \left(\left[\begin{array}{c}\buLtwoLinij\\ \buLtwoGrpij
                          \end{array}\right]
                          -\bmuq{\buLtwoLinij,\buLtwoGrpij)}\right)^T\right\}$
                \item[] \qquad\qquad\qquad$\thickarrow\mbox{$\AUotCicommaj$ component of $\Ssc_4$}$
                \item[] $\lambda_{\qDens(\sigsqeps)}\thickarrow \lambda_{\qDens(\sigsqeps)}
                         +\big\Vert\by_{ij}-\bCgblij\bmu_{\qDens(\bbeta,\buGbl)}
                         -\bCLoneGrpij\bmu_{\qDens(\buLoneLini,\buLoneGrpi)}$
                \item[] \qquad \qquad \qquad $-\bCLtwoGrpij\bmu_{\qDens(\buLtwoLinij,\buLtwoGrpij)}\big\Vert^2$
                \item[] $\lambda_{\qDens(\sigsqeps)}\thickarrow \lambda_{\qDens(\sigsqeps)}
                         +\mbox{tr}(\bCgblij^T\bCgblij\bSigma_{\qDens(\bbeta,\buGbl)})
                         +\mbox{tr}((\bCLoneGrpij)^T\bCLoneGrpij\bSigma_{\qDens(\buLoneLini,\buLoneGrpi)})$
                \item[] $\lambda_{\qDens(\sigsqeps)}\thickarrow \lambda_{\qDens(\sigsqeps)}
                        +\mbox{tr}((\bCLtwoGrpij)^T\bCLtwoGrpij\bSigma_{\qDens(\buLtwoLinij,\buLtwoGrpij)})$
              \item[] $\lambda_{\qDens(\sigsqeps)}\thickarrow \lambda_{\qDens(\sigsqeps)}
                         +2\,\mbox{tr}\left[\bCgrpi^T\bCgbli\,E_{\qDens}\left\{\left(
                         \left[\begin{array}{c}\bbeta \\ \buGbl\end{array}\right]
                         -\bmu_{\qDens(\bbeta,\buGbl)}\right) \right. \right. $
                \item[] \qquad\qquad\qquad\qquad\qquad\qquad\qquad $\left.\left. \times
                         \left(\left[\begin{array}{c}\buLoneLini\\
                         \buLoneGrpi\end{array}\right]
                         -\bmuq{\buLoneLini,\buLoneGrpi)}\right)^T\right\}\right]$
                    \item[] $\lambda_{\qDens(\sigsqeps)}\thickarrow \lambda_{\qDens(\sigsqeps)}
                             +2\,\mbox{tr}\left[(\bCLoneGrpij)^T\bCgblij\,E_{\qDens}\left\{\left(
                             \left[\begin{array}{c}\bbeta \\ \buGbl\end{array}\right]
                             -\bmu_{\qDens(\bbeta,\buGbl)}\right) \right. \right. $
                    \item[] \qquad\qquad\qquad\qquad\qquad\qquad\qquad\qquad $\left.\left. \times
                             \left(\left[\begin{array}{c}\buLtwoLinij\\
                             \buLtwoGrpij\end{array}\right]
                             -\bmuq{\buLtwoLinij,\buLtwoGrpij)}\right)^T\right\}\right]$
                    \item[] $\lambda_{\qDens(\sigsqeps)}\thickarrow \lambda_{\qDens(\sigsqeps)}
                             +2\,\mbox{tr}\left[(\bCLoneGrpij)^T\bCLtwoGrpij\,E_{\qDens}
                            \left\{\left(\left[\begin{array}{c}\buLoneLini\\
                             \buLoneGrpi\end{array}\right]
                             -\bmuq{\buLoneLini,\buLoneGrpi}\right) \right. \right. $
                    \item[] \qquad\qquad\qquad\qquad\qquad\qquad\qquad\qquad $\left.\left. \times
                             \left(\left[\begin{array}{c}\buLtwoLinij\\
                             \buLtwoGrpij\end{array}\right]
                             -\bmu_{\qDens(\buLtwoLinij,\buLtwoGrpij)}\right)^T\right\}\right]$
                    \item[] $\bLambda_{\qDens(\bSigmah)}\thickarrow\bLambda_{\qDens(\bSigmah)}+
                            \bmu_{\qDens(\buLtwoLinij)}\bmu_{\qDens(\buLtwoLinij)}^T+ \bSigma_{\qDens(\buLtwoLinij)}$
                    \item[] $\lambda_{\qDens(\sigmaGrph^{2})} \thickarrow \lambda_{\qDens(\sigmaGrph^{2})} +
                            \Vert \bmu_{\qDens(\buLtwoGrpij)} \Vert^2 + \mbox{tr} \left( \bSigma_{\qDens(\buLtwoGrpij)}
                            \right)$
                  \end{itemize}
                  \item[] $\bLambda_{\qDens(\bSigmag)}\thickarrow\bLambda_{\qDens(\bSigmag)}+
                           \bmu_{\qDens(\buLoneLini)}\bmu_{\qDens(\buLoneLini)}^T+ \bSigma_{\qDens(\buLoneLini)}$
                  \item[] $\lambda_{\qDens(\sigmaGrpg^{2})} \thickarrow \lambda_{\qDens(\sigmaGrpg^{2})} +
                           \Vert \bmu_{\qDens(\buLoneGrpi)} \Vert^2 + \mbox{tr} \left( \bSigma_{\qDens(\buLoneGrpi)}
                           \right)$
                \end{itemize}
              \item[] $\lambda_{\qDens(\sigma^{2}_{\mbox{\rm\tiny gbl}})} \thickarrow \mu_{\qDens(1/a_{\mbox{\rm\tiny gbl}})} +
                       \Vert \bmu_{\qDens(\bu_{\mbox{\rm\tiny gbl}})} \Vert^2 + \mbox{tr} \left(
                       \bSigma_{\qDens(\bu_{\mbox{\rm\tiny gbl}})} \right)$
              \item[] $\muq{1/\sigsqeps} \leftarrow \xi_{\qDens(\sigeps)}/\lambda_{\qDens(\sigsqeps)}$ \ \ \ ; \ \ \
                      $\muq{1/\sigma^{2}_{\mbox{\rm\tiny gbl}}} \leftarrow \xi_{\qDens(\sigma^{2}_{\mbox{\rm\tiny gbl}})}/
                       \lambda_{\qDens(\sigma^{2}_{\mbox{\rm\tiny gbl}})}$
              \item[] $\MqSigmag \leftarrow(\xi_{\qDens(\bSigmag)} - 2 + 1) \bLambda^{-1}_{\qDens(\bSigmag)}$ \ \ \ ; \ \ \
                      $\MqSigmah \leftarrow(\xi_{\qDens(\bSigmah)} - 2 + 1) \bLambda^{-1}_{\qDens(\bSigmah)}$
              \item[] $\muq{1/\sigmaGrpg^{2}} \leftarrow \xi_{\qDens(\sigmaGrpg^{2})}/\lambda_{\qDens(\sigmaGrpg^{2})}$ \ \ \ ; \ \ \
                      $\muq{1/\sigmaGrph^{2}} \leftarrow \xi_{\qDens(\sigmaGrph^{2})}/\lambda_{\qDens(\sigmaGrph^{2})}$
              \item[] $\lambda_{\qDens(a_{\varepsilon})}\thickarrow\muq{1/\sigma_{\varepsilon}^{2}}
                       +1/(\nu_{\varepsilon} s_{\varepsilon}^2)$\ \ \ ;\ \ \
                       $\muq{1/a_{\varepsilon}} \thickarrow \xi_{\qDens(a_{\varepsilon})}/
                       \lambda_{\qDens(a_{\varepsilon})}$
              \item[] $\bM_{\qDens(\ASigmag^{-1})}\thickarrow \xi_{\qDens(\ASigmag)}\bLambda_{\qDens(\ASigmag)}^{-1}$
              \ \ ;\ \ $\bM_{\qDens(\ASigmah^{-1})}\thickarrow \xi_{\qDens(\ASigmah)}\bLambda_{\qDens(\ASigmah)}^{-1}$
              \item[] $\bLambda_{\qDens(\ASigmag)}\thickarrow
               \diag\big\{\mbox{diagonal}\big(\bM_{\qDens(\bSigmag^{-1})}\big)\big\}+\{\nuSigmag\diag(\sSigmagOne^2,\sSigmagTwo^2)\}^{-1}$
              \item[] $\bLambda_{\qDens(\ASigmah)}\thickarrow
               \diag\big\{\mbox{diagonal}\big(\bM_{\qDens(\bSigmah^{-1})}\big)\big\}+\{\nuSigmah\diag(\sSigmahOne^2,\sSigmahTwo^2)\}^{-1}$
              \item[] $\lambda_{\qDens(a_{\mbox{\rm\tiny gbl}})}\thickarrow\muq{1/\sigma_{\mbox{\rm\tiny gbl}}^{2}}
                      +1/(\nu_{\mbox{\rm\tiny gbl}} s_{\mbox{\rm\tiny gbl}}^2)$\ \ \ ;\ \ \
                      $\muq{1/a_{\mbox{\rm\tiny gbl}}} \thickarrow \xi_{\qDens(a_{\mbox{\rm\tiny gbl}})}/
                      \lambda_{\qDens(a_{\mbox{\rm\tiny gbl}})}$
              \item[] $\lambda_{\qDens(\aGrpg)}\thickarrow\muq{1/\sigmaGrpg^{2}}
                       +1/(\nu_{\mbox{\rm\tiny grp, $g$}} s_{\mbox{\rm\tiny grp, $g$}}^2)$\ \ \ ;\ \ \
                      $\muq{1/\aGrpg} \thickarrow \xi_{\qDens(\aGrpg)}/\lambda_{\qDens(\aGrpg)}$
              \item[] $\lambda_{\qDens(\aGrph)}\thickarrow\muq{1/\sigmaGrph^{2}}
                      +1/(\nu_{\mbox{\rm\tiny grp, $h$}} s_{\mbox{\rm\tiny grp, $h$}}^2)$\ \ \ ;\ \ \
                      $\muq{1/\aGrph} \thickarrow \xi_{\qDens(\aGrph)}/\lambda_{\qDens(\aGrph)}$
            \end{itemize}
          \item[] until the increase in $\underline{\pDens}(\by;\qDens)$ is negligible.
          \item[] \textsl{continued on a subsequent page}\ $\ldots$
          \end{itemize}
      \end{small}
    \end{minipage}
  \end{center}
  \caption{\textbf{continued.} \textit{This is a continuation of the description of this algorithm that commences
on a preceding page.}}
\end{algorithm}
%
\setcounter{algorithm}{3}
\begin{algorithm}[!th]
\begin{center}
\begin{minipage}[t]{154mm}
\begin{small}
          \begin{itemize}
          \item[] Outputs: $\bmu_{\qDens(\bbeta,\buGbl)}$,\ $\bSigma_{\qDens(\bbeta,\buGbl)}$,\
                           $\Big\{\bmu_{\qDens(\buLoneLini,\buLoneGrpi)}, \bSigma_{\qDens(\buLoneLini,\buLoneGrpi)},$
          \item[] \qquad\qquad $E_{\qDens}\left\{\left(\left[\begin{array}{c}\bbeta\\ \buGbl\end{array}\right]
                   -\bmu_{\qDens(\bbeta,\buGbl)}\right)
                   \left(\left[\begin{array}{c}\buLoneLini\\ \buLoneGrpi\end{array}\right]
                   -\bmuq{\buLoneLini,\buLoneGrpi)}\right)^T\right\}:1\le i\le m,$
          \item[] \qquad\qquad $ E_{\qDens}\left\{\left( \left[\begin{array}{c}\bbeta \\ \buGbl\end{array}\right]
                  -\bmu_{\qDens(\bbeta,\buGbl)}\right)
                  \left(\left[\begin{array}{c}\buLtwoLinij\\ \buLtwoGrpij\end{array}\right]
                  -\bmuq{\buLtwoLinij,\buLtwoGrpij)}\right)^T\right\}, $
          \item[] \qquad\qquad $E_{\qDens} \left\{\left(\left[\begin{array}{c}\buLoneLini\\
                    \buLoneGrpi\end{array}\right]-\bmuq{\buLoneLini,\buLoneGrpi}\right)
                    \left(\left[\begin{array}{c}\buLtwoLinij\\ \buLtwoGrpij\end{array}\right]
                    -\bmu_{\qDens(\buLtwoLinij,\buLtwoGrpij)}\right)^T\right\},$
          \item[] \qquad \qquad $ \bmu_{\qDens(\buLtwoLinij,\buLtwoGrpij)}, \bSigma_{\qDens(\buLtwoLinij,\buLtwoGrpij)}:
                  1 \le i \le m, 1 \le j \le n_{i} \Big\}, \xi_{\qDens(\sigeps)}, \ \lambda_{\qDens(\sigsqeps)}, \
                  \xi_{\qDens(\sigmaGbl^2)}, $
          \item[] \qquad \qquad $\lambda_{\qDens(\sigmaGbl^2)}, \xi_{\qDens(\bSigmag)}, \
                  \bLambda^{-1}_{\qDens(\bSigmag)}, \ \xi_{\qDens(\bSigmah)}, \ \bLambda^{-1}_{\qDens(\bSigmah)}, \
                  \xi_{\qDens(\sigmaGrpg^2)}, \ \lambda_{\qDens(\sigmaGrpg^2)}, \
                  \xi_{\qDens(\sigmaGrph^2)}, \ \lambda_{\qDens(\sigmaGrph^2)}.$
        \end{itemize}
      \end{small}
    \end{minipage}
  \end{center}
  \caption{\textbf{continued.} \textit{This is a continuation of the description of this algorithm that commences
on a preceding page.}}
\end{algorithm}

Algorithm \ref{alg:threeLevMFVB} makes use of Result \ref{res:threeLevelMFVB}
to facilitate streamlined computation of all variational parameters in 
the three-level group specific curves model.

\begin{figure}[!ht]
\centering
{\includegraphics[width=\textwidth]{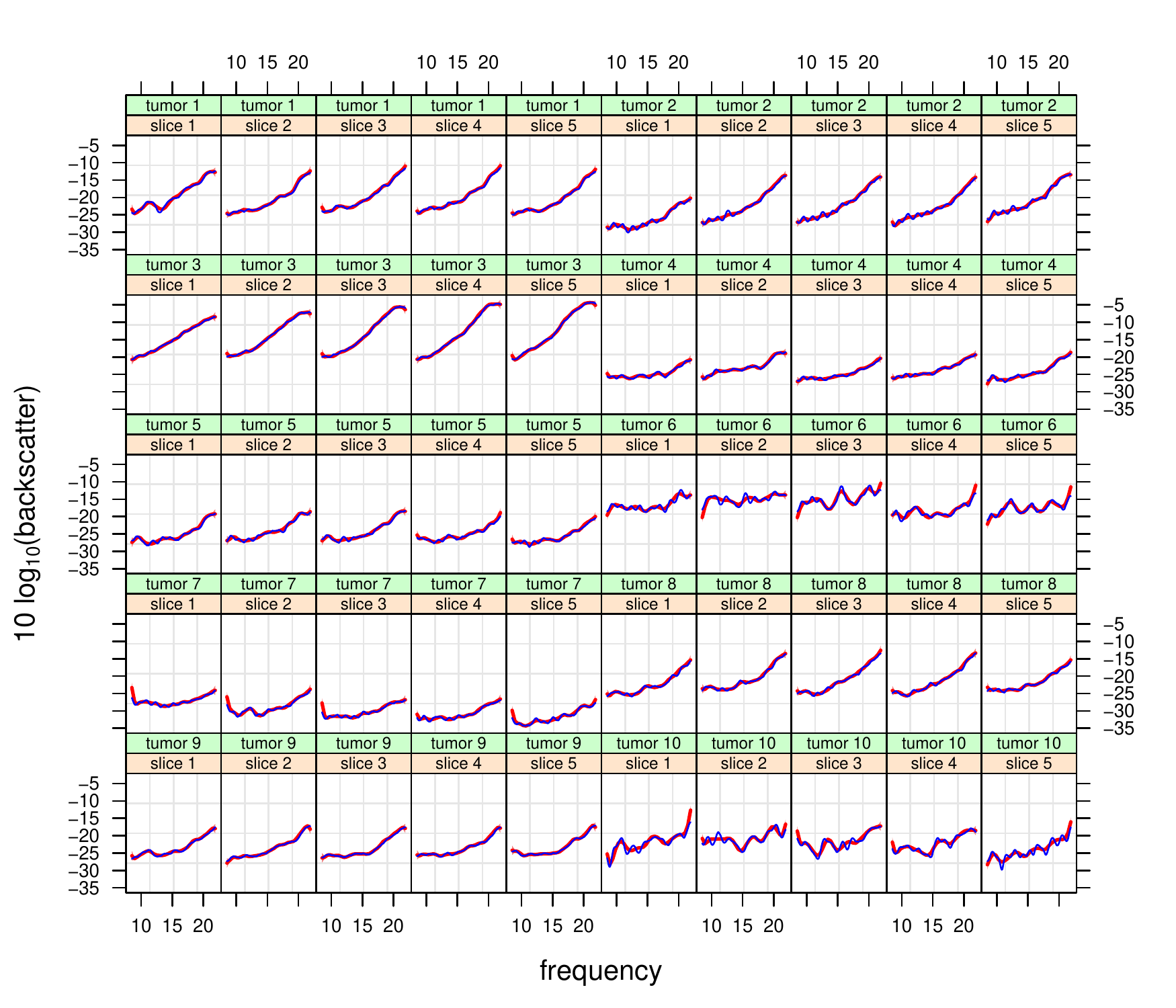}}
\caption{\it Illustrative three-level curve-type data with approximate fitted group
specific curves and corresponding 99\% credible sets based on mean field variational Bayes via Algorithm
\ref{alg:threeLevMFVB}.
The response variable is $10\log_{10}(\mbox{backscatter})$ according to ultrasound technology.
Level 1 corresponds to different ultrasound frequencies and matches the horizontal axes in each panel.
Level 2 corresponds to different slices of a tumor due to differing probe locations. Level 3
corresponds to different tumors with one tumor for each of $10$ laboratory mice.}
\label{fig:MNSWintroWithFit}
\end{figure}

Figure \ref{fig:MNSWintroWithFit} provides illustration of Algorithm \ref{alg:threeLevMFVB}
by showing the fits to the Figure \ref{fig:MNSWintro} ultrasound data. 
Posterior mean curves and (narrow) 99\% pointwise credible intervals are shown.
As discussed in the next section, such fits can be obtained rapidly and accurately
and Algorithm \ref{alg:threeLevMFVB} is scalable to much larger data sets
of the type illustrated by Figures \ref{fig:MNSWintro} and \ref{fig:MNSWintroWithFit}.

\null\vfill\eject
\null\vfill\eject
\null\vfill\eject
\null\vfill\eject

\section{Accuracy and Speed Assessment}\label{sec:accAndSpeed}

In this section we provide some assessment of the accuracy and
speed of the inference delivered by streamlined variational
inference for group-specific curves models.

\subsection{Accuracy Assessment}

\begin{figure}[!ht]
\centering
{\includegraphics[width=\textwidth]{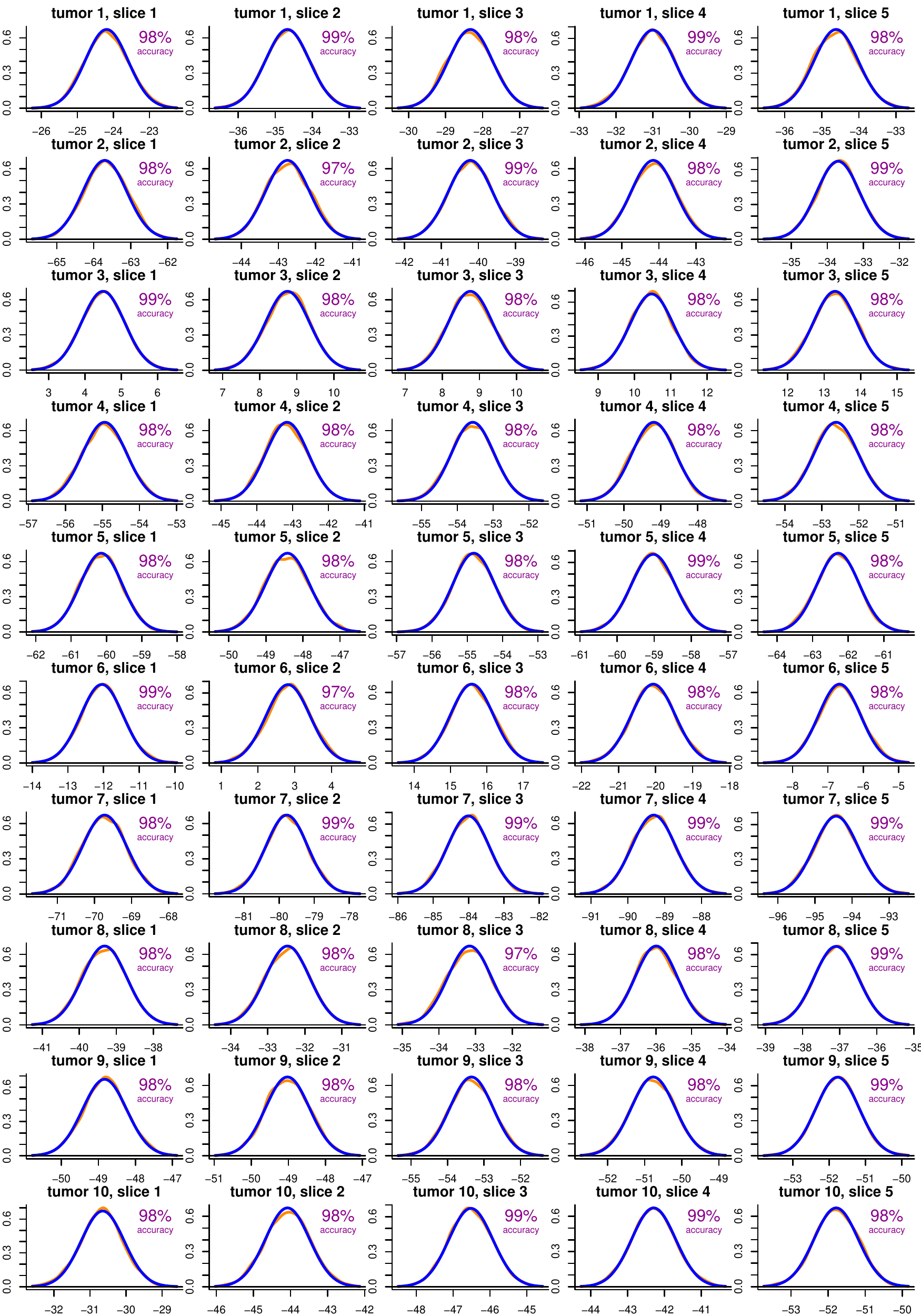}}
\caption{\it Accuracy assessment of Algorithm \ref{alg:threeLevMFVB}. Each panel
displays approximate posterior density functions corresponding to mean function estimates
according to the three-level group specific curve model (\ref{eq:threeLevelfg}).
In each case the estimate is at the median frequency value. The orange
density functions are based on Markov chain Monte Carlo
and the blue density functions are based on mean field
variational Bayes. The accuracy percentage scores are defined by (\ref{eq:accurDefn}).
}
\label{fig:ultrasoundAcc}
\end{figure}

Mean field restrictions such as (\ref{eq:producRestrict}) 
and (\ref{eq:producRestrict3lev}) imply that there is some 
loss of accuracy in inference produced by Algorithms 
\ref{alg:twoLevMFVB} and \ref{alg:threeLevMFVB}.
However, at least for the Gaussian response case treated here, 
approximate parameter orthogonality between the coefficient parameters
and covariance parameters from likelihood theory implies that such 
restrictions are mild and mean field accuracy is high. 
Figure \ref{fig:ultrasoundAcc} corroborates this claim 
by assessing accuracy of the mean function estimates and 95\% credible intervals 
at the median values of frequency for each panel in Figure \ref{fig:MNSWintroWithFit}.
As a benchmark we use Markov chain Monte Carlo-based inference via the 
\textsf{rstan} package (Guo \textit{et al.}, 2018).
After a warmup of size $1,000$ we retained $5,000$ Markov chain Monte Carlo
samples from the mean function and median frequency posterior distributions
and used kernel density estimation to approximate the corresponding posterior
density function. For a generic univariate parameter $\theta$, the accuracy
of an approximation $\qDens(\theta)$ to $\pDens(\theta|\by)$ is defined to be 
\begin{equation}
\mbox{accuracy}\equiv\,
100\left\{1-\smhalf\infint\big|\qDens(\theta)-\pDens(\theta|\by)\big|\,d\theta\right\}\%.
\label{eq:accurDefn}
\end{equation}
The percentages in the top right-hand panel of Figure \ref{fig:ultrasoundAcc}
correspond to (\ref{eq:accurDefn}) with replacement of $\pDens(\theta|\by)$ by 
the above-mentioned kernel density estimate. In this case accuracy is 
seen to be excellent, with accuracy percentages between 97\% and 99\% 
for all 40 curves.

\null\vfill\eject

\subsection{Speed Assessment}

We also conducted some simulation studies to assess the speed of streamlined 
variational higher level group-specific curve models, in terms of both
comparative advantage over na\"{\i}ve implementation and absolute performance.
The focus of these studies was variational inference in the two-level case
and to probe maximal speed potential Algorithm \ref{alg:twoLevMFVB} 
was implemented in the low-level computer language \texttt{Fortran 77}.
An implementation of the na\"{\i}ve counterpart of Algorithm \ref{alg:twoLevMFVB},
involving storage and direct calculations concerning the full 
$\bSigma_{\qDens(\bbeta,\bu)}$ matrix, was also carried out. We then simulated 
data according to model (\ref{eq:twoLevelfg}) with $\sigeps=0.2$,
$$f(x)=3\sqrt{x(1.3-x)}\,\Phi(6x-3)\quad\mbox{and}\quad 
g_i(x)=\alpha_1\alpha_2\sin(2\pi x^{\alpha_3})
$$
where, for each $i$, $\alpha_1$, $\alpha_2$ and $\alpha_3$ are, respectively, 
random draws from the $N(\quarter,\quarter)$ distribution and the sets $\{-1,1\}$ 
and $\{1,2,3\}$. The level-2 sample sizes $n_i$ generated 
randomly from the set $\{30,31,\ldots,60\}$ and the 
level-1 sample sizes $m$ ranging over the set $\{100,200,300,400,500\}$.
All $x_{ij}$ data were generated from a Uniform distribution over the unit interval.
Table \ref{tab:StreamVsNaive} summarizes the timings based on 100 replications
with the number of mean field variational Bayes iterations fixed at 50.
The study was run on a \texttt{MacBook Air} laptop with a 2.2 gigahertz processor
and 8 gigabytes of random access memory.

\begin{table}[!ht]
\begin{center}
\begin{tabular}{cccc}
\hline\\[-1.5ex]
$m$               & na\"{\i}ve           &  streamlined   &  na\"{\i}ve/streamlined\\
\hline\\[-1.5ex]                      
100               &     75     (1.21)    &0.748 (0.0334)  &  100\\
200               &    660     (7.72)    &1.490 (0.0491)  &  442\\
300               &   2210    (22.00)    &2.260 (0.0567)  &  974\\
400               &   5180    (92.20)    &3.040 (0.0718)  & 1700\\
500               &     NA               &3.780 (0.0593)  &   NA\\
\hline
\end{tabular}
\end{center}
\caption{\textit{Average (standard deviation) of elapsed computing times in seconds for 
fitting model (\ref{eq:twoLevelfg}) na\"{\i}vely versus with streamlining
via Algorithm \ref{alg:twoLevMFVB}. The NA entries indicates 
non applicability due to the na\"{\i}ve computations not being feasible.}
}
\label{tab:StreamVsNaive}
\end{table}

For $m$ ranging from $100$ to $400$ we see that the na\"{\i}ve to streamlined
ratios increase from about $100$ to $1,700$. When $m=500$ the na\"{\i}ve
implementation fails to run due to its excessive storage demands.
In contrast, the streamlined fits are produced in about $3$ seconds.
It is clear that streamlined variational inference is to be preferred
and is the only option for large numbers of groups.

We then obtained timings for the streamlined algorithm for $m$ becoming
much larger, taking on values $100$, $500$, $2,500$ and $12,500$.
The iterations in Algorithm \ref{alg:twoLevMFVB} were stopped when
the relative increase in the marginal log-likelihood fell below $10^{-5}$.
The average and standard deviation times in seconds over 100 replications
are shown in Table \ref{tab:StreamAbsolute}. We see that the computational
times are approximately linear in $m$. Even with twelve and a half thousand
groups, Algorithm \ref{alg:twoLevMFVB} is able to deliver fitting
and inference on a contemporary laptop computer in about one and
a half minutes.

\begin{table}[!ht]
\begin{center}
\begin{tabular}{cccc}
\hline\\[-1.5ex]
$m=100$    & $m=500$  &  $m=2,500$  &  $m=12,500$\\
%
%
\hline\\[-1.5ex]
0.635     &  2.900      &   16.90      &    95.00   \\
(0.183)   &  (0.391)    &   (1.92)     &    (4.92)  \\
\hline
\end{tabular}
\end{center}
\caption{\textit{Average (standard deviation) of elapsed computing times in seconds for 
fitting model (\ref{eq:twoLevelfg}) with streamlining via Algorithm \ref{alg:twoLevMFVB}.}
}
\label{tab:StreamAbsolute}
\end{table}

\section*{Acknowledgments}

This research was partially supported by the Australian Research Council
Discovery Project DP140100441. The ultrasound data was provided by the 
Bioacoustics Research Laboratory, Department of Electrical and Computer
Engineering, University of Illinois at Urbana-Champaign, Illinois, U.S.A.

\section*{References}

\bib
Atay-Kayis, A. \myand Massam, H. (2005).
A Monte Carlo method for computing marginal likelihood
in nondecomposable Gaussian graphical models.
\textit{Biometrika}, {\bf 92}, 317--335.

\bib
Bates, D., M\"achler, M., Bolker, B. and Walker, S. (2015).
Fitting linear mixed-effects models using \texttt{lme4}.
\textit{Journal of Statistical Software}, {\bf 67(1)}, 1--48.

\bib
Bishop, C.M. (2006). {\it Pattern Recognition and Machine Learning.}
New York: Springer.

\bib
Brumback, B.A. and Rice, J.A. (1998).
Smoothing spline models for the analysis of nested and crossed
samples of curves (with discussion).
{\it Journal of the American Statistical Association},
{\bf 93}, 961--994.

\bib
Donnelly, C.A., Laird, N.M. and Ware, J.H. (1995).
Prediction and creation of smooth curves for
temporally correlated longitudinal data.
{\it Journal of the American Statistical Association},
{\bf 90}, 984--989.

\bib
Durban, M., Harezlak, J., Wand, M.P. and Carroll, R.J. (2005).
Simple fitting of subject-specific curves for longitudinal data.
{\it Statistics in Medicine}, {\bf 24}, 1153--1167.

\bib
Goldsmith, J., Zipunnikov, V. and Schrack, J. (2015).
Generalized multilevel function-on-scalar regression and
principal component analysis. \textit{Biometrics},
{\bf 71}, 344--353.

\bib
Guo, J., Gabry, J. and Goodrich, B. (2018). \textsf{rstan}: 
\textsf{R} interface to \textsf{Stan}.
\textsf{R} package version 2.18.2.\\
\texttt{http://mc-stan.org}.

\bib
Huang, A. and Wand, M.P. (2013).
Simple marginally noninformative prior distributions 
for covariance matrices. \textit{Bayesian Analysis}, 
{\bf 8}, 439--452.

\bib
Lee, C.Y.Y. and Wand, M.P. (2016).
Variational inference for fitting complex Bayesian mixed
effects models to health data. \textit{Statistics in Medicine},
{\bf 35}, 165--188.

\bib
Nolan, T.H., Menictas, M. and Wand, M.P. (2019).
Streamlined computing for variational inference with higher level 
random effects. Unpublished manuscript submitted
to the arXiv.org e-Print archive; on hold as of 11th March 2019.
Soon to be posted also on \texttt{http://matt-wand.utsacademics.info/statsPapers.html}

\bib
Nolan, T.H. and Wand, M.P. (2018). Solutions
to sparse multilevel matrix problems. Unpublished manuscript
available at \textit{https://arxiv.org/abs/1903.03089}.

\bib
Pinheiro, J.C. and Bates, D.M. (2000).
{\it Mixed-Effects Models in S and S-PLUS}.
New York: Springer.

\bib
Pinheiro, J., Bates, D., DebRoy, S., Sarkar, D. and \textsf{R} Core Team. (2018). 
\textsf{nlme}: Linear and nonlinear mixed effects models. 
\textsf{R} package version 3.1. \\
\texttt{http://cran.r-project.org/package=nlme}.

\bib
Pratt, J.H., Jones, J.J., Miller, J.Z., Wagner, M.A. and Fineberg, N.S.
(1989). Racial differences in aldosterone excretion and plasma aldosterone
concentrations in children. {\it New England Journal of Medicine}, {\bf 321},
1152--1157.

\bib
Robinson, G.K. (1991). That BLUP is a good thing: the estimation
of random effects. {\it Statistical Science}, {\bf 6}, 15--51.

\bib
Trail, J.B., Collins, L.M., Rivera, D.E., Li, R., Piper, M.E. and Baker, T.B. (2014).
Functional data analysis for dynamical system identification of behavioral processes. 
{\it Psychological Methods}, {\bf 19(2)}, 175--187.

\bib
Verbyla, A.P., Cullis, B.R., Kenward, M.G. and Welham, S.J. (1999).
The analysis of designed experiments and longitudinal data
by using smoothing splines (with discussion).
{\it Applied Statistics}, {\bf 48}, 269--312.

\bib
Wahba, G. (1990). {\it Spline Models for Observational Data.}
Philadelphia: Society for Industrial and Applied Mathematics.

\bib
Wand, M.P. and Ormerod, J.T. (2008).
On semiparametric regression with O'Sullivan penalized splines.
{\it Australian and New Zealand Journal of Statistics},
{\bf 50}, 179--198.

\bib
Wand, M.P. and Ormerod, J.T. (2011).
Penalized wavelets: embedding wavelets
into semiparametric regression.
{\it Electronic Journal of Statistics},
{\bf 5}, 1654--1717.

\bib
Wang, Y. (1998). Mixed effects smoothing spline analysis
of variance. {\it Journal of the Royal Statistical
Society, Series B}, {\bf 60}, 159--174.

\bib
Wirtzfeld, L.A., Ghoshal, G., Rosado-Mendez, I.M., Nam, K., Park, Y.,
Pawlicki, A.D., Miller, R.J., Simpson, D.G., Zagzebski, J.A., Oelze, M.I.,
Hall, T.J. and O'Brien, W.D. (2015).
Quantitative ultrasound comparison of MAT and 4T1 mammary tumors in
mice and rates across multiple imaging systems. 
\textit{Journal of Ultrasound Medicine}, {\bf 34}, 1373--1383.

\bib
Zhang, D., Lin, X., Raz, J. and Sowers, M. (1998).
Semi-parametric stochastic mixed models for longitudinal data.
{\it Journal of the American Statistical Association}, {\bf 93}, 710--719.

\vfill\eject
%
%
\renewcommand{\theequation}{S.\arabic{equation}}
\renewcommand{\thesection}{S.\arabic{section}}
\renewcommand{\thetable}{S.\arabic{table}}
\renewcommand{\thealgorithm}{S.\arabic{algorithm}}
\setcounter{equation}{0}
\setcounter{table}{0}
\setcounter{section}{0}
\setcounter{page}{1}
\setcounter{footnote}{0}
\setcounter{algorithm}{0}

\cleardoublepage

\null
\vskip5mm
\centerline{\Large Web-Supplement for:}
\vskip3mm
\centerline{\Large\bf Streamlined Variational Inference for}
\vskip2mm
\centerline{\Large\bf Higher Level Group-Specific Curve Models}
\vskip4mm
\centerline{\normalsize\sc By M. Menictas$\null^1$, T.H. Nolan$\null^1$,
D.G. Simpson$\null^2$ and M.P. Wand$\null^1$}
\vskip5mm
\centerline{\textit{University of Technology Sydney$\null^1$ and University of Illinois$\null^2$}}
\vskip6mm

\section{Derivation of Result \ref{res:twoLevelBLUP}}\label{sec:drvResultOne}
Straightforward algebra can be used to verify that
\begin{equation*}
  \begin{array}{c}
    \bC^T\RBLUP^{-1}\bC+\DBLUP = \bB^T \bB \mbox{  and  } \bC^T\RBLUP^{-1}\by = \bB^T \bb
  \end{array}
\end{equation*}
where $\bB$ and $\bb$ have sparse forms (\ref{eq:BandbForms}) with non-zero sub-blocks
equal to
$$
\bveci\equiv
\left[
\begin{array}{c}
\sigeps^{-1}\by_i\\[1ex]
\bzero \\[1ex]
\bzero \\[1ex]
\bzero \\[1ex]
\end{array}
\right],
\quad
\Bmati\equiv
\left[
\begin{array}{cc}
\sigeps^{-1}\bX_i & \sigeps^{-1}\bZgbli\\[1ex]
\bO & m^{-1/2}\sigmaGbl^{-1}\bI_{\Kgbl}\\[1ex]
\bO & \bO   \\[1ex]
\bO & \bO   \\[1ex]
\end{array}
\right]
\quad\mbox{and}\quad
\Bmatdoti\equiv
\left[
\begin{array}{cc}
\sigeps^{-1}\bX_i  & \sigeps^{-1}\bZgrpi        \\[1ex]
\bO                & \bO            \\[1ex]
\bSigma^{-1/2}     & \bO            \\[1ex]
\bO                & \sigmaGrp^{-1}\bI_{\Kgrp}
\end{array}
\right].
$$
Therefore, in view of (\ref{eq:BLUPandCov}) and (\ref{eq:CovMain}),
$$
\left[\begin{array}{c}
\bbetahat\\
\buhat
\end{array}
\right]=(\bB^T\bB)^{-1}\bB^T\bb
\quad\mbox{and}\quad
\mbox{\rm Cov}\left(\left[\begin{array}{c}
\bbetahat\\
\buhat-\bu
\end{array}
\right]\right)=(\bB^T\bB)^{-1}.
$$
\section{Derivation of Algorithm \ref{alg:twoLevBLUP}}\label{sec:drvAlgOne}
Algorithm \ref{alg:twoLevBLUP} is simply a proceduralization of Result \ref{res:twoLevelBLUP}.

\section{The Inverse G-Wishart and Inverse $\chi^2$ Distributions}\label{sec:IGWandICS}

The Inverse G-Wishart corresponds to the matrix inverses of random matrices that have
a \emph{G-Wishart} distribution (e.g. Atay-Kayis \myand Massam, 2005). 
For any positive integer $d$, let $G$ be an undirected graph with $d$ nodes 
labeled $1,\ldots,d$ and set $E$ consisting of sets of pairs of nodes that 
are connected by an edge. We say that the symmetric $d\times d$ matrix $\bM$ 
\emph{respects} $G$ if 
$$\bM_{ij}=0\quad\mbox{for all}\quad \{i,j\}\notin E.$$
A $d\times d$ random matrix $\bX$ has an Inverse G-Wishart distribution
with graph $G$ and parameters $\xi>0$ and symmetric $d\times d$ 
matrix $\bLambda$, written
$$\bX\sim\mbox{Inverse-G-Wishart}(G,\xi,\bLambda)$$
if and only if the density function of $\bX$ satisfies
$$\pDens(\bX)\propto |\bX|^{-(\xi+2)/2}\exp\{-\smhalf\tr(\bLambda\,\bX^{-1})\}$$
over arguments $\bX$ such that $\bX$  is symmetric and positive definite 
and $\bX^{-1}$ respects $G$. Two important special cases are 
$$G=\Gfull\equiv\mbox{totally connected $d$-node graph},$$
for which the Inverse G-Wishart distribution coincides with the ordinary
Inverse Wishart distribution, and
$$G=\Gdiag\equiv\mbox{totally disconnected $d$-node graph},$$
for which the Inverse G-Wishart distribution coincides with
a product of independent Inverse Chi-Squared random variables.
The subscripts of $\Gfull$ and $\Gdiag$ reflect the 
fact that $\bX^{-1}$ is a full matrix and $\bX^{-1}$ is
a diagonal matrix in each special case.

The $G=\Gfull$ case corresponds to the ordinary Inverse Wishart
distribution. However, with message passing in mind, we will
work with the more general Inverse G-Wishart
family throughout this article.

In the $d=1$ special case the graph $G=\Gfull=\Gdiag$ 
and the Inverse G-Wishart distribution reduces to the
Inverse Chi-Squared distributions. We write
$$x\sim\mbox{Inverse-$\chi^2$}(\xi,\lambda)$$
for this  $\mbox{Inverse-G-Wishart}(\Gdiag,\xi,\lambda)$
special case with $d=1$ and $\lambda>0$ scalar.

\section{Derivation of Result \ref{res:twoLevelMFVB}}\label{sec:drvResultTwo}

It is straightforward to verify that the $\bmu_{\qDens(\bbeta,\bu)}$
and $\bSigma_{\qDens(\bbeta,\bu)}$ updates, given at (\ref{eq:muSigmaMFVBupd}),
may be written as
$$\bmu_{\qDens(\bbeta,\bu)}\thickarrow(\bB^T\bB)^{-1}\bB^T\bb
\quad\mbox{and}\quad
\bSigma_{\qDens(\bbeta,\bu)}\thickarrow(\bB^T\bB)^{-1}
$$
where $\bB$ and $\bb$ have the forms (\ref{eq:BandbForms})
with
$$\bveci\equiv\left[\begin{array}{c}
\mu_{\qDens(1/\sigeps^2)}^{1/2}\by_i\\[2ex]
m^{-1/2}\bSigma_{\bbeta}^{-1/2}\bmu_{\bbeta}\\[2ex]
\bzero\\[1ex]
\bzero\\[1ex]
\bzero
\end{array}
\right],
\quad\Bmati\equiv\left[\begin{array}{cc}
\mu_{\qDens(1/\sigeps^2)}^{1/2}\bX_i & \mu_{\qDens(1/\sigeps^2)}^{1/2}\bZgbli\\[2ex]
m^{-1/2}\bSigma_{\bbeta}^{-1/2}& \bO \\[2ex]
\bO & m^{-1/2}\mu_{\qDens(1/\sigmaGbl^2)}^{1/2}\bI_{\Kgbl}\\[2ex]
\bO                            &  \bO        \\[2ex]
\bO                            &  \bO        \\[2ex]
\end{array}
\right]
$$
\quad\mbox{\it and}\quad
$$
\Bmatdoti\equiv
\left[\begin{array}{cc}
\mu_{\qDens(1/\sigeps^2)}^{1/2}\bX_i & \mu_{\qDens(1/\sigeps^2)}^{1/2}\bZgrpi  \\[2ex]
\bO                             &   \bO                              \\[2ex]
\bO                             &   \bO                              \\[2ex]
\bM_{\qDens(\bSigma^{-1})}^{1/2}     &   \bO                              \\[2ex]
\bO                             &  \mu_{\qDens(1/\sigmaGrp^2)}^{1/2}\bI_{\Kgrp}
\end{array}
\right].
$$
Result \ref{res:twoLevelMFVB} immediately follows from Theorem 2 of Nolan \myand Wand (2018).
\section{Derivation of Algorithm \ref{alg:twoLevMFVB}}\label{sec:drvAlgTwo}
We provide expressions for the $\qDens$-densities for mean
field variational Bayesian inference for the parameters in (\ref{eq:twoLevBayes}),
with product density restriction (\ref{eq:producRestrict}).
Arguments analogous to those given in, for example, Appendix C of
Wand \myand Ormerod (2011) lead to:
$$\qDens(\bbeta,\bu)\ \mbox{is a $N(\bmu_{\qDens(\bbeta,\bu)},\bSigma_{\qDens(\bbeta,\bu)})$
density function}$$
where
$$\bSigma_{\qDens(\bbeta,\bu)}=(\bC^T\RMFVB^{-1}\bC+\DMFVB)^{-1}
\quad
\mbox{and}
\quad
\bmu_{\qDens(\bbeta,\bu)}=\bSigma_{\qDens(\bbeta,\bu)}(\bC^T\RMFVB^{-1}\by + \oMFVB)
$$
with $\RMFVB$, $\DMFVB$ and $\oMFVB$ defined via (\ref{eq:MFVBmatDefns}),
$$\qDens(\sigsqeps)\ \mbox{is an $\mbox{Inverse-$\chi^2$}
\left(\xi_{\qDens(\sigsqeps)},\lambda_{\qDens(\sigsqeps)}\right)$ density function}
$$
where $\xi_{\qDens(\sigsqeps)}=\nuEps+\sumim n_i$ and
\begin{eqnarray*}
\lambda_{\qDens(\sigsqeps)}&=&\mu_{\qDens(1/\aeps)}+\sumim\,
E_{\qDens}\left\{ \Bigg \Vert \by_{i} - \bCgbli \left[ \begin{array}{c}
      \bbeta \\[1ex] \buGbl \end{array} \right] - \bCgrpi \left[ \begin{array}{c}
            \buLini \\[1ex] \buGrpi \end{array} \right] \Bigg \Vert^2 \right\} \\[1ex]
& = & \mu_{\qDens(1/\aeps)}+\sumim\,\left[ \Bigg \Vert\,E_{\qDens} \left( \by_{i}
    - \bCgbli \left[ \begin{array}{c} \bbeta \\[1ex] \buGbl \end{array} \right]
    - \bCgrpi \left[ \begin{array}{c} \buLini \\[1ex] \buGrpi \end{array} \right]
    \right) \Bigg \Vert^2 \right.
\\[1ex]
& & \left. \qquad \qquad \qquad \qquad
    + \mbox{tr} \left\{\mbox{\rm Cov}_{\qDens} 
    \left(\bCgbli\left[\begin{array}{c}
    \bbeta \\[1ex] \buGbli \end{array} \right] 
    + \bCgrpi \left[ \begin{array}{c}
                     \buLini \\[1ex] 
                     \buGrpi \end{array} 
               \right] \right)  
              \right\} \right] \\[1ex]
& = & \mu_{\qDens(1/\aeps)} + \displaystyle{\sum_{i=1}^{m}} \left\{
      \Bigg \Vert E_{\qDens} \left( \by_{i} - \bCgbli \left[ \begin{array}{c}
      \bbeta \\[1ex] \buGbl \end{array} \right] - \bCgrpi \left[ \begin{array}{c}
      \buLini \\[1ex] \buGrpi \end{array} \right] \right) \Bigg \Vert^2  \right.
\\[3ex]
& & \qquad \qquad \qquad \quad \left. + \mbox{tr}(\bCgbli^T\bCgbli\bSigma_{\qDens(\bbeta,\buGbl)})
    + \mbox{tr}(\bCgrpi^T\bCgrpi\bSigma_{\qDens(\buLini,\buGrpi)}) \right.
\\[1ex]
& & \qquad \qquad \qquad \quad \left. + 2 \, \mbox{tr}\left[\bCgrpi^T\bCgbli\,E_{\qDens}\left\{\left(
        \left[\begin{array}{c}\bbeta \\ \buGbl\end{array}\right]
        -\bmu_{\qDens(\bbeta,\buGbl)}\right) \times \right. \right. \right.
\\[1ex]
& & \left. \left. \left. \qquad \qquad \qquad \qquad \qquad \qquad \qquad \quad \quad
    \left(\left[\begin{array}{c}\buLini\\
    \buGrpi\end{array}\right] -\bmuq{\buLini,\buGrpi}\right)^T\right\}\right] \right\}
\end{eqnarray*}
where $\bCgbli \equiv [ \bX_{i} \, \bZgbli ]$, $\bCgrpi \equiv [ \bX_{i} \, \bZgrpi ]$, and
with reciprocal moment $\mu_{\qDens(1/\sigsqeps)}=\xi_{\qDens(\sigsqeps)}/\lambda_{\qDens(\sigsqeps)},$
$$\qDens(\sigmaGbl^2)\ \mbox{is an $\mbox{Inverse-$\chi^2$}
\left(\xi_{\qDens(\sigmaGbl^2)},\lambda_{\qDens(\sigmaGbl^2)}\right)$ density function}
$$
where $\xi_{\qDens(\sigmaGbl^2)}=\nuGbl+\Kgbl$ and
$$
\lambda_{\qDens(\sigmaGbl^2)}= \mu_{\qDens(1/{\aGbl})} + \Vert \bmu_{\qDens(\buGbl)} \Vert^2
 + \mbox{tr} \left( \bSigma_{\qDens(\buGbl)} \right),
$$
with reciprocal moment $\mu_{\qDens(1/{\sigmaGbl^2})}=\xi_{\qDens(\sigmaGbl^2)} /
\lambda_{\qDens(\sigmaGbl^2)},$
$$\qDens(\sigmaGrp^2)\ \mbox{is an $\mbox{Inverse-$\chi^2$}
\left(\xi_{\qDens(\sigmaGrp^2)},\lambda_{\qDens(\sigmaGrp^2)}\right)$ density function}
$$
where $\xi_{\qDens(\sigmaGrp^2)}=\nuGrp+m\Kgrp$ and
$$
\lambda_{\qDens(\sigmaGrp^2)} = \mu_{\qDens(1/{\aGrp})} + \sum_{i=1}^m \left\{ \Vert \bmu_{\qDens(\buGrpi)} \Vert^2
 + \mbox{tr} \left( \bSigma_{\qDens(\buGrpi)} \right) \right\},
$$
with reciprocal moment $\mu_{\qDens(1/{\sigmaGrp^2})}=\xi_{\qDens(\sigmaGrp^2)} /
\lambda_{\qDens(\sigmaGrp^2)},$
$$\qDens(\bSigma)\ \mbox{is an $\mbox{Inverse-G-Wishart}
\left(\Gfull,\xi_{\qDens(\bSigma)},\bLambda_{\qDens(\bSigma)}\right)$ density function}
$$
where $\xi_{\qDens(\bSigma)}=\nuSigma+2+m$
$$\bLambda_{\qDens(\bSigma)}=\bM_{\qDens(\bA_{\bSigma}^{-1})}
+\sumim\left(\bmu_{\qDens(\buLini)}\bmu_{\qDens(\buLini)}\trans + \bSigma_{\qDens(\buLini)}\right),$$
with inverse moment $\bM_{\qDens(\bSigma^{-1})}=(\xi_{\qDens(\bSigma)}-1)\bLambda_{\qDens(\bSigma)}^{-1}$,
$$\qDens(\aeps)\ \mbox{is an $\mbox{Inverse-$\chi^2$}
(\xi_{\qDens(\aeps)},\lambda_{\qDens(\aeps)})$ density function}$$
where $\xi_{\qDens(\aeps)}=\nuEps+1$,
$$\lambda_{\qDens(\aeps)}=\mu_{\qDens(1/\sigsqeps)}+1/(\nuEps\sEps^2)$$
with reciprocal moment $\mu_{\qDens(1/\aeps)}=\xi_{\qDens(\aeps)}/\lambda_{\qDens(\aeps)}$,
$$\qDens(\aGbl)\ \mbox{is an $\mbox{Inverse-$\chi^2$}
(\xi_{\qDens(\aGbl)},\lambda_{\qDens(\aGbl)})$ density function}$$
where $\xi_{\qDens(\aGbl)}=\nuGbl+1$,
$$\lambda_{\qDens(\aGbl)}=\mu_{\qDens(1/\sigmaGbl^2)}+1/(\nuGbl\sGbl^2)$$
with reciprocal moment $\mu_{\qDens(1/\aGbl)}=\xi_{\qDens(\aGbl)}/\lambda_{\qDens(\aGbl)}$,
$$\qDens(\aGrp)\ \mbox{is an $\mbox{Inverse-$\chi^2$}
(\xi_{\qDens(\aGrp)},\lambda_{\qDens(\aGrp)})$ density function}$$
where $\xi_{\qDens(\aGrp)}=\nuGrp+1$,
$$\lambda_{\qDens(\aGrp)}=\mu_{\qDens(1/\sigmaGrp^2)}+1/(\nuGrp\sGrp^2)$$
with reciprocal moment $\mu_{\qDens(1/\aGrp)}=\xi_{\qDens(\aGrp)}/\lambda_{\qDens(\aGrp)}$ and
$$\qDens(\ASigma)\ \mbox{is an $\mbox{Inverse-G-Wishart}
\left(\Gdiag,\xi_{\qDens(\ASigma)},\bLambda_{\qDens(\ASigma)}\right)$ density function}
$$
where $\xi_{\qDens(\ASigma)}=\nuSigma+2$,
$$\bLambda_{\qDens(\ASigma)}=\diag\big\{\mbox{diagonal}\big(\bM_{\qDens(\bSigma^{-1})}\big)\big\}
+\bLambda_{\ASigma}$$
with inverse moment $\bM_{\qDens(\ASigma^{-1})}=\xi_{\qDens(\ASigma)}\bLambda_{\qDens(\ASigma)}^{-1}$.
%
%
%
%
\section{Marginal Log-Likelihood Lower Bound and Derivation}\label{sec:lowerBound}
\noindent The expression for the lower bound on the marginal log-likelihood for Algorithm
\ref{alg:twoLevMFVB} is \\
\begin{equation}
  \begin{array}{l}
    \log \underline{\pDens}(\by; \qDens) =
    \\[2ex]
    -\frac{1}{2} \log (\pi) \displaystyle{\sum_{i=1}^{m}} n_{i} - \smhalf \log |\bSigma_{\bbeta}|
    - \smhalf \mbox{tr} \left( \bSigma_{\bbeta}^{-1} \left\{
    \left( \bmu_{\qDens (\bbeta)} - \bmu_{\bbeta} \right) \left( \bmu_{\qDens (\bbeta)}
    - \bmu_{\bbeta} \right)^T + \bSigma_{\qDens (\bbeta)} \right\} \right)
    \\[1ex]
    -\smhalf \mbox{tr} \left( \bM_{\qDens (\bSigma^{-1})} \left\{ \displaystyle{\sum_{i=1}^{m}} \left(
    \bmu_{\qDens (\buLini} \bmu_{\qDens (\buLini)}^T + \bSigma_{\qDens (\buLini)} \right) \right\} \right)
    + \smhalf \{ 2 + \Kgbl + m (2 + \Kgrp) \}
    \\[1ex]
    - \smhalf \mu_{\qDens(1/{\sigmaGbl^2})} \left\{
    \Vert \bmu_{\qDens ( \buGbl )} \Vert^2 + \mbox{tr} ( \bSigma_{\qDens(\buGbl)} ) \right\}
    - \smhalf \mu_{\qDens ( 1/{\sigmaGrp^2} )} \displaystyle{\sum_{i=1}^{m}} \left\{
    \Vert \bmu_{\qDens ( \buGrpi ))} \Vert^2 + \mbox{tr} (\bSigma_{\qDens(\buGrpi)}) \right\}
    \\[1ex]
    +  \smhalf \log |\bSigma_{\bbeta}|
    + \{ \nuSigma + m + 1 + \smhalf (\nuEps + \nuGbl + \Kgbl + \nuGrp + m \Kgrp) \} \log(2)
    - \log \Gamma (\frac{\nuEps}{2})
    \\[1ex]
    - \smhalf \mu_{\qDens(1/{\aeps})} \mu_{\qDens (1/{\sigsqeps})}
    - \smhalf \xi_{\qDens (\sigsqeps)} \log ( \lambda_{\qDens(\sigsqeps)}) + \log \{
    \Gamma (\smhalf \xi_{\qDens (\sigsqeps)}) \} + \smhalf \lambda_{\qDens(\sigsqeps)}
    \mu_{\qDens(1/\sigsqeps)} - \smhalf \log (\nuEps \sEps^2 )
    \\[1ex]
    - 3 \log \{ \Gamma ( \smhalf ) \}
    - \frac{1}{2 \nuEps \sEps^2} \mu_{\qDens (1/\aeps)} - \smhalf \xi_{\qDens (\aeps)}
    \log (\lambda_{\qDens (\aeps)}) + \log \{ \Gamma (\smhalf \xi_{\qDens(\aeps)}) \}
    + \smhalf \lambda_{\qDens (\aeps)} \mu_{\qDens(1/ \aeps)}
    \\[1ex]
    - \log \Gamma (\frac{\nuGbl}{2}) - \smhalf \mu_{\qDens(1/{\aGbl})} \mu_{\qDens (1/{\sigmaGbl^2})}
    - \smhalf \xi_{\qDens (\sigmaGbl^2)} \log ( \lambda_{\qDens(\sigmaGbl^2)}) + \log \{
    \Gamma (\smhalf \xi_{\qDens (\sigmaGbl^2)}) \} - \smhalf \log (\nuGbl \sGbl^2 )
    \\[1ex]
    + \smhalf \lambda_{\qDens(\sigmaGbl^2)} \mu_{\qDens(1/\sigmaGbl^2)}
     - \{ 1/(2 \nuGbl \sGbl^2) \} \mu_{\qDens (1/\aGbl)}
    - \smhalf \xi_{\qDens (\aGbl)} \log (\lambda_{\qDens (\aGbl)})
    - \smhalf \mu_{\qDens(1/{\aGrp})} \mu_{\qDens (1/{\sigmaGrp^2})}
    \\[1ex]
    + \log \{ \Gamma (\smhalf \xi_{\qDens(\aGbl)}) \} + \smhalf \lambda_{\qDens (\aGbl)}
    \mu_{\qDens(1/ \aGbl)} - \log \Gamma (\frac{\nuGrp}{2}) + \log \{
    \Gamma (\smhalf \xi_{\qDens (\sigmaGrp^2)}) \} - \smhalf \log (\nuGrp \sGrp^2 )
    \\[1ex]
    - \smhalf \xi_{\qDens (\sigmaGrp^2)} \log ( \lambda_{\qDens(\sigmaGrp^2)})
    + \smhalf \lambda_{\qDens(\sigmaGrp^2)} \mu_{\qDens(1/\sigmaGrp^2)}
    - \{ 1/(2 \nuGrp \sGrp^2) \} \mu_{\qDens (1/\aGrp)} - \smhalf \xi_{\qDens (\aGrp)}
    \log (\lambda_{\qDens (\aGrp)})
    \\[1ex]
     + \log \{ \Gamma (\smhalf \xi_{\qDens(\aGrp)}) \} + \smhalf \lambda_{\qDens (\aGrp)}
     \mu_{\qDens(1/ \aGrp)} - \smhalf \mbox{tr} ( \bM_{\qDens(\bASigma^{-1})}
     \bM_{\qDens(\bSigma^{-1})} ) + \smhalf \mbox{tr} (\bLambda_{\qDens(\bSigma)}
     \bM_{\qDens(\bSigma^{-1})})
    \\[1ex]
    + \displaystyle{\sum_{j=1}^{2}} \log \Gamma (\smhalf (\xi_{\qDens(\bASigma)} + 2 - j))
    - \displaystyle{\sum_{j=1}^{2}} \log \Gamma (\smhalf (\nuSigma + 4 - j))
    - \smhalf (\xi_{\qDens(\bSigma)} - 1) \log |\bLambda_{\qDens(\bSigma)}|
    \\[1ex]
    + \displaystyle{\sum_{j=1}^{2}} \log \Gamma (\smhalf (\xi_{\qDens(\bSigma)} + 2 - j))
    - \smhalf \displaystyle{\sum_{j=1}^{2}} 1/(\nuSigma \sSigmaj^2)
    \left( \bM_{\qDens(\bASigma^{-1})} \right)_{jj}
    - \displaystyle{\sum_{j=1}^{2}} \log \Gamma (\smhalf (3-j))
    \\[1ex]
    - \smhalf (\xi_{\qDens(\bASigma)} - 1) \log |\bLambda_{\qDens(\bASigma)}|
    + \smhalf \mbox{tr} (\bLambda_{\qDens(\bASigma)} \bM_{\qDens(\bASigma^{-1})})
    \\[1ex]
    - \frac{1}{2} \mu_{\qDens (1/{\sigsqeps})} \displaystyle{\sum_{i=1}^{m}} \left\{
    \Bigg \Vert E_{\qDens} \left( \by_{i} - \bCgbli \left[ \begin{array}{c}
                \bbeta \\[1ex] \buGbli \end{array} \right] - \bCgrpi \left[ \begin{array}{c}
                \buLini \\[1ex] \buGrpi \end{array} \right] \right) \Bigg \Vert^2  \right.
    \\[3ex]
    \qquad \qquad \qquad \quad \left. + \mbox{tr}(\bCgbli^T\bCgbli\bSigma_{\qDens(\bbeta,\buGbl)})
    +\mbox{tr}(\bCgrpi^T\bCgrpi\bSigma_{\qDens(\buLini,\buGrpi)}) \right.
    \\[1ex]
    \qquad \qquad \qquad \quad \left. + 2 \, \mbox{tr}\left[\bCgrpi^T\bCgbli\,E_{\qDens}\left\{\left(
    \left[\begin{array}{c}\bbeta \\ \buGbl\end{array}\right]
    -\bmu_{\qDens(\bbeta,\buGbl)}\right) \times \right. \right. \right.
    \\[1ex]
    \qquad \qquad \qquad \qquad \qquad \qquad \qquad \qquad \left. \left. \left.
    \left(\left[\begin{array}{c}\buLini\\ \buGrpi\end{array}\right]
    -\bmuq{\buLini,\buGrpi)}\right)^T\right\}\right] \right\}.
  \end{array}
  \label{eq:lowerBound}
\end{equation}
\vskip4mm
\noindent \textit{Derivation:} The lower-bound on the marginal log-likelihood is achieved
through the following expression:
\begin{equation*}
  \begin{array}{lcl}
    \log \underline{\pDens}(\by; \qDens) & = & E_{\qDens} \{ \log \pDens (\by, \bbeta, \bu, \sigsqeps, \aeps,
    \sigmaGbl^2, \aGbl, \sigmaGrp^2, \aGrp, \bSigma, \ASigma )
    \\[1ex]
    & & - \, \log \qDens^{*} ( \bbeta, \bu, \sigsqeps, \aeps,
    \sigmaGbl^2, \aGbl, \sigmaGrp^2, \aGrp, \bSigma, \ASigma ) \}
  \end{array}
\end{equation*}
\begin{equation}
  \begin{array}{lcl}
    & = & E_{\qDens} \{ \log \pDens (\by \, | \, \bbeta, \bu, \sigsqeps) \}
    \\[1ex]
    & & + \, E_{\qDens} \{ \log \pDens (\bbeta, \bu \, | \, \sigmaGbl^2, \sigmaGrp^2, \bSigma) \} - E_{\qDens} \{ \log \qDens^{*} (\bbeta, \bu) \}
    \\[1ex]
    & & + \, E_{\qDens} \{ \log \pDens (\sigsqeps \, | \, \aeps) \} - E_{\qDens} \{ \log \qDens^{*} (\sigsqeps) \}
        + E_{\qDens} \{ \log \pDens (\aeps) \} - E_{\qDens} \{ \log \qDens^{*} (\aeps) \}
    \\[1ex]
    & & + \, E_{\qDens} \{ \log \pDens (\sigmaGbl^2 \, | \, \aGbl) \} - E_{\qDens} \{ \log \qDens^{*} (\sigmaGbl^2) \}
        + E_{\qDens} \{ \log \pDens (\aGbl) \} - E_{\qDens} \{ \log \qDens^{*} (\aGbl) \}
    \\[1ex]
    & & + \, E_{\qDens} \{ \log \pDens (\sigmaGrp^2 \, | \, \aGrp) \} - E_{\qDens} \{ \log \qDens^{*} (\sigmaGrp^2) \}
        + E_{\qDens} \{ \log \pDens (\aGrp) \} - E_{\qDens} \{ \log \qDens^{*} (\aGrp) \}
    \\[1ex]
    & & + \, E_{\qDens} \{ \log \pDens (\bSigma \, | \, \bA_{\mbox{\rm\tiny $\bSigma$}}) \} - E_{\qDens} \{ \log \qDens^{*} (\bSigma) \}
        + E_{\qDens} \{ \log \pDens (\bA_{\mbox{\rm\tiny $\bSigma$}}) \} - E_{\qDens} \{ \log \qDens^{*} (\bA_{\mbox{\rm\tiny $\bSigma$}})\}.
  \end{array}
  \label{eq:lowerBoundSetup}
\end{equation}
First we note that
$$
\log \pDens (\by \, | \, \bbeta, \bu, \sigsqeps)
     =  -\frac{1}{2} \log (2 \pi) \displaystyle{\sum_{i=1}^{m}} n_{i}
    - \frac{1}{2} \log (\sigsqeps) \displaystyle{\sum_{i=1}^{m}} n_{i}
    - \frac{1}{2 \sigsqeps} \displaystyle{\sum_{i=1}^{m}} \Vert \by - \bX \bbeta - \bZ \bu \Vert^2
$$
where
\begin{equation*}
  \begin{array}{l}
    \Vert \by - \bX \bbeta - \bZ \bu \Vert^2
    \\[1ex]
    \quad = \vast \Vert \left[ \begin{array}{c}
                    \by_{1} \\
                    \vdots \\
                    \by_{m}
                 \end{array}
           \right] -
           \left[ \begin{array}{c}
                    \bX_{1} \\
                    \vdots \\
                    \bX_{m}
                  \end{array}
            \right] \bbeta -
            \left[ \begin{array}{c}
                      \bZgblo \\
                      \vdots \\
                      \bZgblm
                    \end{array}
            \right] \buGbl
       - \displaystyle \blockdiag{1\le i\le m}([\bX_i\ \bZgrpi])
            \left[ \begin{array}{c}
                      \buLini\\
                      \buGrpi
                    \end{array}
            \right]_{1\le i\le m} \vast \Vert^2
    \\[5ex]
    \quad = \displaystyle \sum_{i=1}^{m} \Vert \by_{i} - \bX_{i} \bbeta - \bZgbli \buGbli
          - \bX_{i} \buLini - \bZgrpi \buGrpi \Vert^2
    \\[3ex]
    \quad = \displaystyle \sum_{i=1}^{m} \Bigg \Vert \by_{i} - \bCgbli \left[ \begin{array}{c}
          \bbeta \\[1ex] \buGbli \end{array} \right] - \bCgrpi \left[ \begin{array}{c}
                \buLini \\[1ex] \buGrpi \end{array} \right] \Bigg \Vert^2
  \end{array}
\end{equation*}
and
\begin{equation*}
  \begin{array}{c}
    \bCgbli \equiv [ \bX_{i} \, \bZgbli ], \quad \bCgrpi \equiv [ \bX_{i} \, \bZgrpi ].
  \end{array}
\end{equation*}
Therefore,
\begin{equation*}
  \begin{array}{l}
    E_{\qDens} \{ \log \pDens (\by \, | \, \bbeta, \bu, \sigsqeps) \}
    \\[1ex]
    \quad = -\frac{1}{2} \log (2 \pi) \displaystyle{\sum_{i=1}^{m}} n_{i}
    - \frac{1}{2} E_{\qDens} \{ \log (\sigsqeps) \} \displaystyle{\sum_{i=1}^{m}} n_{i}
    \\[1ex]
    \quad \quad - \frac{1}{2} \mu_{\qDens (1/{\sigsqeps})} \displaystyle{\sum_{i=1}^{m}} \left\{
    \Bigg \Vert E_{\qDens} \left( \by_{i} - \bCgbli \left[ \begin{array}{c}
                \bbeta \\[1ex] \buGbli \end{array} \right] - \bCgrpi \left[ \begin{array}{c}
                \buLini \\[1ex] \buGrpi \end{array} \right] \right) \Bigg \Vert^2  \right.
    \\[3ex]
    \quad \quad \quad \left. + \mbox{tr}(\bCgbli^T\bCgbli\bSigma_{\qDens(\bbeta,\buGbl)})
    +\mbox{tr}(\bCgrpi^T\bCgrpi\bSigma_{\qDens(\buLini,\buGrpi)}) \right.
    \\[1ex]
    \quad \quad \quad \left. + 2 \, \mbox{tr}\left[\bCgrpi^T\bCgbli\,E_{\qDens}\left\{\left(
            \left[\begin{array}{c}\bbeta \\ \buGbl\end{array}\right]
            -\bmu_{\qDens(\bbeta,\buGbl)}\right)\left(\left[\begin{array}{c}\buLini\\
            \buGrpi\end{array}\right]
           -\bmuq{\buLini,\buGrpi)}\right)^T\right\}\right] \right\}
  \end{array}
\end{equation*}
The remainder of the expectations in (\ref{eq:lowerBoundSetup}) are expressed as:
\begin{equation*}
  \begin{array}{lcl}
    E_{\qDens} \{ \log \pDens (\bbeta, \bu \, | \, \sigmaGbl^2, \sigmaGrp^2, \bSigma) \} & = &
    - \smhalf \{ 2 + \Kgbl + m (2 + \Kgrp) \} \log (2\pi) - \smhalf \log |\bSigma_{\bbeta}|
    \\[1ex]
    & & -\frac{\Kgbl}{2} E_{\qDens} \{ \log ( \sigmaGbl^2 ) \}
    - \frac{m}{2} E_{\qDens} \{ \log |\bSigma| \}
    - \frac{m \Kgrp}{2} E_{\qDens} \{ \log ( \sigmaGrp^2 ) \}
    \\[1ex]
    & & - \smhalf \mbox{tr} \left( \bSigma_{\bbeta}^{-1} \left\{
    \left( \bmu_{\qDens (\bbeta)} - \bmu_{\bbeta} \right) \left( \bmu_{\qDens (\bbeta)} - \bmu_{\bbeta} \right)^T
    + \bSigma_{\qDens (\bbeta)} \right\} \right)
    \\[1ex]
    & & - \smhalf \mu_{\qDens(1/{\sigmaGbl^2})} \left\{
    \Vert \bmu_{\qDens ( \buGbl )} \Vert^2 + \mbox{tr} ( \bSigma_{\qDens(\buGbl)} ) \right\}
    \\[1ex]
    & & -\smhalf \mbox{tr} \left( \bM_{\qDens (\bSigma^{-1})} \left\{ \displaystyle{\sum_{i=1}^{m}} \left(
    \bmu_{\qDens (\buLini} \bmu_{\qDens (\buLini)}^T + \bSigma_{\qDens (\buLini)} \right) \right\} \right)
    \\[1ex]
    & & - \smhalf \mu_{\qDens ( 1/{\sigmaGrp^2} )} \displaystyle{\sum_{i=1}^{m}} \left\{
    \Vert \bmu_{\qDens ( \buGrpi ))} \Vert^2 + \mbox{tr} (\bSigma_{\qDens(\buGrpi)}) \right\}
  \end{array}
\end{equation*}
\begin{equation*}
  \begin{array}{lcl}
    E_{\qDens} \{ \log \qDens^{*} (\bbeta, \bu) \} & = & - \smhalf \{ 2 + \Kgbl + m (2 + \Kgrp) \}
    - \smhalf \{ 2 + \Kgbl + m (2 + \Kgrp) \} \log (2\pi)
    \\[1ex]
    & & - \smhalf \log |\bSigma_{\bbeta}|
  \end{array}
\end{equation*}
\begin{equation*}
  \begin{array}{lcl}
    E_{\qDens} \{ \log \pDens (\sigsqeps \, | \, \aeps) \} & = &
    -\smhalf \nuEps E_{\qDens} \{ \log (2 \aeps) \}
    - \log \Gamma (\nuEps/2) - (\smhalf \nuEps + 1) E_{\qDens} \{ \log(\sigsqeps) \}
    \\[1ex]
    & & - \smhalf \mu_{\qDens(1/{\aeps})} \mu_{\qDens (1/{\sigsqeps})}
  \end{array}
\end{equation*}
\begin{equation*}
  \begin{array}{lcl}
    E_{\qDens} \{ \log \qDens^{*} (\sigsqeps) \}
    & = & \smhalf \xi_{\qDens (\sigsqeps)} \log ( \lambda_{\qDens(\sigsqeps)}/2) - \log \{
    \Gamma (\smhalf \xi_{\qDens (\sigsqeps)}) \} - (\smhalf \xi_{\qDens(\sigsqeps)} + 1) E_{\qDens} \{
    \log (\sigsqeps) \}
    \\[1ex]
    & & - \smhalf \lambda_{\qDens(\sigsqeps)} \mu_{\qDens(1/\sigsqeps)}
  \end{array}
\end{equation*}
\begin{equation*}
  \begin{array}{lcl}
    E_{\qDens} \{ \log \pDens (\aeps) \} & = &
    -\smhalf \log (2 \nuEps \sEps^2 ) - \log \{ \Gamma ( \smhalf ) \} - (\smhalf + 1) E_{\qDens} \{
    \log (\aeps) \}
    \\[1ex]
    & & - \{ 1/(2 \nuEps \sEps^2) \} \mu_{\qDens (1/\aeps)}
  \end{array}
\end{equation*}
\begin{equation*}
  \begin{array}{lcl}
    E_{\qDens} \{ \log \qDens^{*} (\aeps) \} & = &
    \smhalf \xi_{\qDens (\aeps)} \log (\lambda_{\qDens (\aeps)}/2) -
    \log \{ \Gamma (\smhalf \xi_{\qDens(\aeps)}) \} - (\smhalf \xi_{\qDens (\aeps)} + 1) E_{\qDens} \{
    \log (\aeps) \}
    \\[1ex]
    & & - \smhalf \lambda_{\qDens (\aeps)} \mu_{\qDens(1/ \aeps)}
  \end{array}
\end{equation*}
\begin{equation*}
  \begin{array}{lcl}
    E_{\qDens} \{ \log \pDens (\sigmaGbl^2 \, | \, \aGbl) \} & = &
    -\smhalf \nuGbl  E_{\qDens} \{ \log (2 \aGbl) \}
    - \log \Gamma (\nuGbl/2) - (\smhalf \nuGbl + 1) E_{\qDens} \{ \log(\sigmaGbl^2) \}
    \\[1ex]
    & & - \smhalf \mu_{\qDens(1/{\aGbl})} \mu_{\qDens (1/{\sigmaGbl^2})}
  \end{array}
\end{equation*}
\begin{equation*}
  \begin{array}{lcl}
    E_{\qDens} \{ \log \qDens^{*} (\sigmaGbl^2) \}
    & = & \smhalf \xi_{\qDens (\sigmaGbl^2)} \log ( \lambda_{\qDens(\sigmaGbl^2)}/2) - \log \{
    \Gamma (\smhalf \xi_{\qDens (\sigmaGbl^2)}) \} - (\smhalf \xi_{\qDens(\sigmaGbl^2)} + 1) E_{\qDens} \{
    \log (\sigmaGbl^2) \}
    \\[1ex]
    & & - \smhalf \lambda_{\qDens(\sigmaGbl^2)} \mu_{\qDens(1/\sigmaGbl^2)}
  \end{array}
\end{equation*}
\begin{equation*}
  \begin{array}{lcl}
    E_{\qDens} \{ \log \pDens (\aGbl) \} & = &
    -\smhalf \log (2 \nuGbl \sGbl^2 ) - \log \{ \Gamma ( \smhalf ) \} - (\smhalf + 1) E_{\qDens} \{
    \log (\aGbl) \}
    \\[1ex]
    & & - \{ 1/(2 \nuGbl \sGbl^2) \} \mu_{\qDens (1/\aGbl)}
  \end{array}
\end{equation*}
\begin{equation*}
  \begin{array}{lcl}
    E_{\qDens} \{ \log \qDens^{*} (\aGbl) \} & = &
    \smhalf \xi_{\qDens (\aGbl)} \log (\lambda_{\qDens (\aGbl)}/2) -
    \log \{ \Gamma (\smhalf \xi_{\qDens(\aGbl)}) \} - (\smhalf \xi_{\qDens (\aGbl)} + 1) E_{\qDens} \{
    \log (\aGbl) \}
    \\[1ex]
    & & - \smhalf \lambda_{\qDens (\aGbl)} \mu_{\qDens(1/ \aGbl)}
  \end{array}
\end{equation*}
\begin{equation*}
  \begin{array}{lcl}
    E_{\qDens} \{ \log \pDens (\sigmaGrp^2 \, | \, \aGrp) \} & = &
    -\smhalf \nuGrp  E_{\qDens} \{ \log (2 \aGrp) \}
    - \log \Gamma (\nuGrp/2) - (\smhalf \nuGrp + 1) E_{\qDens} \{ \log(\sigmaGrp^2) \}
    \\[1ex]
    & & - \smhalf \mu_{\qDens(1/{\aGrp})} \mu_{\qDens (1/{\sigmaGrp^2})}
  \end{array}
\end{equation*}
\begin{equation*}
  \begin{array}{lcl}
    E_{\qDens} \{ \log \qDens^{*} (\sigmaGrp^2) \}
    & = & \smhalf \xi_{\qDens (\sigmaGrp^2)} \log ( \lambda_{\qDens(\sigmaGrp^2)}/2) - \log \{
    \Gamma (\smhalf \xi_{\qDens (\sigmaGrp^2)}) \} - (\smhalf \xi_{\qDens(\sigmaGrp^2)} + 1) E_{\qDens} \{
    \log (\sigmaGrp^2) \}
    \\[1ex]
    & & - \smhalf \lambda_{\qDens(\sigmaGrp^2)} \mu_{\qDens(1/\sigmaGrp^2)}
  \end{array}
\end{equation*}
\begin{equation*}
  \begin{array}{lcl}
    E_{\qDens} \{ \log \pDens (\aGrp) \} & = &
    -\smhalf \log (2 \nuGrp \sGrp^2 ) - \log \{ \Gamma ( \smhalf ) \} - (\smhalf + 1) E_{\qDens} \{
    \log (\aGrp) \}
    \\[1ex]
    & & - \{ 1/(2 \nuGrp \sGrp^2) \} \mu_{\qDens (1/\aGrp)}
  \end{array}
\end{equation*}
\begin{equation*}
  \begin{array}{lcl}
    E_{\qDens} \{ \log \qDens^{*} (\aGrp) \} & = &
    \smhalf \xi_{\qDens (\aGrp)} \log (\lambda_{\qDens (\aGrp)}/2) -
    \log \{ \Gamma (\smhalf \xi_{\qDens(\aGrp)}) \} - (\smhalf \xi_{\qDens (\aGrp)} + 1) E_{\qDens} \{
    \log (\aGrp) \}
    \\[1ex]
    & & - \smhalf \lambda_{\qDens (\aGrp)} \mu_{\qDens(1/ \aGrp)}
  \end{array}
\end{equation*}
\begin{equation*}
  \begin{array}{lcl}
E_{\qDens}[\log\{\pDens (\bSigma | \bASigma)\}]
& = & - \smhalf (\nuSigma + 1) E_{\qDens} \{ \log | \bASigma | \}
    - \smhalf (\nuSigma + 4) E_{\qDens} \{ \log | \bSigma | \} - \smhalf \log(\pi)
    \\
    & & - \smhalf \mbox{tr} ( \bM_{\qDens(\bASigma^{-1})} \bM_{\qDens(\bSigma^{-1})} )
    - (\nuSigma+3) \log(2)
     - \sum_{j=1}^{2} \log \Gamma (\smhalf (\nuSigma + 4 - j))
   \end{array}
 \end{equation*}
\begin{equation*}
  \begin{array}{lcl}
E_{\qDens} [\log\{\qDens(\bSigma)\}]
&=& \smhalf (\xi_{\qDens(\bSigma)} - 1) \log |\bLambda_{\qDens(\bSigma)}|
    - \smhalf (\xi_{\qDens(\bSigma)} + 2) E_{\qDens} \{ \log |\bSigma| \}
    - \smhalf \mbox{tr} (\bLambda_{\qDens(\bSigma)} \bM_{\qDens(\bSigma^{-1})})
    \\
    & & - (\xi_{\qDens(\bSigma)} + 1) \log (2) - \smhalf \log(\pi)
    - \sum_{j=1}^{2} \log \Gamma (\smhalf (\xi_{\qDens(\bSigma)} + 2 - j))
  \end{array}
\end{equation*}
\begin{equation*}
  \begin{array}{lcl}
E_{\qDens}[\log\{\pDens (\bASigma)\}]
&=& - \frac{3}{2} E_{\qDens} \{ \log |\bASigma| \}
    - \smhalf \sum_{j=1}^{2} 1/(\nuSigma \sSigmaj^2) \left( \bM_{\qDens(\bASigma^{-1})} \right)_{jj}
    - 2 \log (2) - \smhalf \log(\pi)
\\
& & - \sum_{j=1}^{2} \log \Gamma (\smhalf (3-j))
\end{array}
\end{equation*}
\begin{equation*}
  \begin{array}{lcl}
E_{\qDens}[\log\{\qDens(\bASigma)\}]
&=& \smhalf (\xi_{\qDens(\bASigma)} - 1) \log |\bLambda_{\qDens(\bASigma)}|
    - \smhalf (\xi_{\qDens(\bASigma)} + 2) E_{\qDens} \{ \log |\bASigma| \}
    - \smhalf \mbox{tr} (\bLambda_{\qDens(\bASigma)} \bM_{\qDens(\bASigma^{-1})})
    \\
    & & - (\xi_{\qDens(\bASigma)} + 1) \log (2) - \smhalf \log(\pi)
    - \sum_{j=1}^{2} \log \Gamma (\smhalf (\xi_{\qDens(\bASigma)} + 2 - j))
  \end{array}
\end{equation*}
In the summation of each of these $\log\pDensUnder(\by;\qDens)$ terms, note that the coefficient of
$E_{\qDens} \{\log(\sigsqeps)\}$ is
$$-\smhalf \sum_{i=1}^{m} n_{i} -\smhalf\nuEps-1+\smhalf\xi_{\qDens(\sigsqeps)}+1=
  -\smhalf \sum_{i=1}^{m} n_{i} -\smhalf\nuEps-1+\smhalf(\nuEps+\sum_{i=1}^{m} n_{i})+1=0.$$
The coefficient of $E_{\qDens} \{ \log (\sigmaGbl^2) \}$ is
$$ -\smhalf \Kgbl - \smhalf \nuGbl - 1 + \smhalf \xi_{\qDens (\sigmaGbl^2)} + 1 =
  -\smhalf \Kgbl - \smhalf \nuGbl - 1 + \smhalf (\nuGbl + \Kgbl) + 1 = 0.$$
The coefficient of $E_{\qDens} \{ \log (\sigmaGrp^2) \}$ is
$$ -\smhalf m \Kgrp - \smhalf \nuGrp - 1 + \smhalf \xi_{\qDens (\sigmaGrp^2)} + 1 =
  -\smhalf m \Kgrp - \smhalf \nuGrp - 1 + \smhalf (\nuGrp + m \Kgrp) + 1 = 0.$$
The coefficient of $E_{\qDens}\{\log|\bSigma|\}$ is
$$-\frac{m}{2} - \smhalf(\nuSigma + 4) + \smhalf(\xi_{\qDens(\bSigma)} + 2) =
  -\smhalf(m + \nuSigma + 4) + \smhalf(m + \nuSigma + 4)=0.$$
The coefficient of $E_{\qDens} \{\log(\aeps)\}$ is
$$-\smhalf\nuEps-\smhalf-1+\smhalf\xi_{\qDens(\aeps)}+1=-\smhalf\nuEps-\smhalf-1
  +\smhalf(\nuEps+1)+1=0.$$
The coefficient of $E_{\qDens} \{ \log (\aGbl) \}$ is
$$-\smhalf\nuGbl-\smhalf-1+\smhalf\xi_{\qDens(\aGbl)}+1=-\smhalf\nuGbl-\smhalf-1
  +\smhalf(\nuGbl+1)+1=0.$$
The coefficient of $E_{\qDens} \{ \log (\aGrp) \}$ is
$$-\smhalf\nuGrp-\smhalf-1+\smhalf\xi_{\qDens(\aGrp)}+1=-\smhalf\nuGrp-\smhalf-1
  +\smhalf(\nuGrp+1)+1=0.$$
The coefficient of $E_{\qDens}\{\log|\bASigma|\}$ is
$$- \smhalf(\nuSigma + 1) - \frac{3}{2} + \smhalf(\xi_{\qDens(\bASigma)} + 2) =
  -\smhalf(\nuSigma + 2) + \smhalf(\nuSigma + 2)=0.$$
Therefore these terms can be dropped and then the cancellations led by the above expectations
leads to the lower bound expression in (\ref{eq:lowerBound}).
\section{Derivation of Result \ref{res:threeLevelBLUP}}\label{sec:drvResultThree}
If $\bB$ and $\bb$ have the same forms given by equation (7) in Nolan \myand Wand (2018) with
$$
\bvecij \equiv \left[
  \begin{array}{c}
    \sigeps^{-1}\by_{ij}\\[1ex]
    \bzero \\[1ex]
    \bzero \\[1ex]
    \bzero \\[1ex]
    \bzero \\[1ex]
    \bzero \\[1ex]
  \end{array} \right],
\qquad
\Bmatij\equiv \left[
\begin{array}{cc}
  \sigeps^{-1}\bX_{ij} & \sigeps^{-1}\bZgblij\\[1ex]
  \bO &\ndotmh \sigmaGbl^{-1}\bI_{\Kgbl} \\[1ex]
  \bO & \bO   \\[1ex]
  \bO & \bO   \\[1ex]
  \bO & \bO   \\[1ex]
  \bO & \bO   \\[1ex]
\end{array} \right],
$$
$$
\Bmatdotij\equiv \left[
   \begin{array}{cc}
     \sigeps^{-1}\bX_{ij} & \sigeps^{-1}\bZLoneGrpij \\[1ex]
     \bO & \bO \\[1ex]
     n_{i}^{-1/2}\bSigmag^{-1/2} & \bO \\[1ex]
     \bO & n_{i}^{-1/2} \sigmaGrpg^{-1} \bI_{\Kgrpg} \\[1ex]
     \bO & \bO \\[1ex]
     \bO & \bO \\[1ex]
   \end{array}\right]
\quad and\quad
\Bmatdotdotij\equiv \left[
\begin{array}{cc}
  \sigeps^{-1}\bX_{ij} & \sigeps^{-1}\bZLtwoGrpij \\[1ex]
  \bO & \bO \\[1ex]
  \bO & \bO \\[1ex]
  \bO & \bO \\[1ex]
  \bSigmah^{-1/2} & \bO \\[1ex]
  \bO & \sigmaGrph^{-1}\bI_{\Kgrph}
\end{array}\right],
$$
then straightforward algebra leads to
\begin{equation*}
  \begin{array}{c}
    \bB^T \bB = \bC^T\RBLUP^{-1}\bC+\DBLUP \mbox{  and  } \bB^T \bb = \bC^T\RBLUP^{-1}\by
  \end{array}
\end{equation*}
where
\begin{equation}
  \begin{array}{c}
    \bC\equiv[\bX\ \bZ],\quad\DBLUP\equiv\left[
    \begin{array}{cc}
    \bO & \bO               \\[1ex]
    \bO & \bG^{-1}
    \end{array}
    \right]\quad\mbox{and}\quad\RBLUP\equiv\sigeps^2\bI,
  \end{array}
\end{equation}
and $\bG$ as defined in (\ref{eqn:threeLevBLUPCov}). The remainder of the derivation
of Result 3 is analogous to that of Result 1.
\section{Derivation of Algorithm \ref{alg:threeLevBLUP}}\label{sec:drvAlgThree}
Algorithm \ref{alg:threeLevBLUP} is simply a proceduralization of Result 3.
\section{Derivation of Result \ref{res:threeLevelMFVB}}\label{sec:drvResultFour}
It is straightforward to verify that the $\bmu_{\qDens(\bbeta,\bu)}$
and $\bSigma_{\qDens(\bbeta,\bu)}$ updates, given at (\ref{eq:muSigmaMFVBupd}) but with $\DMFVB$ as given
in (\ref{eqn:DMFVB}), may be written as
$$\bmu_{\qDens(\bbeta,\bu)}\thickarrow(\bB^T\bB)^{-1}\bB^T\bb
\quad\mbox{and}\quad
\bSigma_{\qDens(\bbeta,\bu)}\thickarrow(\bB^T\bB)^{-1}
$$
where $\bB$ and $\bb$ have the forms given by equation (7) in Nolan \myand Wand (2018)
with
$$
\bb_{ij}\equiv
\left[
\begin{array}{c}
\mu_{\qDens(1/\sigeps^2)}^{1/2}\by_{ij}\\[1ex]
\ndotmh \bSigma_{\bbeta}^{-1/2} \bmu_{\bbeta}\\[1ex]
\bzero \\[1ex]
\bzero \\[1ex]
\bzero \\[1ex]
\bzero \\[1ex]
\bzero \\[1ex]
\end{array}
\right],
\ \
\bB_{ij}\equiv
\left[
\begin{array}{cc}
\mu_{\qDens(1/\sigeps^2)}^{1/2}\bX_{ij} & \mu_{\qDens(1/\sigeps^2)}^{1/2}\bZgblij\\[1ex]
\ndotmh \bSigma_{\bbeta}^{-1/2} & \bO \\[1ex]
\bO &\ndotmh \mu_{\qDens(1/\sigmaGbl^2)}^{1/2}\bI_{\Kgbl}\\[1ex]
\bO & \bO   \\[1ex]
\bO & \bO   \\[1ex]
\bO & \bO   \\[1ex]
\bO & \bO   \\[1ex]
\end{array}
\right],
$$
$$
\bBdot_{ij}\equiv
\left[
\begin{array}{cc}
\mu_{\qDens(1/\sigeps^2)}^{1/2}\bX_{ij}  & \mu_{\qDens(1/\sigeps^2)}^{1/2}\bZLoneGrpij  \\[1ex]
\bO                & \bO            \\[1ex]
\bO                & \bO            \\[1ex]
n_i^{-1/2}\bM_{\qDens(\bSigmag^{-1})}^{1/2}     & \bO            \\[1ex]
\bO                            &n_i^{-1/2}\mu_{\qDens(1/\sigmaGrpg^2)}^{1/2}\bI_{\Kgrpg}\\[1ex]
\bO                & \bO            \\[1ex]
\bO                & \bO            \\
\end{array}
\right]
\ \ and\ \
\bBdotdot_{ij}\equiv
\left[
\begin{array}{cc}
\mu_{\qDens(1/\sigeps^2)}^{1/2}\bX_{ij}  & \mu_{\qDens(1/\sigeps^2)}^{1/2}\bZLtwoGrpij  \\[1ex]
\bO                & \bO            \\[1ex]
\bO                & \bO            \\[1ex]
\bO                & \bO            \\[1ex]
\bO                & \bO            \\[1ex]
\bM_{\qDens(\bSigmah^{-1})}^{1/2}   & \bO            \\[1ex]
\bO                &\mu_{\qDens(1/\sigmaGrph^2)}^{1/2}\bI_{\Kgrph}\\
\end{array}
\right].
$$
Result \ref{res:threeLevelMFVB} immediately follows from Theorem 4 of Nolan \myand Wand (2018).
\section{Derivation of Algorithm \ref{alg:threeLevMFVB}}\label{sec:drvAlgFour}
We provide expressions for the $\qDens$-densities for mean field variational Bayesian
inference for the parameters in (\ref{eq:threeLevBayes}) with product density restriction
(\ref{eq:producRestrict3lev}).
$$\qDens(\bbeta,\bu)\ \mbox{is a $N(\bmu_{\qDens(\bbeta,\bu)},\bSigma_{\qDens(\bbeta,\bu)})$
density function}$$
where
$$\bSigma_{\qDens(\bbeta,\bu)}=(\bC^T\RMFVB^{-1}\bC+\DMFVB)^{-1}
\quad
\mbox{and}
\quad
\bmu_{\qDens(\bbeta,\bu)}=\bSigma_{\qDens(\bbeta,\bu)}(\bC^T\RMFVB^{-1}\by + \oMFVB)
$$
with $\RMFVB\equiv\mu_{\qDens(1/\sigeps^2)}^{-1}\bI$,
$\oMFVB\equiv\left[
  \begin{array}{c}
  \bSigma_{\bbeta}^{-1}\bmu_{\bbeta}\\[1ex]
  \bzero
  \end{array}
  \right]$ and $\DMFVB$ as given in (\ref{eqn:DMFVB}).
$$\qDens(\sigsqeps)\ \mbox{is an $\mbox{Inverse-$\chi^2$}
\left(\xi_{\qDens(\sigsqeps)},\lambda_{\qDens(\sigsqeps)}\right)$ density function}
$$
where $\xi_{\qDens(\sigsqeps)}=\nuEps+\sumim \sum_{j=1}^{n_{i}} o_{ij}$ and
\begin{eqnarray*}
\lambda_{\qDens(\sigsqeps)}&=&\mu_{\qDens(1/\aeps)}+\sumim \sum_{j=1}^{n_{i}} \,
E_{\qDens}\left\{ \Bigg \Vert \by_{ij} - \bCgblij \left[ \begin{array}{c}
      \bbeta \\[1ex] \buGbl \end{array} \right] - \bCLoneGrpij \left[ \begin{array}{c}
            \buLoneLini \\[1ex] \buLoneGrpi  \end{array} \right] -
            \bCLtwoGrpij \left[ \begin{array}{c} \buLtwoLinij \\[1ex] \buLtwoGrpij
            \end{array} \right] \Bigg \Vert^2 \right\}
\end{eqnarray*}
\begin{eqnarray*}
& = & \mu_{\qDens(1/\aeps)}+\sumim \sum_{j=1}^{n_{i}} \, \left[ \Bigg \Vert\,E_{\qDens} \left( \by_{ij}
    - \bCgblij \left[ \begin{array}{c} \bbeta \\[1ex] \buGbl \end{array} \right]
    - \bCLoneGrpij \left[ \begin{array}{c} \buLoneLini \\[1ex] \buLoneGrpi \end{array} \right]
    - \bCLtwoGrpij \left[ \begin{array}{c} \buLtwoLinij \\[1ex] \buLtwoGrpij \end{array} \right]
    \right) \Bigg \Vert^2 \right.
\\[1ex]
& & \left. \qquad \qquad \qquad \qquad
    + \mbox{tr} \left\{ \mbox{\rm Cov}_{\qDens} \left( \bCgblij \left[ \begin{array}{c}
    \bbeta \\[1ex] \buGbl \end{array} \right] + \bCLoneGrpij \left[ \begin{array}{c}
    \buLoneLini \\[1ex] \buLoneGrpi \end{array} \right] + \bCLtwoGrpij \left[ \begin{array}{c}
    \buLtwoLinij \\[1ex] \buLtwoGrpij \end{array} \right] \right)  \right\} \right] \\[1ex]
\end{eqnarray*}
\begin{eqnarray*}
& = & \mu_{\qDens(1/\aeps)} + \displaystyle{\sum_{i=1}^{m} \sum_{j=1}^{n_{i}}} \left\{
    \Bigg \Vert\,E_{\qDens} \left( \by_{ij}
    - \bCgblij \left[ \begin{array}{c} \bbeta \\[1ex] \buGbl \end{array} \right]
    - \bCLoneGrpij \left[ \begin{array}{c} \buLoneLini \\[1ex] \buLoneGrpi \end{array} \right]
    - \bCLtwoGrpij \left[ \begin{array}{c} \buLtwoLinij \\[1ex] \buLtwoGrpij \end{array} \right]
    \right) \Bigg \Vert^2  \right.
\\[3ex]
& & \quad \left. + \mbox{tr}(\bCgblij^T\bCgblij \bSigma_{\qDens(\bbeta,\buGbl)})
    + \mbox{tr}((\bCLoneGrpij)^T\bCLoneGrpij \bSigma_{\qDens(\buLoneLini,\buLoneGrpi)})
    + \mbox{tr}((\bCLtwoGrpij)^T\bCLtwoGrpij \bSigma_{\qDens(\buLtwoLinij,\buLtwoGrpij)}) \right.
\\[1ex]
& & \quad \left. + 2 \, \mbox{tr}\left[(\bCLoneGrpij)^T\bCgblij\,E_{\qDens}\left\{\left(
        \left[\begin{array}{c}\bbeta \\ \buGbl\end{array}\right]
        -\bmu_{\qDens(\bbeta,\buGbl)}\right) \left(\left[\begin{array}{c}\buLoneLini\\
    \buLoneGrpi\end{array}\right] -\bmuq{\buLoneLini,\buLoneGrpi}\right)^T\right\}\right] \right.
\\[1ex]
& & \quad \left. + 2 \, \mbox{tr}\left[(\bCLtwoGrpij)^T\bCgblij\,E_{\qDens}\left\{\left(
        \left[\begin{array}{c}\bbeta \\ \buGbl\end{array}\right]
        -\bmu_{\qDens(\bbeta,\buGbl)}\right) \left(\left[\begin{array}{c}\buLtwoLinij\\
    \buLtwoGrpij\end{array}\right] -\bmuq{\buLtwoLinij,\buLtwoGrpij}\right)^T\right\}\right] \right.
\\[1ex]
& & \quad \left. + 2 \, \mbox{tr}\left[(\bCLoneGrpij)^T\bCLtwoGrpij\,E_{\qDens}\left\{\left(
    \left[\begin{array}{c}\buLoneLini\\
    \buLoneGrpi\end{array}\right] -\bmuq{\buLoneLini,\buLoneGrpi}\right) \left(\left[\begin{array}{c}\buLtwoLinij\\
    \buLtwoGrpij\end{array}\right] -\bmuq{\buLtwoLinij,\buLtwoGrpij}\right)^T\right\}\right] \right\}
\end{eqnarray*}
where $\bCgblij \equiv [ \bX_{ij} \, \bZgblij ]$, $\bCLoneGrpij \equiv [ \bX_{ij} \, \bZLoneGrpij ]$,
$\bCLtwoGrpij \equiv [ \bX_{ij} \, \bZLtwoGrpij ]$ and
with reciprocal moment $\mu_{\qDens(1/\sigsqeps)}=\xi_{\qDens(\sigsqeps)}/\lambda_{\qDens(\sigsqeps)},$
$$\qDens(\sigmaGbl^2)\ \mbox{is an $\mbox{Inverse-$\chi^2$}
\left(\xi_{\qDens(\sigmaGbl^2)},\lambda_{\qDens(\sigmaGbl^2)}\right)$ density function}
$$
where $\xi_{\qDens(\sigmaGbl^2)}=\nuGbl+\Kgbl$ and
$$
\lambda_{\qDens(\sigmaGbl^2)}= \mu_{\qDens(1/{\aGbl})} + \Vert \bmu_{\qDens(\buGbl)} \Vert^2
 + \mbox{tr} \left( \bSigma_{\qDens(\buGbl)} \right),
$$
with reciprocal moment $\mu_{\qDens(1/{\sigmaGbl^2})}=\xi_{\qDens(\sigmaGbl^2)} /
\lambda_{\qDens(\sigmaGbl^2)},$
$$\qDens(\sigmaGrpg^2)\ \mbox{is an $\mbox{Inverse-$\chi^2$}
\left(\xi_{\qDens(\sigmaGrpg^2)},\lambda_{\qDens(\sigmaGrpg^2)}\right)$ density function}
$$
where $\xi_{\qDens(\sigmaGrpg^2)}=\nuGrpg+m\Kgrpg$ and
$$
\lambda_{\qDens(\sigmaGrpg^2)} = \mu_{\qDens(1/{\aGrpg})} + \sum_{i=1}^m \left\{ \Vert \bmu_{\qDens(\buLoneGrpi)} \Vert^2
 + \mbox{tr} \left( \bSigma_{\qDens(\buLoneGrpi)} \right) \right\},
$$
with reciprocal moment $\mu_{\qDens(1/{\sigmaGrpg^2})}=\xi_{\qDens(\sigmaGrpg^2)} /
\lambda_{\qDens(\sigmaGrpg^2)},$
$$\qDens(\bSigmag)\ \mbox{is an $\mbox{Inverse-G-Wishart}
\left(\Gfull,\xi_{\qDens(\bSigmag)},\bLambda_{\qDens(\bSigmag)}\right)$ density function}
$$
where $\xi_{\qDens(\bSigmag)}=\nuSigmag+2+m$ and
$$\bLambda_{\qDens(\bSigmag)}=\bM_{\qDens(\bA_{\bSigmag}^{-1})}
+\sumim\left(\bmu_{\qDens(\buLoneLini)}\bmu_{\qDens(\buLoneLini)}\trans + \bSigma_{\qDens(\buLoneLini)}\right),$$
with inverse moment $\bM_{\qDens(\bSigmag^{-1})}=(\xi_{\qDens(\bSigmag)}-1)\bLambda_{\qDens(\bSigmag)}^{-1}$,
$$\qDens(\sigmaGrph^2)\ \mbox{is an $\mbox{Inverse-$\chi^2$}
\left(\xi_{\qDens(\sigmaGrph^2)},\lambda_{\qDens(\sigmaGrph^2)}\right)$ density function}
$$
where $\xi_{\qDens(\sigmaGrph^2)}=\nuGrph+\Kgrph \sum_{i=1}^{m} n_{i}$ and
$$
\lambda_{\qDens(\sigmaGrph^2)} = \mu_{\qDens(1/{\aGrph})} + \sum_{i=1}^m \sum_{j=1}^{n_{i}} \left\{
  \Vert \bmu_{\qDens(\buLtwoGrpij)} \Vert^2
  + \mbox{tr} \left( \bSigma_{\qDens(\buLtwoGrpij)} \right) \right\},
$$
with reciprocal moment $\mu_{\qDens(1/{\sigmaGrph^2})}=\xi_{\qDens(\sigmaGrph^2)} /
\lambda_{\qDens(\sigmaGrph^2)},$
$$\qDens(\bSigmah)\ \mbox{is an $\mbox{Inverse-G-Wishart}
\left(\Gfull,\xi_{\qDens(\bSigmah)},\bLambda_{\qDens(\bSigmah)}\right)$ density function}
$$
where $\xi_{\qDens(\bSigmah)}=\nuSigmah+2+\sum_{i=1}^m n_{i}$ and
$$\bLambda_{\qDens(\bSigmah)}=\bM_{\qDens(\bA_{\bSigmah}^{-1})}
+\sumim \sum_{j=1}^{n_{i}} \left(\bmu_{\qDens(\buLtwoLinij)}\bmu_{\qDens(\buLtwoLinij)}\trans
+ \bSigma_{\qDens(\buLtwoLinij)}\right),$$
with inverse moment $\bM_{\qDens(\bSigmah^{-1})}=(\xi_{\qDens(\bSigmah)}-1)\bLambda_{\qDens(\bSigmah)}^{-1}$,
$$\qDens(\aeps)\ \mbox{is an $\mbox{Inverse-$\chi^2$}
(\xi_{\qDens(\aeps)},\lambda_{\qDens(\aeps)})$ density function}$$
where $\xi_{\qDens(\aeps)}=\nuEps+1$,
$$\lambda_{\qDens(\aeps)}=\mu_{\qDens(1/\sigsqeps)}+1/(\nuEps\sEps^2)$$
with reciprocal moment $\mu_{\qDens(1/\aeps)}=\xi_{\qDens(\aeps)}/\lambda_{\qDens(\aeps)}$,
$$\qDens(\aGbl)\ \mbox{is an $\mbox{Inverse-$\chi^2$}
(\xi_{\qDens(\aGbl)},\lambda_{\qDens(\aGbl)})$ density function}$$
where $\xi_{\qDens(\aGbl)}=\nuGbl+1$,
$$\lambda_{\qDens(\aGbl)}=\mu_{\qDens(1/\sigmaGbl^2)}+1/(\nuGbl\sGbl^2)$$
with reciprocal moment $\mu_{\qDens(1/\aGbl)}=\xi_{\qDens(\aGbl)}/\lambda_{\qDens(\aGbl)}$,
$$\qDens(\aGrpg)\ \mbox{is an $\mbox{Inverse-$\chi^2$}
(\xi_{\qDens(\aGrpg)},\lambda_{\qDens(\aGrpg)})$ density function}$$
where $\xi_{\qDens(\aGrpg)}=\nuGrpg+1$,
$$\lambda_{\qDens(\aGrpg)}=\mu_{\qDens(1/\sigmaGrpg^2)}+1/(\nuGrpg\sGrpg^2)$$
with reciprocal moment $\mu_{\qDens(1/\aGrpg)}=\xi_{\qDens(\aGrpg)}/\lambda_{\qDens(\aGrpg)}$ and
$$\qDens(\ASigmag)\ \mbox{is an $\mbox{Inverse-G-Wishart}
\left(\Gdiag,\xi_{\qDens(\ASigmag)},\bLambda_{\qDens(\ASigmag)}\right)$ density function}
$$
where $\xi_{\qDens(\ASigmag)}=\nuSigmag+2$,
$$\bLambda_{\qDens(\ASigmag)}=\diag\big\{\mbox{diagonal}\big(\bM_{\qDens(\bSigmag^{-1})}\big)\big\}
+\bLambda_{\ASigmag}$$
with inverse moment $\bM_{\qDens(\ASigmag^{-1})}=\xi_{\qDens(\ASigmag)}\bLambda_{\qDens(\ASigmag)}^{-1}$,
$$\qDens(\aGrph)\ \mbox{is an $\mbox{Inverse-$\chi^2$}
(\xi_{\qDens(\aGrph)},\lambda_{\qDens(\aGrph)})$ density function}$$
where $\xi_{\qDens(\aGrph)}=\nuGrph+1$,
$$\lambda_{\qDens(\aGrph)}=\mu_{\qDens(1/\sigmaGrph^2)}+1/(\nuGrph\sGrph^2)$$
with reciprocal moment $\mu_{\qDens(1/\aGrph)}=\xi_{\qDens(\aGrph)}/\lambda_{\qDens(\aGrph)}$ and
$$\qDens(\ASigmah)\ \mbox{is an $\mbox{Inverse-G-Wishart}
\left(\Gdiag,\xi_{\qDens(\ASigmah)},\bLambda_{\qDens(\ASigmah)}\right)$ density function}
$$
where $\xi_{\qDens(\ASigmah)}=\nuSigmah+2$
$$\bLambda_{\qDens(\ASigmah)}=\diag\big\{\mbox{diagonal}\big(\bM_{\qDens(\bSigmah^{-1})}\big)\big\}
+\bLambda_{\ASigmah}$$
with inverse moment $\bM_{\qDens(\ASigmah^{-1})}=\xi_{\qDens(\ASigmah)}\bLambda_{\qDens(\ASigmah)}^{-1}$.

\section{The \textsc{\normalsize SolveTwoLevelSparseLeastSquares}\ Algorithm}\label{sec:Solve2Lev}

The \SolveTwoLevelSparseLeastSquares\ is listed in Nolan {\it et al.} (2018) and
based on Theorem 2 of Nolan \myand Wand (2018). Given its centrality to 
Algorithms \ref{alg:twoLevBLUP} and \ref{alg:twoLevMFVB} we list it again here.
The algorithm solves a sparse version of the the least squares problem:
$$\min_{\bx}\Vert\bb-\bB\bx\Vert^2$$
which has solution $\bx=\bA^{-1}\bB^T\bb$ where $\bA=\bB^T\bB$ where $\bB$ 
and $\bb$ have the following structure:
\begin{equation}
\bB\equiv
\left[
\arraycolsep=2.2pt\def\arraystretch{1.6}
\begin{array}{c|c|c|c|c}
\setstretch{4.5}
\Bmato            &\Bmatdoto           &\bO   &\cdots&\bO\\ 
\hline
\Bmatt            &\bO              &\Bmatdott&\cdots&\bO\\ 
\hline
\vdots            &\vdots           &\vdots           &\ddots&\vdots\\
\hline
\Bmatm &\bO       &\bO              &\cdots           &\Bmatdotm
\end{array}
\right]
\quad\mbox{and}\quad
\bb=\left[
\arraycolsep=2.2pt\def\arraystretch{1.6}
\begin{array}{c}
\setstretch{4.5}
\bveco  \\ 
\hline
\bvect \\ 
\hline
\vdots \\
\hline
\bvecm \\
\end{array}
\right].
\label{eq:BandbFormsReprise}
\end{equation}
The sub-matrices corresponding to the non-zero blocks of $\AtLev$
are labelled according to:
\begin{equation}
\AtLev^{-1}=
\left[
\arraycolsep=2.2pt\def\arraystretch{1.6}
\begin{array}{c|c|c|c|c}
\setstretch{4.5}
\AUoo     & \AUotCo & \AUotCt  &\ \ \cdots\ \ &\AUotCm \\
\hline
\AUotCoT & \AUttCo & \bigX      & \cdots   & \bigX    \\
\hline
\AUotCtT & \bigX     & \AUttCt  & \cdots   & \bigX    \\ 
\hline
\vdots    & \vdots  & \vdots   & \ddots   & \vdots   \\
\hline
\AUotCmT & \bigX     & \bigX      & \cdots   &\AUttCm \\ 
\end{array}
\right].
\label{eq:AtLevInv}
\end{equation}
with $\bigX$ denoting sub-blocks that are not of interest.
The \SolveTwoLevelSparseLeastSquares\ algorithm is given
in Algorithm \ref{alg:SolveTwoLevelSparseLeastSquares}.

\begin{algorithm}[!th]
\begin{center}
\begin{minipage}[t]{154mm}
\begin{small}
\begin{itemize}
\setlength\itemsep{4pt}
\item[] Inputs: $\big\{\big(\bveci(\nadj_i\times1),
\ \Bmati(\nadj_i\times p),\ \Bmatdoti(\nadj_i\times q)\big):\ 1\le i\le m\big\}$
\item[] $\bomega_3\thickarrow\mbox{NULL}$\ \ \ ;\ \ \ $\bOmega_4\thickarrow\mbox{NULL}$
\item[] For $i=1,\ldots,m$:
\begin{itemize}
\setlength\itemsep{4pt}
\item[] Decompose $\Bmatdoti=\bQ_i\left[\begin{array}{c}   
\bR_i\\
\bzero
\end{array}
\right]$ such that $\bQ_i^{-1}=\bQ_i^T$ and $\bR_i$ is upper-triangular.
\item[] $\cveczi\thickarrow\bQ_i^T\bveci\ \ \ ;\ \ \ \Cmatzi\thickarrow\bQ_i^T\Bmati$
\item[] $\cvecoi\thickarrow\mbox{first $q$ rows of}\ \cveczi$\ \ \ ;\ \ \ 
$\cvecti\thickarrow\mbox{remaining rows of}\ \cveczi$\ \ \ ;\ \ \ 
$\bomega_3\thickarrow
\left[
\begin{array}{c}
\bomega_3\\
\cvecti
\end{array}
\right]$
\item[]$\Cmatoi\thickarrow\mbox{first $q$ rows of}\ \Cmatzi$\ \ \ ;\ \ \ 
$\Cmatti\thickarrow\mbox{remaining rows of}\ \Cmatzi$\ \ \ ;\ \ \  
$\bOmega_4\thickarrow
\left[
\begin{array}{c}
\bOmega_4\\
\Cmatti
\end{array}
\right]$
\end{itemize}
\item[] Decompose $\OmegaAtwoTwo=\bQ\left[\begin{array}{c}   
\bR\\
\bzero
\end{array}
\right]$ such that $\bQ^{-1}=\bQ^T$ and $\bR$ is upper-triangular.
\item[] $\bc\thickarrow\mbox{first $p$ rows of $\bQ^T\bomega_3$}$
\ \ \ ;\ \ \ $\xveco\thickarrow\bR^{-1}\bc$\ \ \ ;\ \ \ 
$\AUoo\thickarrow\bR^{-1}\bR^{-T}$
\item[] For $i=1,\ldots,m$:
\begin{itemize}
\setlength\itemsep{4pt}
\item[] $\xvectCi\thickarrow\bR_i^{-1}(\bc_{1i}-\Cmatoi\xveco)$\ \ \ ;\ \ \ 
$\AUotCi\thickarrow\,-\AUoo(\bR_i^{-1}\Cmatoi)^T$
\item[] $\AUttCi\thickarrow\bR_i^{-1}(\bR_i^{-T} - \Cmatoi\AUotCi)$
\end{itemize}
\item[] Output: $\Big(\xveco,\AUoo,\big\{\big(\xvectCi,\AUttCi,\AUotCi):\ 1\le i\le m\big\}\Big)$
\end{itemize}
\end{small}
\end{minipage}
\end{center}
\caption{\SolveTwoLevelSparseLeastSquares\ \textit{for solving the two-level sparse matrix
least squares problem: minimise $\Vert\bb-\bB\,\bx\Vert^2$ in $\bx$ and sub-blocks of $\bA^{-1}$
corresponding to the non-zero sub-blocks of $\bA=\bB^T\bB$. The sub-block notation is
given by (\ref{eq:BandbFormsReprise}) and (\ref{eq:AtLevInv}).}}
\label{alg:SolveTwoLevelSparseLeastSquares} 
\end{algorithm}

\null\vfill\eject
\null\vfill\eject

\section{The \textsc{\normalsize SolveThreeLevelSparseLeastSquares}\ Algorithm}\label{sec:Solve3Lev}

The \SolveThreeLevelSparseLeastSquares,\ listed in Nolan {\it et al.} (2018) 
is a proceduralization of Theorem 4 of Nolan \myand Wand (2018). Since it is central to
Algorithms \ref{alg:threeLevBLUP} and \ref{alg:threeLevMFVB} we list it here.
The \SolveThreeLevelSparseLeastSquares\ algorithm is concerned with solving
the sparse three-level version of
$$\min_{\bx}\Vert\bb-\bB\bx\Vert^2$$
with the solution $\bx=\bA^{-1}\bB^T\bb$ where $\bA=\bB^T\bB$ where $\bB$ 
and $\bb$ have the following structure:
\begin{equation}
\bB\equiv\Big[\stack{1\le i\le m}\Big\{\stack{1\le j\le n_i}(\B{ij})\Big\}\ \Big\vert
\blockdiag{1\le i\le m}\Big\{\big[\stack{1\le j\le n_i}(\dB{ij})\ \big\vert
\  \blockdiag{1\le j\le n_i}(\ddB{ij})\big]\Big\} \Big]
\label{eq:catweasleOne}
\end{equation}
and
\begin{equation}
\bb\equiv\stack{1\le i\le m}\Big\{\stack{1\le j\le n_i}(\bb_{ij})\Big\}.
\label{eq:catweasleTwo}
\end{equation}
The three-level sparse matrix inverse problem involves determination of the sub-blocks
of $\bA^{-1}$ corresponding to the non-zero sub-blocks of $\bA$. Our notation for
these sub-blocks is illustrated by 
\begin{equation}
\bA^{-1} =
\left[\arraycolsep=2.2pt\def\arraystretch{1.6}
   \begin{array}{c | c | c | c | c | c | c | c}
 \setstretch{4.5}
   \Ainv{11} & \Ainv{12,1} & \Ainv{12,11} & \Ainv{12,12} & \Ainv{12,2} & \Ainv{12,21} & \Ainv{12,22} & \Ainv{12,23} \\
   \hline
   \ATinv{12,1} & \Ainv{22,1} & \Ainv{12,1,1} & \Ainv{12,1,2} & \bigX & \bigX & \bigX & \bigX \\
   \hline
   \ATinv{12,11} & \ATinv{12,1,1} & \Ainv{22,11} & \bigX & \bigX & \bigX & \bigX & \bigX \\
   \hline
   \ATinv{12,12} & \ATinv{12,1,2} & \bigX & \Ainv{22,12} & \bigX & \bigX & \bigX & \bigX \\
   \hline
   \ATinv{12,2} & \bigX & \bigX & \bigX & \Ainv{22,2} & \Ainv{12,2,1} & \Ainv{12,2,2} & \Ainv{12,2,3} \\
   \hline
   \ATinv{12,21} & \bigX & \bigX & \bigX & \ATinv{12,2,1} & \Ainv{22,21} & \bigX & \bigX \\
   \hline
   \ATinv{12,22} & \bigX & \bigX & \bigX & \ATinv{12,2,2} & \bigX & \Ainv{22,22} & \bigX \\
   \hline
   \ATinv{12,23} & \bigX & \bigX & \bigX & \ATinv{12,2,3} & \bigX & \bigX & \Ainv{22,23}
\end{array} \right]
\label{eq:catweasleThree}
\end{equation}
for the $m=2$, $n_1=2$ and $n_2=3$ case. The $\bigX$ symbol denotes sub-blocks that are not of interest.
The \SolveThreeLevelSparseLeastSquares\ algorithm is given
in Algorithm \ref{alg:SolveThreeLevelSparseLeastSquares}.

\begin{algorithm}[!th]
\begin{center}
\begin{minipage}[t]{155mm}
\begin{small}
\begin{itemize}
\setlength\itemsep{4pt}
\item[] Inputs: $\big\{\big(\bb_{ij}(\oadj_{ij}\times1),
\ \bB_{ij}(\oadj_{ij}\times p),\ \bBdot_{ij}(\oadj_{ij}\times q_1),
\ \bBdotdot_{ij}(\oadj_{ij}\times q_2)\big):\ 1\le i\le m,\ 1\le j\le n_i\big\}$
\item[] $\bomega_7\thickarrow\mbox{NULL}$\ \ \ ;\ \ \ $\bOmega_8\thickarrow\mbox{NULL}$
\item[] For $i=1,\ldots,m$:
\begin{itemize}
\setlength\itemsep{4pt}
\item[]  $\bomega_9\thickarrow\mbox{NULL}$\ \ \ ;\ \ \ $\bOmega_{10}\thickarrow\mbox{NULL}$
\ \ \ ;\ \ \ $\bOmega_{11}\thickarrow\mbox{NULL}$
\item[] For $j=1,\ldots,n_i$:
\begin{itemize}
\setlength\itemsep{4pt}
\item[] Decompose $\bBdotdot_{ij}=\bQ_{ij}\left[\begin{array}{c}   
\bR_{ij}\\
\bzero
\end{array}
\right]$ such that $\bQ_{ij}^{-1}=\bQ_{ij}^T$ and $\bR_{ij}$ is upper-triangular.
\item[] $\bd_{0ij}\thickarrow\bQ_{ij}^T\bb_{ij}\ \ \ ;\ \ \ \bD_{0ij}\thickarrow\bQ_{ij}^T\bB_{ij}
\ \ \ ;\ \ \ \bDdot_{0ij}\thickarrow\bQ_{ij}^T\bBdot_{ij}
$
\item[] $\bd_{1ij}\thickarrow\mbox{1st $q_2$ rows of}\ \bd_{0ij}$\ \ ;\ \ 
$\bd_{2ij}\thickarrow\mbox{remaining rows of}\ \bd_{0ij}$\ \ ;\ \ 
$\bomega_9\thickarrow
\left[
\begin{array}{c}
\bomega_9\\
\bd_{2ij}
\end{array}
\right]$
\item[]$\bD_{1ij}\thickarrow\mbox{1st $q_2$ rows of}\ \bD_{0ij}$\ ;\  
$\bD_{2ij}\thickarrow\mbox{remaining rows of}\ \bD_{0ij}$\ ;\   
$\bOmega_{10}\thickarrow
\left[
\begin{array}{c}
\bOmega_{10}\\
\bD_{2ij}
\end{array}
\right]$
\item[]$\bDdot_{1ij}\thickarrow\mbox{1st $q_2$ rows of}\ \bDdot_{0ij}$\ ;\ 
$\bDdot_{2ij}\thickarrow\mbox{remaining rows of}\ \bDdot_{0ij}$\ ;\  
$\bOmega_{11}\thickarrow
\left[
\begin{array}{c}
\bOmega_{11}\\
\bDdot_{2ij}
\end{array}
\right]$
\end{itemize}
\item[] Decompose $\bOmega_{11}=\bQ_i\left[\begin{array}{c}   
\bR_i\\
\bzero
\end{array}
\right]$ such that $\bQ_i^{-1}=\bQ_i^T$ and $\bR_i$ is upper-triangular.
\item[] $\bc_{0i}\thickarrow\bQ_i^T\bomega_9\ \ \ ;\ \ \ \bC_{0i}\thickarrow\bQ_i^T\bOmega_{10}$
\item[] $\bc_{1i}\thickarrow\mbox{1st $q_1$ rows of}\ \bc_{0i}$\ \ ;\ \ 
$\bc_{2i}\thickarrow\mbox{remaining rows of}\ \bc_{0i}$\ \ ;\ \ 
$\bomega_7\thickarrow
\left[
\begin{array}{c}
\bomega_7\\
\bc_{2i}
\end{array}
\right]$
\item[] $\bC_{1i}\thickarrow\mbox{1st $q_1$ rows of}\ \bC_{0i}$\ \ ;\ \ 
$\bC_{2i}\thickarrow\mbox{remaining rows of}\ \bC_{0i}$\ \ ;\ \ 
$\bOmega_8\thickarrow
\left[
\begin{array}{c}
\bOmega_8\\
\bC_{2i}
\end{array}
\right]$
\end{itemize}
\item[] Decompose $\bOmega_8=\bQ\left[\begin{array}{c}   
\bR\\
\bzero
\end{array}
\right]$ so that $\bQ^{-1}=\bQ^T$ and $\bR$ is upper-triangular.
\item[] $\bc\thickarrow\mbox{first $p$ rows of $\bQ^T\bomega_7$}$
\ \ \ ;\ \ \ $\xveco\thickarrow\bR^{-1}\bc$\ \ \ ;\ \ \ 
$\AUoo\thickarrow\bR^{-1}\bR^{-T}$
\item[] For $i=1,\ldots,m$:
\begin{itemize}
\setlength\itemsep{4pt}
\item[] $\xvectCi\thickarrow\bR_i^{-1}(\bc_{1i}-\bC_{1i}\xveco)$\ \ \ ;\ \ \ 
$\AUotCi\thickarrow\,-\AUoo(\bR_i^{-1}\Cmatoi)^T$
\item[] $\AUttCi\thickarrow\bR_i^{-1}(\bR_i^{-T} - \Cmatoi\AUotCi)$
\item[] For $j=1,\ldots,n_i$:
\begin{itemize}
\item[] $\bx_{2,ij}\leftarrow\bR_{ij}^{-1} (\bd_{1ij} - \bD_{1ij} \bx_1 - \bDdot_{1ij} \bx_{2,i})$
\item[] $\Ainv{12,ij}\leftarrow - \left\{ \bR_{ij}^{-1}(\bD_{1ij} \Ainv{11} + \bDdot_{1ij} \ATinv{12,i}) \right\}^T$
\item[] $\Ainv{12,\iCOMMAj}\leftarrow - \left\{\bR_{ij}^{-1}(\bD_{1ij} \Ainv{12,i} + \bDdot_{1ij} \Ainv{22,i}) \right\}^T$
\item[] $\Ainv{22,ij}\leftarrow\bR_{ij}^{-1}\big(\bR_{ij}^{-T}-\bD_{1ij}\Ainv{12,ij}-\bDdot_{1ij}\Ainv{12,\iCOMMAj}               \big)$

\end{itemize}
\end{itemize}
\item[] Output: $\Big(\xveco,\AUoo,\big\{\big(\xvectCi,\AUttCi,\AUotCi):\ 1\le i\le m\big\}\Big)$\\
$\null\qquad\qquad\big\{\big(\bx_{2,ij},\bA^{22,ij},\bA^{12,ij},\bA^{12,\iCOMMAj}
\big):\ 1\le i\le m,\ 1\le j\le n_i\big\}\Big)$
\end{itemize}
\end{small}
\end{minipage}
\end{center}
\caption{\SolveThreeLevelSparseLeastSquares\ \textit{for solving the three-level sparse matrix
least squares problem: minimise $\Vert\bb-\bB\,\bx\Vert^2$ in $\bx$ and sub-blocks of $\bA^{-1}$
corresponding to the non-zero sub-blocks of $\bA=\bB^T\bB$. The sub-block notation is
given by (\ref{eq:catweasleOne}), (\ref{eq:catweasleTwo}) and (\ref{eq:catweasleThree}).}}
\label{alg:SolveThreeLevelSparseLeastSquares} 
\end{algorithm}

\end{document}